\documentclass[longauth]{aa}       
\usepackage{graphicx}
\usepackage{txfonts}
\usepackage{multirow}
\usepackage{longtable}
\usepackage{lscape}
\usepackage[]{natbib}
\usepackage{supertabular}
\bibpunct{(}{)}{;}{a}{}{,}

\begin{document} 
  \title{X-shooter and ALMA spectroscopy of GRB\,161023A}
  \subtitle{A study of metals and molecules in the line of sight towards a luminous GRB\thanks{Based on observations collected at the European Organisation for Astronomical Research in the Southern Hemisphere under ESO programmes 098.A-0055, 098.D-0710 and 0100.D-0649. This paper makes use of the following ALMA data: ADS/JAO.ALMA\#2016.1.00862.T. ALMA is a partnership of ESO (representing its member states), NSF (USA) and NINS (Japan), together with NRC (Canada) and NSC and ASIAA (Taiwan) and KASI (Republic of Korea), in cooperation with the Republic of Chile. The Joint ALMA Observatory is operated by ESO, AUI/NRAO and NAOJ. This work is based in part on observations made with the Spitzer Space Telescope, which is operated by the Jet Propulsion Laboratory, California Institute of Technology under a contract with NASA. Support for this work was provided by NASA through an award issued by JPL/Caltech.}\thanks{Tables A.1 and A.2 are only available in electronic form at the CDS via anonymous ftp to cdsarc.u-strasbg.fr (130.79.128.5) or via http://cdsweb.u-strasbg.fr/cgi-bin/qcat?J/A+A/}}

  \author{A. de Ugarte Postigo\inst{\ref{af:iaa},\ref{af:dark}}
     \and
           C.~C.~Th\"one\inst{\ref{af:iaa}}
     \and 
           J.~Bolmer\inst{\ref{af:esochile}, \ref{af:garching}}
     \and
           S.~Schulze\inst{\ref{af:weizmann}}
     \and
           S.~Mart\'in\inst{\ref{af:esochile},\ref{af:alma}}
     \and  
           D.~A.~Kann\inst{\ref{af:iaa}}
     \and
           V.~D'Elia\inst{\ref{af:inafoar},\ref{af:asi}}
     \and
           J.~Selsing\inst{\ref{af:dark}}
     \and  
           A.~Martin-Carrillo\inst{\ref{af:ucd}}
     \and
           D.~A.~Perley\inst{\ref{af:liverpool}} 
     \and       
           S.~Kim\inst{\ref{af:uc}}  
     \and
           L.~Izzo\inst{\ref{af:iaa}} 
     \and
           R.~S\'anchez-Ram\'irez\inst{\ref{af:inafiaps}}
     \and
           C.~Guidorzi\inst{\ref{af:ferrara}} 
     \and
           A.~Klotz\inst{\ref{af:cnrs}} 
     \and
           K.~Wiersema\inst{\ref{af:warwick},\ref{af:leicester}} 
     \and
           F.~E.~Bauer\inst{\ref{af:uc},\ref{af:mil},\ref{af:sci}}
     \and
           K.~Bensch\inst{\ref{af:iaa}} 
     \and
           S.~Campana\inst{\ref{af:oab}}
     \and
           Z.~Cano\inst{\ref{af:iaa}} 
     \and
           S.~Covino\inst{\ref{af:oab}}
     \and
           D.~Coward\inst{\ref{af:uwa}} 
     \and
           A.~De~Cia\inst{\ref{af:esogermany}} %
     \and
           I.~de~Gregorio-Monsalvo\inst{\ref{af:esochile},\ref{af:alma}} 
     \and  
           M.~De~Pasquale\inst{\ref{af:tur}} 
     \and
           J.~P.~U.~Fynbo\inst{\ref{af:dark},\ref{af:dawn}} 
     \and
           J.~Greiner\inst{\ref{af:garching}} 
     \and
           A.~Gomboc\inst{\ref{af:slov}} 
     \and
           L.~Hanlon\inst{\ref{af:ucd}} 
     \and 
           M.~Hansen\inst{\ref{af:dark}} 
     \and
           D.~H.~Hartmann\inst{\ref{af:clemson}}
     \and
           K.~E.~Heintz\inst{\ref{af:iceland},\ref{af:dawn}}
     \and
           P.~Jakobsson\inst{\ref{af:iceland}} 
     \and
           S.~Kobayashi\inst{\ref{af:liverpool}} 
     \and
           D.~B.~Malesani\inst{\ref{af:dark},\ref{af:dawn}} 
     \and
           R.~Martone\inst{\ref{af:ferrara}} 
     \and
           P.~J.~Meintjes\inst{\ref{af:sa}} 
     \and
           M.~J.~Micha\l owski\inst{\ref{af:poznan}}
     \and
           C.~G.~Mundell\inst{\ref{af:bath}} 
     \and
           D.~Murphy\inst{\ref{af:ucd}} 
     \and
           S.~Oates\inst{\ref{af:warwick}} 
     \and
           L.~Salmon\inst{\ref{af:ucd}}  
     \and
           B.~van~Soelen\inst{\ref{af:sa}} 
     \and
           N.~R.~Tanvir\inst{\ref{af:leicester}} 
     \and
           D.~Turpin\inst{\ref{af:cnrs}} 
     \and
           D.~Xu\inst{\ref{af:xu}} 
     \and
           T.~Zafar\inst{\ref{af:aao}} 
          }

  \institute{Instituto de Astrof\'isica de Andaluc\'ia (IAA-CSIC), Glorieta de la Astronomía, s/n, E18008 Granada, Spain\\
              \email{deugarte@iaa.es}\label{af:iaa}
         \and
             Dark Cosmology Centre, Niels Bohr Institute, University of Copenhagen, 2100 Copenhagen \O, Denmark\label{af:dark}
         \and 
             European Southern Observatory, Alonso de C\'{o}rdova 3107, Vitacura, Casilla 19001, Santiago 19, Chile\label{af:esochile}
         \and
             Max-Planck-Institut f\"ur extraterrestrische Physik, Giessenbachstraße, 85748 Garching, Germany\label{af:garching}
         \and
             Department of Particle Physics and Astrophysics, Weizmann Institute of Science, 234 Herzl Street, Rehovot, 761000, Israel\label{af:weizmann}
         \and
             Joint ALMA Observatory, Alonso de C\'ordova 3107, Vitacura 763 0355, Santiago, Chile\label{af:alma}
         \and
             INAF – Osservatorio Astronomico di Roma, Via Frascati 33, I-00040 Monteporzio Catone, Italy\label{af:inafoar}
         \and
             ASI-Space Science Data Centre, Via del Politecnico snc, I-00133 Rome, Italy\label{af:asi}
         \and
             Space Science Group, School of Physics, University College Dublin, Belfield, Dublin 4, Ireland\label{af:ucd}
         \and
             Astrophysics Research Institute, Liverpool John Moores University, IC2, Liverpool Science Park, 146 Brownlow Hill, Liverpool L3 5RF, UK\label{af:liverpool}
         \and
             Instituto de Astrof\'isica y Centro de Astroingenier\'ia, Facultad de F\'isica, Pontificia Universidad Cat\'olica de Chile, Casilla 306, Santiago 22, Chile\label{af:uc}
         \and
             INAF, Istituto Astrofisica e Planetologia Spaziali, Via Fosso del Cavaliere 100, I-00133 Roma, Italy\label{af:inafiaps}
         \and
             Department of Physics and Earth Science, University of Ferrara, via Saragat 1, I-44122 Ferrara, Italy\label{af:ferrara}
         \and
              IRAP, Universit\'e de Toulouse, CNRS, UPS, CNES, Toulouse, France\label{af:cnrs}
         \and
              Department of Physics, University of Warwick, Coventry, CV4 7AL, UK\label{af:warwick}
         \and
             Department of Physics and Astronomy, University of Leicester, University Road, Leicester, LE1 7RH, UK\label{af:leicester}
         \and
             Millennium Institute of Astrophysics (MAS), Nuncio Monse{\~{n}}or S{\'{o}}tero Sanz 100, Providencia, Santiago, Chile\label{af:mil}
         \and
             Space Science Institute, 4750 Walnut Street, Suite 205, Boulder, Colorado 80301, U.S.A.\label{af:sci}
         \and
             INAF - Osservatorio Astronomico di Brera, via Bianchi 46, 23807, Merate (LC), Italy\label{af:oab}
         \and
             School of Physics, University of Western Australia, M013, Crawley, WA 6009, Australia\label{af:uwa}
         \and
             European Southern Observatory, Karl-Schwarzschild Str. 2, 85748 Garching bei M\"unchen, Germany\label{af:esogermany}
         \and
             Department of Astronomy and Space Sciences, Istanbul University, 34119 Beyazıt, Istanbul, Turkey\label{af:tur}
         \and
             Cosmic Dawn Center, Niels Bohr Institute, University of Copenhagen, Juliane Maries Vej 30, 2100 Copenhagen \O, Denmark\label{af:dawn}
         \and
             Centre for Astrophysics and Cosmology, University of Nova Gorica, Vipavska 11c, Ajdov\v s\v cina 5270, Slovenia\label{af:slov}
         \and
             Department of Physics and Astronomy, Clemson University, Clemson, SC 29634, USA\label{af:clemson}
         \and
             Centre for Astrophysics and Cosmology, Science Institute, University of Iceland, Dunhagi 5, 107 Reykjav\'ik, Iceland\label{af:iceland}
         \and
             Department of Physics, University of the Free State, 9300, Bloemfontein, South Africa\label{af:sa}
         \and
             Astronomical Observatory Institute, Faculty of Physics, Adam Mickiewicz University, ul.~S{\l}oneczna 36, 60-286 Pozna{\'n}, Poland\label{af:poznan}
         \and
             Department of Physics, University of Bath, Claverton Down, Bath, BA2 7AY, UK\label{af:bath}
         \and
             Indian Institute of Space Science \& Technology, Trivandrum 695547, India\label{af:india}
         \and
             CAS Key Laboratory of Space Astronomy and Technology, National Astronomical Observatories, Chinese  Academy of Sciences, Beijing, 100012, P.R. China\label{af:xu}
         \and
             Australian Astronomical Observatory, PO Box 915, North Ryde, NSW 1670, Australia\label{af:aao}
             }

  \date{Received 25 March 2018; accepted 21 July 2018}

  \abstract
   {Long gamma-ray bursts (GRBs) are produced during the dramatic deaths of massive stars with very short lifetimes, meaning that they explode close to the birth place of their progenitors. Over a short period they become the most luminous objects observable in the Universe, being perfect beacons to study high-redshift star-forming regions.}
   {We aim to use the afterglow of GRB\,161023A at a redshift $z=2.710$ as a background source to study the environment of the explosion and the intervening systems along its line of sight.}
   {For the first time, we complement ultraviolet (UV), optical and near-infrared (NIR) spectroscopy with millimetre spectroscopy using the Atacama Large Millimeter Array (ALMA), which allows us to probe the molecular content of the host galaxy. The X-shooter spectrum shows a plethora of absorption features including fine-structure and metastable transitions of Fe, Ni, Si, C, and O. We present photometry ranging from 43 s to over 500 days after the burst.}
   {We infer a host-galaxy metallicity of [Zn/H] $=-1.11\pm0.07$, which, corrected for dust depletion, results in [X/H] $=-0.94\pm0.08$. 
   We do not detect molecular features in the ALMA data, but we derive limits on the molecular content of $log(N_{CO}/cm^{-2})<15.7$ and $log(N_{HCO+}/cm^{-2})<13.2$, which are consistent with those that we obtain from the optical spectra, $log(N_{H_2}/cm^{-2})<15.2$ and $log(N_{CO}/cm^{-2})<14.5$. Within the host galaxy, we detect three velocity systems through UV, optical andNIR absorption spectroscopy, all with levels that were excited by the GRB afterglow. We determine the distance from these systems to the GRB to be in the range between 0.7 and 1.0 kpc. The sight line to GRB\,161023A shows nine independent intervening systems, most of them with multiple components.}
   {Although no molecular absorption was detected for GRB\,161023A, we show that GRB millimetre spectroscopy is now feasible and is opening a new window on the study of molecular gas within star-forming galaxies at all redshifts. The most favoured lines of sight for this purpose will be those with high metallicity and dust.}
   
\keywords{gamma ray bursts: individual: GRB161023A; Techniques: spectroscopic; ISM: abundances; ISM: molecules}

\maketitle


\section{Introduction}

Long-duration gamma-ray bursts (GRBs) are produced by the deaths of very massive stars. As such, the lives of their progenitors are short and these are expected to explode within or near the star-forming regions in which they were formed. Their early emission outshines the rest of the Universe in gamma-rays for a few seconds. In optical they can also reach extreme luminosities during the first seconds, having reached a record absolute magnitude of -38.7 mag (in the rest-frame $U$-band; \citealt{blo09}). Their afterglows are produced as material that has been ejected through jets at ultra-relativistic velocities collides with the interstellar medium, decelerating and emitting synchrotron radiation. This long-lasting emission can be usually detected at optical wavelengths for several days after the burst onset, and for weeks or even months at radio frequencies.

   \begin{figure*}
   \centering
   \includegraphics[width=15cm]{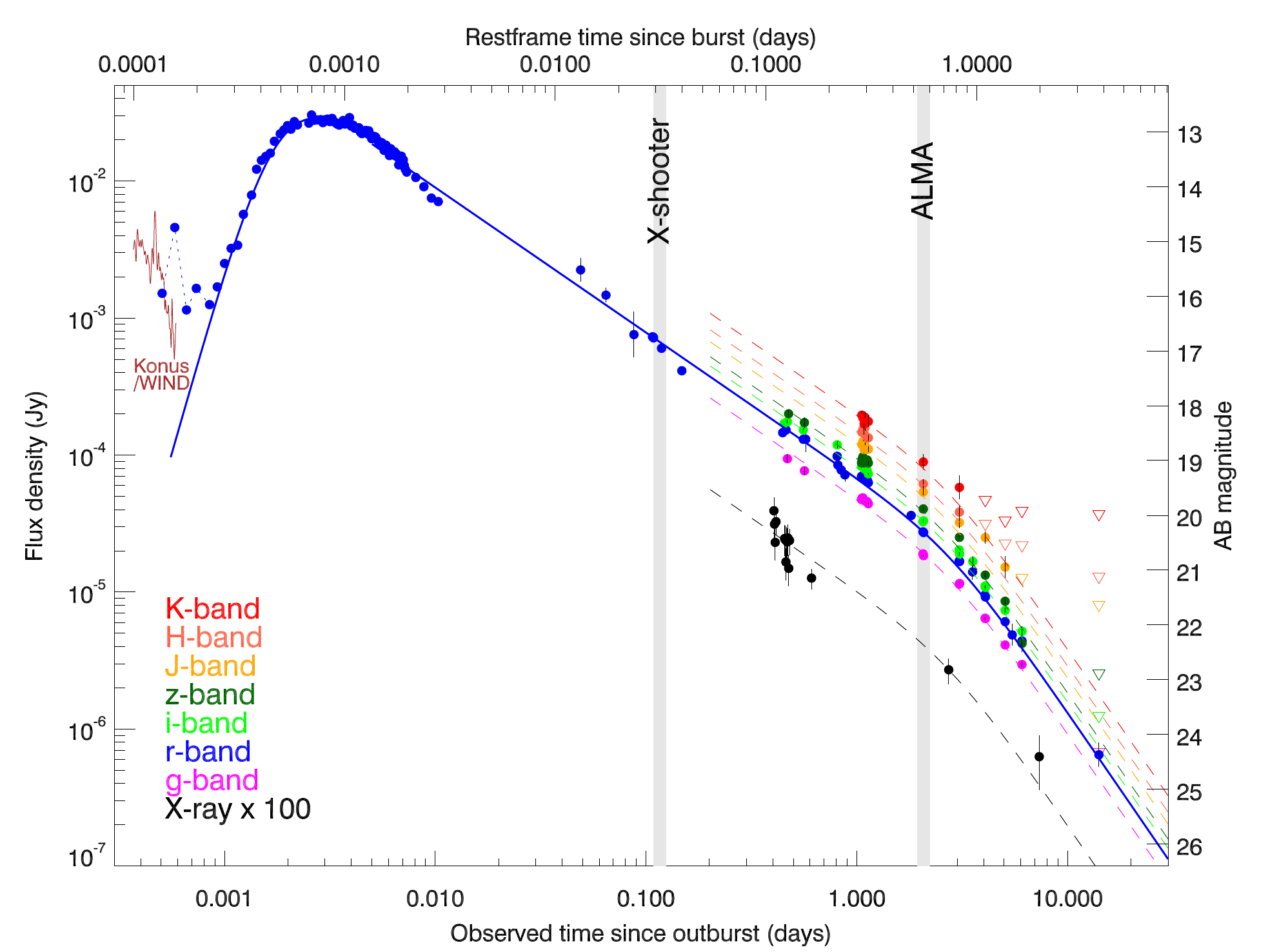}
      \caption{Light curve of GRB\,161023A in optical and near-infrared bands. Downward pointing triangles are 3-$\sigma$ upper limits. The grey vertical lines mark the times of the X-shooter and ALMA observations. The solid and dashed lines show the fits of the photometric data to broken power laws, as explained in Sect.~\ref{sect:SED}. Plotted in brown is the tail of the gamma-ray emission detected by Konus-{\it Wind}.
              }
         \label{Fig:lc}
   \end{figure*}

GRBs are consequently not only interesting sources for the study of some of the most extreme physics in the Universe, but can also be used as extremely luminous beacons that shine from the heart of star-forming regions. These beacons of light allow us to probe the region in which the GRB formed, its host galaxy, the intergalactic medium, and any intervening systems along the line of sight \citep[e.g.][]{pol09}. In fact, they are one of the only means of studying star-forming galaxies at very high redshift from the inside, independently of their size and luminosity \citep{fyn08}. However, when studying the closest environment of the GRB, we have to consider that the intense radiation released during the GRB explosion has a detectable effect on the surroundings of the GRB. Afterglow emission has been shown to excite spectral lines up to distances of the order of 1 kpc within the host galaxy (e.g. \citealt{vre07}, \citealt{del09}).

The most significant drawback when studying GRBs and using them as light beacons is that they also fade very rapidly. This implies that, even using the largest telescopes, we need to have a very fast reaction time and use instrumentation that allows us to observe a broad spectral range simultaneously. This is the reason why instruments like the Gamma-ray Burst Optical/Near-infrared Detector \citep[GROND;][]{gre08}, with its seven-band simultaneous imaging capability, or X-shooter, capable of obtaining spectroscopy from 3000 to 24800 {\AA} in a single shot, were built. To observe the very early light curve and probe the phase when the gamma-ray emission is still ongoing, even faster approaches, such as robotic observatories \citep[i.e.][]{fre04,boe99}, are needed.

The advances in millimetre and submillimetre instrumentation are giving us a new way of observing astronomical sources. For GRB studies, these wavelengths have the advantage that the peak emission is normally reached within hours to days after the burst, unlike the gamma-ray emission that peaks within seconds or the optical emission that reaches maximum within tens of seconds. In particular the Atacama Large Millimeter Array (ALMA) is now capable of responding to target-of-opportunity alerts with reaction times of the order of a day and with unprecedented sensitivities. These wavelengths are interesting because they normally cover the location of the peak synchrotron frequency during the first day and also cover some of the most prominent molecular transitions.

In this paper we present imaging and spectroscopy of the afterglow of GRB\,161023A, including for the first time millimetre spectroscopy in search of several molecular features. This burst, at a redshift of $z=2.710$, is amongst the group of most luminous afterglows detected to date, in both the optical and millimetre wavelengths. 

Section 2 presents the observations and some details on the data reduction of the different data sets. Section 3 presents the results derived from these observations. In Sect. 4 we discuss these results and place them into the context of previous observations. Finally, in Sect. 5 we provide our conclusions. Throughout the paper we use the convention where F$_{\nu}\propto t^{\alpha}\nu^{\beta}$, and a cosmology with H$_0=71$ km s$^{-1}$ Mpc$^{-1}$, $\Omega_M=0.3$, and $\Omega_V=0.7$ \citep{spe03}.


\section{Observations and data reduction}

\subsection{High-energy observations}

GRB\,161023A was first identified by {\it INTEGRAL} (INTErnational Gamma-Ray Astrophysics Laboratory) as a long gamma-ray burst that triggered the IBIS (Imager on-Board the INTEGRAL Satellite) detector at 22:38:40 UT on 23 October 2016 \citep{mer16}, which we use as the trigger reference time T$_0$ throughout this paper. The GRB had a T$_{90}$ duration of $\sim80$ s in the IBIS observations. Only lower limits on the fluence of the burst were given using the IBIS data, as the brightness of the event resulted in saturation of the detectors.
The burst was also observed by Konus-{\it Wind} at 22:39:12.321 UT (the later detection with respect to {\it INTEGRAL} is due to the lower sensitivity of Konus with respect to IBIS) which measured a duration of $\sim 50$ s. The burst had a fluence of $(4.2 \pm 0.6)\times10^{-5}$ erg cm$^{-2}$ in the 20 keV - 10 MeV observer-frame energy range and a peak energy of E$_{{\rm p}}$ $= 163_{-28}^{+37}$ keV \citep{fre16}. Finally, the burst was also detected by {\it AstroSat} CZTI (Cadmium Zinc Telluride) in the 40-200 keV range with a T$_{90}$ duration of 44 s \citep{sha16}. The light curve showed a single peak at 22:39:10.00 UT.

Using the Konus-{\it Wind} fluence and spectral fit, together with the redshift of $z=2.710$ (\citealt{tan16}, refined in this paper), we determine the isotropic-equivalent released energy in the 1 keV - 10 MeV rest-frame band to be E$_{{\rm iso,\gamma}}=(6.3\pm0.9)\times10^{53}$ erg, which is toward the upper end of the typical energy release of GRBs \citep{but07,ama09,fer13}. We also use the redshift to transform the peak energy to rest frame, E$_{{\rm p, rest}}=605_{-104}^{+137}$ keV. These values lie well within the expected values for the E$_{{\rm p}}$ - E$_{{\rm iso}}$ correlation for long GRBs \citep{ama02}.

Follow-up observations of the burst were performed by the Neil Gehrels {\it Swift} Observatory \citep{geh04}, which detected an X-ray counterpart starting 34.8 ks after the trigger \citep{dai16}. The analysis of the X-ray afterglow data \citep{eva09} showed a spectral slope of $\beta=-1.05_{-0.18}^{+0.26}$, and a decay with a slope of $\alpha=-1.23_{-0.21}^{+0.16}$ (considering a single power-law decay). The intrinsic equivalent hydrogen column density derived from the X-ray attenuation is consistent with the Galactic value of N$_H=3.18\times10^{20}{\,\rm cm}^{-2}$, allowing us to only derive a 3-$\sigma$ upper limit at the redshift of the host galaxy of N$_H<3.0\times10^{22}{\,\rm cm}^{-2}$.

\subsection{Optical and infrared photometry}

A bright optical counterpart to GRB\,161023A was quickly identified by the MASTER II (Mobile Astronomical System of TElescope Robots) robotic telescope \citep{gor16}, starting at 22:39:24 UT (T$_0+44$ s). The afterglow brightness reached a maximum $\sim3$ min after the burst, with a magnitude of $\sim12.5$. Early follow-up was also obtained by the Watcher robotic telescope which started observing 43 s after the trigger \citep{mar16}, using 5 s images for the first ten minutes and then switching to 60 s exposures. The evolution as seen by Watcher also had an early rise in brightness, reaching a maximum 200 s after the burst at $r^{\prime}=12.8$ mag, after which it decayed to $r^{\prime}=14.4$ mag 893 s after the burst. Later observations were reported by TAROT (T\'elescopes \`a Action Rapide pour les Objets Transitoires), ranging from 64 to 143 min after the burst \citep{klo16b}, the Faulkes Telescope South 10.6 hr after the burst \citep{gui16}, the Zadko telescope 0.55 days after the GRB \citep{klo16a}, and the GROND detector \citep{gre08} mounted on the 2.2 m MPG (Max-Planck-Gesellschaft) telescope, at ESO's (European Southern Observatory) La Silla observatory, Chile, for which the first epoch, starting 25.2 hr after the burst, was reported by \citet{kru16}. 

In this paper we present the observations performed with the 0.4 m Watcher telescope \citep{fre04}, the 0.25 m TAROT La Silla telescope, the 2.0 m Faulkes South telescope, the 1.0 m Zadko Gingin telescope, and the 2.2 m MPG telescope with GROND. Our photometric data set also includes an $r^{\prime}$-band observation obtained during the acquisition of the X-shooter spectrum at ESO's Very Large Telescope (VLT). We have added observations obtained using the FORS2 (FOcal Reducer/low dispersion Spectrograph 2) instrument at the VLT as part of a different programme (Wiersema et al. in prep.). Finally, we include a deep observation obtained with FORS2 in March 2018 in search of the host galaxy, which remains undetected down to a strong 3-$\sigma$ limit of 26.5 mag. The photometry of all these data was obtained based on Sloan magnitudes of secondary standard stars based on GROND observations of SDSS (Sloan Digital Sky Survey) fields. In the case of FORS2 the observations performed with the $R_{\rm special}$ filter were calibrated directly with the Sloan $r^{\prime}$. The photometric data are presented in Table~\ref{tab:phot} and shown in Fig.~\ref{Fig:lc}.

The field of GRB\,161023A was also observed at late times with the Infrared Array Camera \citep[IRAC;][]{faz04} on the {\it Spitzer} Space Telescope \citep{wer04}, as part of the extended {\it Swift} GRB HOst gAlaxy Legacy Survey \citep[SHOALS;][]{per16a}.  We acquired 108 dithered images of 100 s each for a total exposure time of 3.0 hours, all using the 3.6 $\mu$m filter (channel 1). The post-basic calibrated data image was downloaded from the {\it Spitzer} legacy archive; no source is detected at the position of the GRB to a limit of >25.3 mag.

\addtocounter{table}{1}

\subsection{X-shooter spectroscopy}

The X-shooter spectrograph \citep{ver11}, installed on ESO's Very Large Telescope (Chile) observed the optical counterpart of GRB\,161023A at a mean epoch 2.795 hrs after the burst.The observation consisted of $2\times600$ s exposures covering the spectral range between 3000 and 24800 {\AA}.

The spectral resolution is slightly different in the three arms. Furthermore, when the seeing is smaller than the slit width, the actual resolution is better than the nominal resolution for the given arm and slit. We determine the real spectral resolution following the procedure outlined in \cite{sel18} and find resolutions of 38.5 km s$^{-1}$, 31.4 km s$^{-1}$, and 35.5 km s$^{-1}$ for the UVB (UV-Blue), VIS (Visible), and NIR (near-IR) arms, respectively.

\begin{figure*}
   \centering
   \includegraphics[width=17cm]{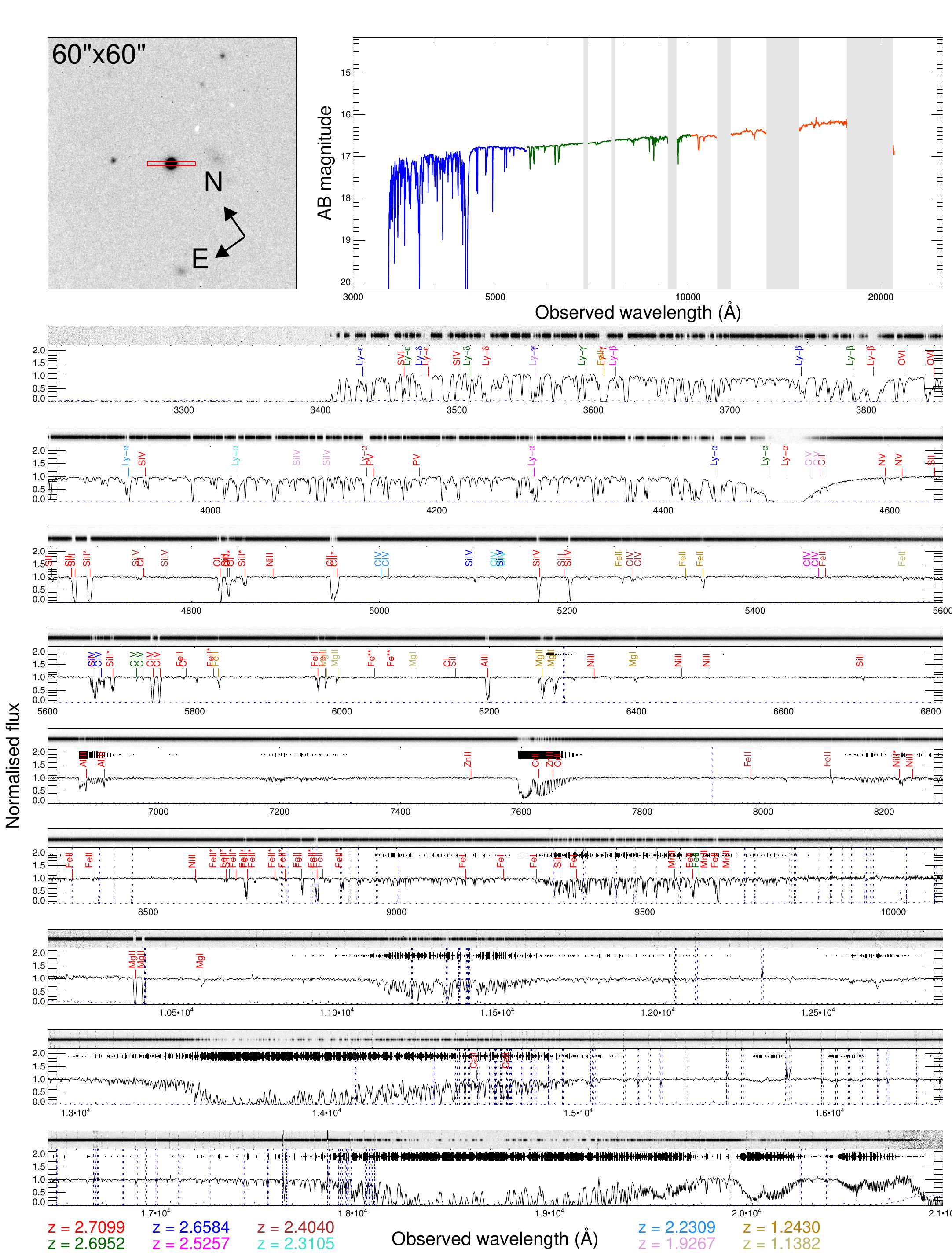}
      \caption{X-shooter spectrum of GRB\,161023A. Top left: Acquisition image indicating the position of the slit. Top right: Flux-calibrated spectrum. Each colour represents one of the X-shooter arms. Horizontal gray bars mask the regions affected by strong telluric absorption bands. Below: Normalised spectrum together with the 2D spectrum. Each of the different absorption systems are marked in a different colour, as indicated at the bottom. At the top of the 1D spectra we mark the telluric features with black lines, their widths being proportional to their intensity, as done in \cite{deu10}.}
         \label{fig:xs_spec}
   \end{figure*}

In the X-shooter spectrum we identify numerous absorption features (see Table~\ref{tab:ew} and Fig.~\ref{fig:xs_spec}) including the hydrogen Lyman series down to the Lyman limit, \ion{S}{II}, \ion{Si}{II}, \ion{Si}{IV}, \ion{C}{IV}, \ion{Al}{II}, \ion{Fe}{II}, \ion{Ni}{II}, \ion{Zn}{II}, \ion{Mg}{II}, \ion{Mg}{I}, \ion{Mn}{II} and \ion{O}{I}, many of which have been reported by \citet{tan16}. Several elements also show fine-structure lines and metastable transitions, at a redshift of $z=2.710$, which are expected to be excited by the radiation of the GRB, confirming that this is indeed the redshift of the GRB. The absorption features also show several velocity components, as will be discussed in the following section. In addition, the spectrum shows multiple intervening systems along the line of sight ranging from $\sim$1200 km\,s$^{-1}$ bluewards of the GRB redshift, down to $z=1.14$ (see Sect. \ref{sec:int}). This data set is available through the GRBSpec database\footnote{http://grbspec.iaa.es} \citep{deu14}.

\subsection{ALMA observations}

Millimetre observations were carried out in ALMA Bands 3 and 7 two days after the burst, from 25 October 2016 23:03:11.0 to 26 October 2016 00:22:33.3 UT, and from 26 October 2016 00:44:44.1 to 26 October 2016 01:42:38.4 UT, respectively. The configuration used 38 array elements with baselines ranging from 18~m to 1.2~km, equivalent to 5-374~$k\lambda$ (Band 3) and 21-1300~$k\lambda$ (Band 7). Integration time on source was 51 minutes (with PWV~1.0 mm) and 38 minutes (with PWV~0.74 mm), respectively.
In Band 3, the correlator was set up in two low-resolution (15.6 MHz bins) 1.875~GHz wide spectral windows centred at $\sim$106~GHz and $\sim$108~GHz, and three high-resolution (one 0.122~MHz and two 0.244~MHz bins) 468.75~MHz wide spectral windows centred at 93.5, 96.4 and 96.8~GHz, respectively. In Band 7, the correlator was set up in four low-resolution (15.6~MHz) 1.875~GHz windows centred at 337.5, 339.4, 347.5, and 349.5~GHz.

   \begin{figure}[t!]
   \centering
   \includegraphics[width=6.5cm]{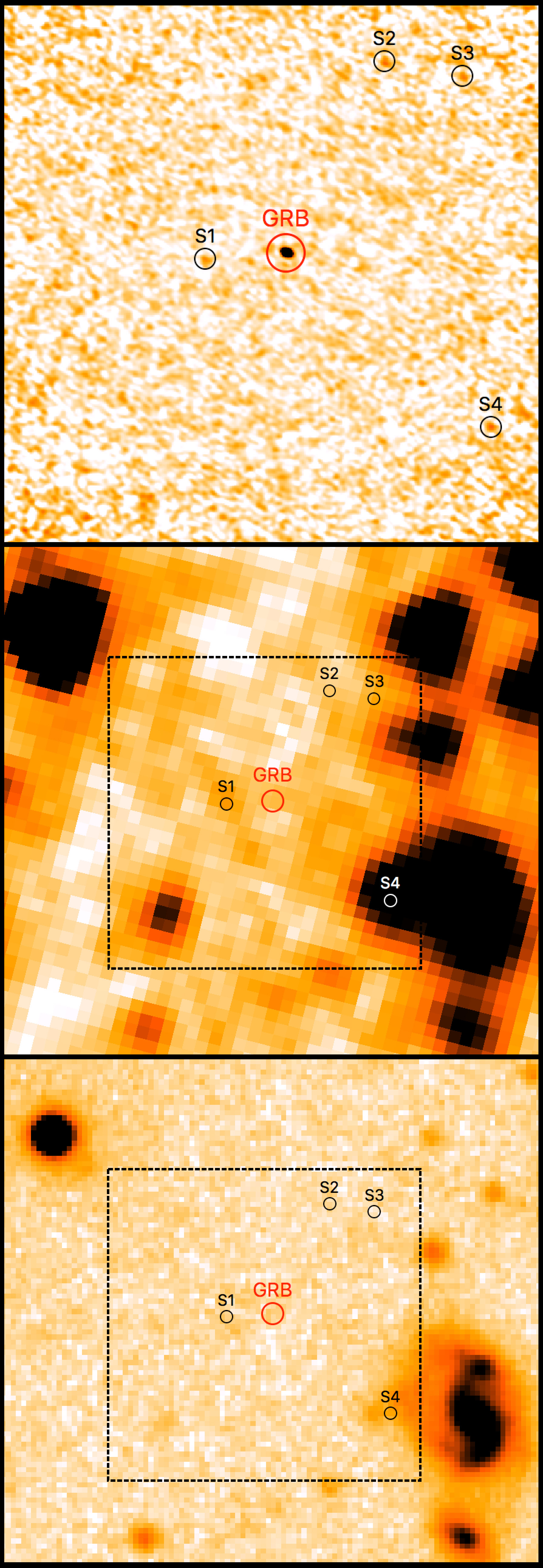}
      \caption{Top: Continuum image obtained with ALMA at 350 GHz. The GRB afterglow is clearly seen in the centre of the image. Several other faint objects can be seen within the field-of-view, which measures $15^{\prime\prime}\times15^{\prime\prime}$. North is up and East is to the left. Centre: {\it Spitzer} image of the field in the 3.6~$\mu$m band, showing a slightly larger region. The box indicates the area of the upper figure. Sources S1 and S4 are detected. Bottom: Late deep image obtained with FORS2 at VLT in the $R_{\rm special}$ band.  Whereas ALMA sources S2, and S3 have no corresponding optical counterparts, S4 could be part of a nearby spiral galaxy to the South West and S1 is marginally detected.
              }
         \label{FigALMA}
   \end{figure}
   
The data were calibrated with CASA (Common Astronomy Software Applications, version 4.7, \citealt{mcm07}) using the calibration script provided in the ALMA archive, and subsequently images were produced by using the combined emission in each spectral window. The resolution achieved using  natural weighting was
$0\farcs77\times0\farcs63$
(P.A. 50$^\circ$) and 
$0\farcs24\times0\farcs17$
(P.A. 70$^\circ$) for Bands 3 and 7, respectively.
Moreover, data cubes were produced for the individual spectral windows that were imaged at the original resolution for the low-resolution windows and binned to 12~MHz for the high-resolution windows.
The flux calibration was based on observations of the quasars J1924-2914 and J2056-4714 for Bands 3 and 7, respectively. In Table~\ref{table:alma} we show the flux measurements obtained for the afterglow of GRB\,161023A in each of the different spectral windows.

\begin{table}[t!]
\caption{Flux measurements for the afterglow of GRB\,161023A.}            
\label{table:alma}     
\centering                       
\begin{tabular}{c c c c c}       
\hline\hline
T-T$_0$ & Frequency & ALMA  & Band Width    & F$_{\nu}$         \\
(day)   & (GHz)     & Band  & (GHz)         & (mJy)             \\
\hline 
2.045   & 93.24041  & 3     & 0.469         & 2.87$\pm$0.07   \\
2.045   & 96.20021  & 3     & 0.469         & 2.83$\pm$0.07   \\
2.045   & 96.59022  & 3     & 0.469         & 2.96$\pm$0.07   \\
2.045   & 105.9892  & 3     & 1.875         & 3.00$\pm$0.07   \\
2.045   & 107.9890  & 3     & 1.875         & 3.04$\pm$0.04   \\
2.108   & 336.4946  & 7     & 1.875         & 2.24$\pm$0.12     \\
2.108   & 338.4321  & 7     & 1.875         & 2.01$\pm$0.11     \\
2.108   & 348.4946  & 7     & 1.875         & 1.88$\pm$0.11     \\
2.108   & 350.4946  & 7     & 1.875         & 2.09$\pm$0.12     \\
\hline                                
\end{tabular}
\end{table}

From the ALMA observations we derive the most precise coordinates (J2000) for the afterglow of GRB\,161023A:\\\\
RA: 20:44:05.1718 ($\pm0.\!\!^{\rm s}0003$)\\
Dec.: $-$47:39:47.921 ($\pm0\farcs005$)\\

In the 350 GHz (Band 7) image obtained by ALMA, we identify four additional tentative sources detected at $\gtrsim3\sigma$, as shown in Fig.~\ref{FigALMA}. The coordinates and flux densities of these sources are given in Table~\ref{table:sources}. In the deep {\it Spitzer} image, shown in the middle panel, sources S2 and S3 are not detected. However, S4 is part of a bright extended galaxy and S1 is also detected at $24.9\pm0.3$ mag. The host galaxy of GRB\,161023A is not detected down to a $3-\sigma$ limit of $>25.3$ mag at 3.6 $\mu$m. Within the deep FORS2 image (bottom panel), we do not detect any optical counterparts to sources S2 or S3. As in the {\it Spitzer} image, S4 is detected in the outskirts of a bright spiral galaxy. S1 is faintly detected at a magnitude of $r^{\prime}=26.27\pm0.24$. The 3-$\sigma$ limit for the host galaxy is $r^{\prime}>26.5$ mag.

\begin{table}[t!]
\caption{Sources tentatively detected in the neighbourhood of GRB\,161023A in the 350 GHz band, with typical positional errors of 0\farcs03.}            
\label{table:sources}     
\centering                       
\begin{tabular}{c c c c}       
\hline\hline
Object  & R.A.          & Dec.              & F$_{\nu}$ ($\mu Jy$)  \\
\hline 
S1      & 20:44:05.398  & -47:39:48.15      & $190\pm70$            \\
S2      & 20:44:04.892  & -47:39:42.55      & $320\pm70$            \\
S3      & 20:44:04.676  & -47:39:42.95      & $230\pm70$            \\
S4      & 20:44:04.598  & -47:39:52.88      & $260\pm80$            \\
\hline                                
\end{tabular}
\end{table}

\section{Results}

\subsection{Spectral energy distribution and light curves in the context of the fireball model}
\label{sect:SED}

The standard fireball model \citep{sar98,sar99} describes the temporal evolution of the afterglow light curve as a series of power laws. Similarly, the spectra of the afterglow are expected to be synchrotron radiation, which is described by several spectral power laws that meet at characteristic break frequencies: $\nu_m$ is the typical frequency of the electrons, $\nu_a$ the self-absorption frequency, and $\nu_c$ is the cooling frequency. Depending on several parameters that characterise the properties of the ejecta and the media with which the shock interacts, we obtain different spectral and temporal slopes for the different observed bands. The most relevant parameters that define these slopes are the electron index $p$, which describes the distribution of shock-accelerated electrons in the circumburst medium, and the density profile of the external medium with which the ejecta interacts. The density profile is often found to be uniform $\propto r^0$ (as for an interstellar medium), but in some cases the data have been seen to be fit better with a wind medium, $\propto r^{-2}$. This standard fireball is often complemented with the presence of a reverse-shock component at early times, with a sharper temporal rise and decay than expected for the forward shock of the fireball \citep{sar99b,kob00a,kob00b}.

To make a rough analysis in the context of this model we perform temporal and spectral fits of our data. In this analysis we ignore the first four Watcher $r'$-band datapoints, which are likely linked to the tail of the prompt gamma-ray emission, and behave in a similar way to that of the afterglow of GRB\,080603A \citep{Gui11}. We then fit the early rise and subsequent decay with a smoothly broken power law:


\begin{equation}
F_k(t)=m_k\left(\left(\frac{t_b}{t}\right)^{\alpha_1n}+\left(\frac{t_b}{t}\right)^{\alpha_2n}\right)^{-\frac{1}{n}}
\end{equation}

Where $m_k$ is the normalisation for each band (left as a free and non-shared parameter), $\alpha_{1,2}$ are the temporal slopes, $t_b$ is the break time, and $n$ is the smoothness of the break (see e.g. \citealt{zeh06} for more details). While for the early fit (between 0.001 and 0.1 d) we only use the $r^\prime$ band, as we have no other colour information, the ``normal'' decay (after 0.1 d) and the late steep decay are fit in all bands with another smoothly broken power law, where we share all parameters in the different bands, except for the different normalisations. Such a fit takes the maximum amount of available data into account, leading to a more tightly constrained Spectral Energy Distribution (SED). We note that the ``normal'' decay slope result is identical in both fits. From these fits, we derive the following parameters for the optical light curve (see Fig.~\ref{Fig:lc}): the initial rise to peak $\alpha_{rise}=5.44\pm0.13$, the break from rise to decay $t_{b,1}=0.00181\pm0.00002$ days ($156\pm2$ s), the ``normal'' decay slope $\alpha_1=-1.063\pm0.002$, the break from normal to steep decay $t_{b,2}=2.81\pm0.30$ days, and the post-break decay slope $\alpha_2=-2.25\pm0.15$. The break smoothness for the peak was left as a parameter free to vary and we find $n_1=0.605\pm0.025$, a very smooth rollover. The second break is also not sharp, we find $n_2=2.41\pm0.60$. As we find no evidence for a host-galaxy contribution to the light curve (especially considering the late deep limit from VLT/FORS2), we do not include a host value in the fit.

   \begin{figure}
   \includegraphics[width=\hsize]{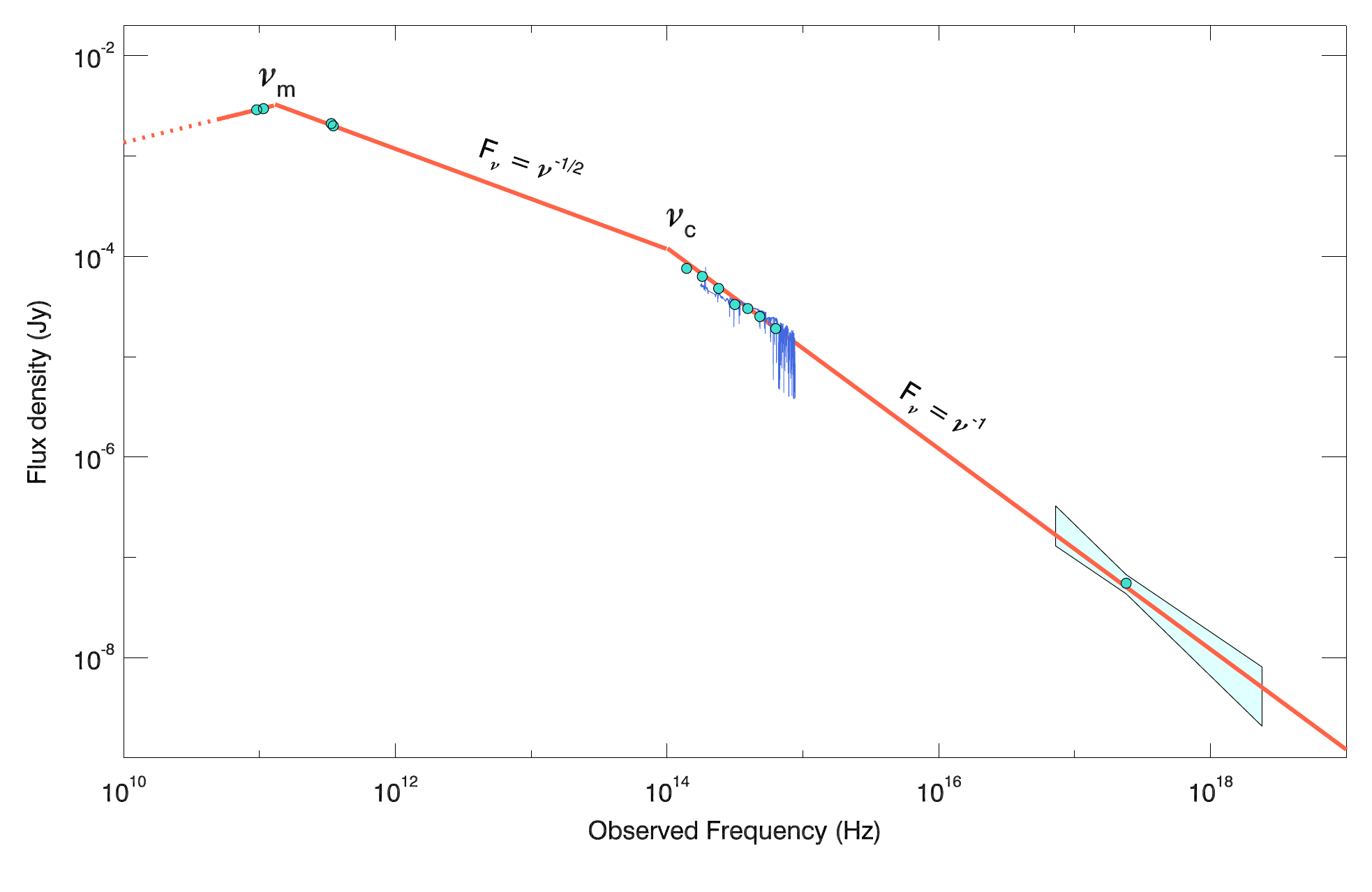}
      \caption{Spectral energy distribution of GRB\,161023A at the time of the ALMA observations, 50 hrs after the GRB.
              }
         \label{Fig:SED}
   \end{figure}

We have created an SED at the time of the ALMA observation ($t-t_0=50$ hr) spanning from the millimetre data all the way to the X-ray domain, observed with {\it Swift}/XRT, which is shown in Fig.~\ref{Fig:SED}. For this fit we interpolate the X-ray band to 50 hr using the fits of the previous section. The optical and the X-ray data seem to be on the same synchrotron spectral segment: The optical to X-ray spectral slope, $\beta_{OX} = -1.021\pm0.015$, is consistent with the X-ray index, $\beta_X=-1.05^{+0.26}_{-0.18}$ (a value obtained from the {\it Swift}/XRT pages\footnote{http://www.swift.ac.uk} that was calculated by fitting all the X-ray data), and is also very similar to the one obtained from the GROND photometry when performing a direct fit without assuming any host galaxy extinction ($\beta_O=-1.03\pm0.05$). The ALMA data are below the extrapolation, meaning that a spectral break must lie in-between the NIR and submillimetre wavelengths. This scenario is consistent with a standard fireball expanding in a medium with an electron index of $p=2.03$, which implies $\beta=-1.015$ at frequencies higher than $\nu_c$ and $\beta=-0.515$ between $\nu_m$ and $\nu_c$. However, the spectral slope between the two ALMA observations ($\beta_{ALMA}=-0.29\pm0.03$) is shallower than $\beta=-0.515$, but is not compatible with $\beta=+0.3$, which would be expected below the $\nu_m$ frequency. This indicates that the peak spectral frequency is between the two bands. This is confirmed when we look at the spectral slope defined within the side bands of Band 3 which, thanks to their high signal-to-noise ratio and in spite of the short wavelength range allow us to determine a slope of $\beta_{Band3} = +0.39\pm0.13$ consistent with the expected $\beta=+1/3$. The larger errors in Band 7 prevent us from determining the spectral slope, although a flux density decreasing toward shorter wavelengths is favoured.

By using the light-curve slopes together with the spectral slopes determined in the previous paragraphs, we can constrain the model even more. We use the closure relations \citep{gao13} to determine that the fireball was expanding in an environment with an electron index of $p=2.06\pm0.04$. According to these assumptions, the spectral slopes are expected to be $\beta_{model}=-0.53\pm0.02$ for the case $\nu_m<\nu<\nu_c$ and $\beta_{model}=-1.03\pm0.2$ for $\nu_c<\nu$ (consistent with the observed $\beta_{OX} = -1.021\pm0.015$). The temporal slopes for $\nu>\nu_c$, where we expect to have both optical and X-ray data, would be  $\alpha_{model}=-1.04\pm0.03$ before the jet break (consistent with the observed $\alpha_1=-1.063\pm0.002$), and $\alpha_{model}=-2.06\pm0.04$ after the jet break (consistent with the observed $\alpha_2=-2.25\pm0.15$).

\begin{figure}
   \includegraphics[width=\hsize]{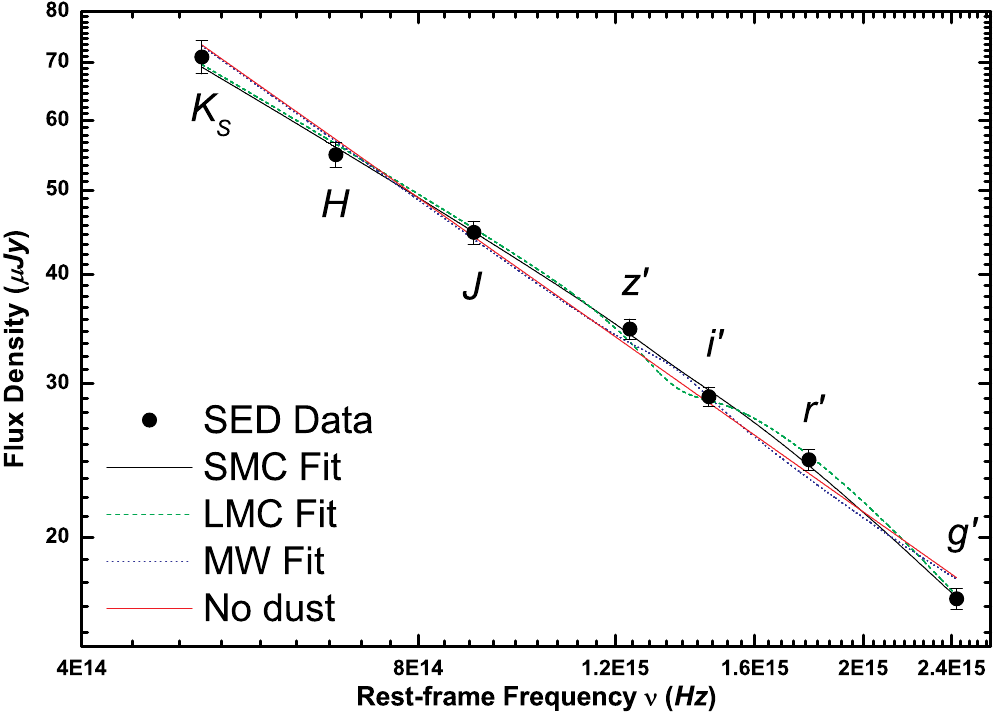}
      \caption{Fits to the $g^\prime r^\prime i^\prime z^\prime JHK_S$ SED of the GRB 161023A afterglow. We use the analytical expressions for the dust attenuation curves for MW (dotted blue line), LMC (dashed green line), and SMC (solid black line) dust from \cite{pei92}. A fit with no dust is also shown (red solid line). SMC and LMC dust both give good fits with a small amount of extinction, whereas MW dust or the complete absence of line-of-sight extinction is ruled out. The flux density is determined at the time of the light-curve break assuming $n=\infty$, but since this is a multi-band joint fit, scaling is essentially arbitrary (the evolution is achromatic).
              }
         \label{Fig:SEDopt}
   \end{figure}

The joint fit of all data after 0.1 days, as described above, allows us to derive a very precise $g^\prime r^\prime i^\prime z^\prime JHK_S$ SED that takes all multi-colour data into account. We find no evidence for colour evolution; this would manifest itself in a high $\chi^2$ value as well as significant residuals from the fit, which we do not see. Fitting the SED with dust models from \cite{pei92}, considering the filter response curves, we find that the best fit is given by SMC (Small Magellanic Cloud) -like dust (in accordance with the majority of 
GRB sightlines, \citealt{kan10}, \citealt{greiner11}, \citealt{bolmer18}), although an LMC (Large Magellanic Cloud) -like extinction profile can not be ruled out. For SMC dust, we find small but significant line-of-sight extinction $A_V=0.09\pm0.03$ mag and an intrinsic spectral slope $\beta=-0.66\pm0.08$ (see Fig.~\ref{Fig:SEDopt}). 

Finally, using the light-curve break time we can estimate the half-opening angle of the jet. The observed break time of t$_{b,obs}=2.81\pm0.30$ days implies a rest-frame break time of $t_{b,rest}=0.758\pm0.082$ day. Using the method described by \citet{rac09} based on that of \citet{fra01}, the jet half-opening angle (in radians) is defined as

\begin{equation}
\theta_j=0.057t_j^{3/8}\left(\frac{3.5}{1+z}\right)^{3/8}\left(\frac{\eta_\gamma}{0.1}\right)^{1/8}\left(\frac{n}{E_{\gamma,iso,53}}\right)^{1/8}
\end{equation}

where $t_j$ is the break time in days, $z$ is the redshift, and $E_{\gamma,iso,53}$ is the rest-frame isotropic equivalent energy radiated in gamma-rays between 1 keV and 10 MeV in units of $10^{53}$ erg. We assume a uniform interstellar medium (ISM) density of $n=1$ cm$^{-3}$ and a radiative efficiency of $n_\gamma=0.1$, and we note that the dependence on these values is only 1/8, therefore having a small effect on $\theta_j$. Using this equation and these assumptions, we estimate a jet half-opening angle of $\theta_j=3.74\pm0.09$ deg (the errors are only derived from the measurements of $t_{b}$ and E$_{\gamma,{\mathrm{iso}}}$, so there could be an additional systematic error due to our parameter assumptions but, due to the power of 1/8 dependencies, they are expected to be small). This can be compared to the sample of \citet{rac09}, who found a mean aperture of $<\theta_j>\,=6.5$ deg. A narrow opening angle allows GRB\,161023A to have the large luminosity that we see while constraining the overall energetics of the event.

We note that the spectral slope derived from the optical and NIR data fit is in conflict with the picture derived before from the light curves and broadband SED. A possibility would be that, due to the proximity of the cooling break (see Fig.~\ref{Fig:SED}), the curvature is due to the intrinsic shape of the spectrum and not only due to extinction (see \citealt{fil11} for a good example). This would imply that the extinction would strictly be an upper limit. Furthermore, the proximity of the cooling frequency $\nu_c$ at the time of the broadband SED implies that we would have expected it to cross the optical bands shortly before the time of the SED. This would result in a shallower decay slope at earlier times, which the light curve does not seem to show. We would also expect a colour change to be present in such a case, although the lack of early multi-band photometry prevents us from doing a thorough analysis of this. We also notice that the light-curve slope towards the early optical peak is too steep and thus inconsistent with the expected values for a forward shock. Therefore, it is possible that the early peak that we are seeing is, in fact, dominated by a reverse shock. The effect of this reverse shock, combined with other possible contributions such as flares, or energy injections, often seen in GRB light curves, may be contributing to mask the pure fireball evolution expected at early times. A more detailed analysis of the broadband model describing the GRB emission is beyond the scope of this paper.

\subsection{Luminosity of the afterglow of GRB\,161023A}
\label{Sect:lum}

   \begin{figure}
   \centering
   \includegraphics[width=9cm]{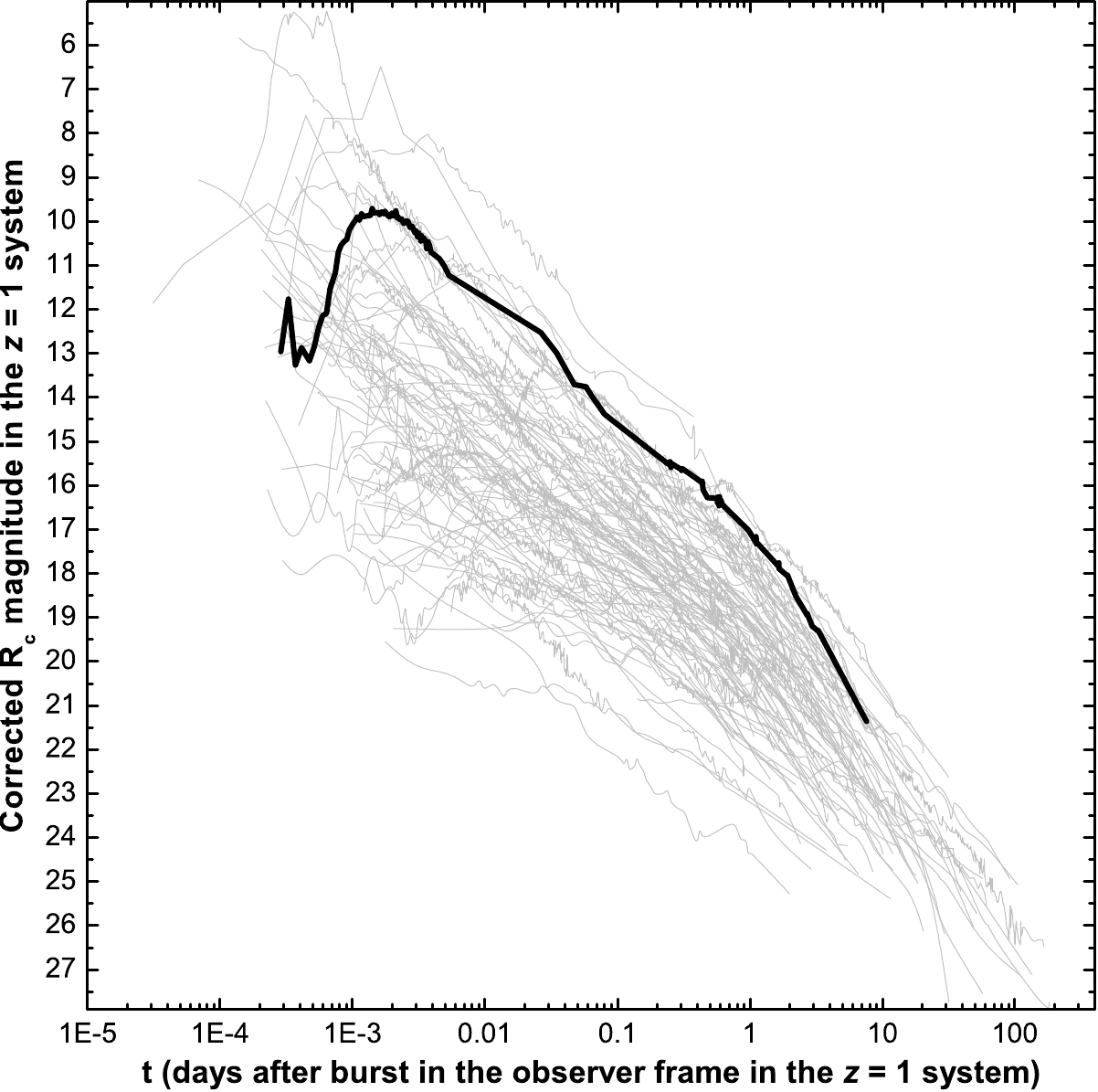}
   \includegraphics[width=9cm]{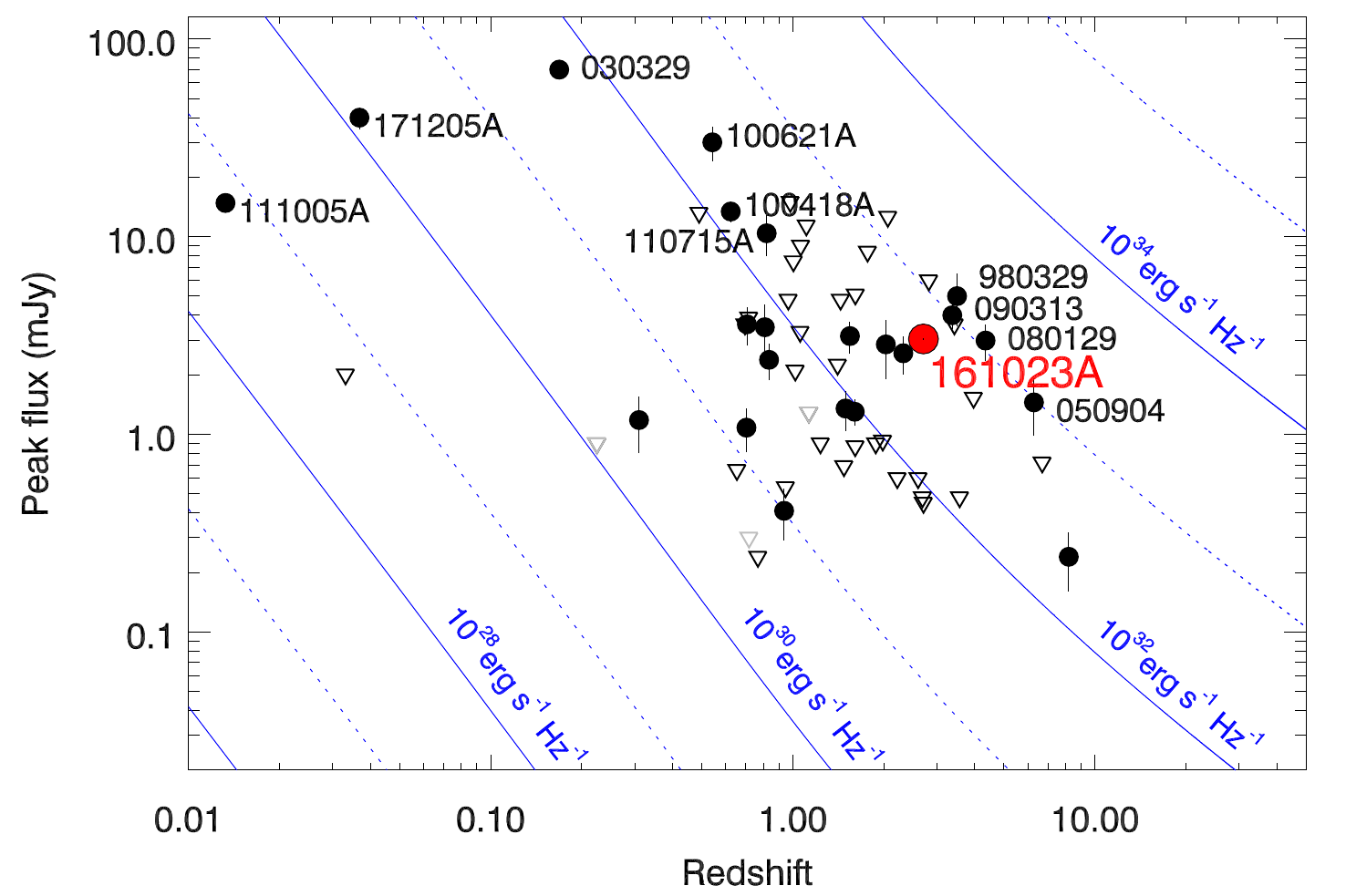}
      \caption{Top: Comparison of the light curves of a large sample of GRB afterglows shifted to a common redshift of $z=1$ (adapted from \citealt{kan17}) The afterglow of GRB\,161023A is amongst the most luminous afterglows observed so far. More details can be found in the text. Bottom: Comparison of the observed peak fluxes in the mm/submm bands (adapted from \citealt{deu12a}).
              }
         \label{Fig:kann}
         \label{Figmm}
   \end{figure}

Peaking at $r^\prime=12.6$ mag (AB system, corrected for Galactic foreground extinction), the afterglow of GRB\,161023A is one of the brightest that has been detected so far. Combining the information we derived for the spectral slopes and the temporal evolution in the previous section, together with the redshift, we can use the method of \cite{kan06} to place the afterglow of GRB\,161023A into the luminosity context of a large afterglow sample \citep{kan06,kan10,kan11,kan17}. We derive a magnitude shift from the original $z=2.71$ frame (i.e as observed) to $z=1$ of $dR_C=-2.77^{+0.07}_{-0.09}$ mag. We show the light curves in Fig. \ref{Fig:kann}, highlighting that of GRB\,161023A. This plot represents how GRB afterglows would be observed if they were completely unaffected by extinction and were all at $z=1$. Following the very early rapid variability, the afterglow rises rapidly to a slow rollover peak. For $\alpha_{rise}\neq-\alpha_{decay}$, the peak time of the light curve is not identical to the light-curve break time as determined in Sect. \ref{sect:SED}. Following \cite{mol07}, we derive a peak time $t_{peak}=0.00274$ days (64 s post-GRB in the rest frame),
and a peak $m_{R_C}=9.75\pm0.13$ mag at $z=1$, equivalent to an absolute magnitude of M$_U=-34.43\pm0.13$ mag. The very early afterglow of GRB\,161023A ($t<0.01$ days) is not among the most luminous early peaks (Fig. \ref{Fig:kann}), but it still shows one of the most luminous peaks that roll over slowly and are likely not completely dominated by a reverse shock, since the post-peak decay is slower than one would expect for such a case \citep{kob00a}; most other early optical peaks are clearly linked to reverse-shock flashes or central-engine activity (e.g. \citealt{ake99}, \citealt{boe06}, \citealt{kan07}, \citealt{rac08}, \citealt{blo09}, \citealt{mar14}) -- we note that the luminosity is similar to that of the prompt flash of GRB 130427A \citep{ves14}. The only other afterglows that show peaks that evolve similarly to that of GRB\,161023A but that are more luminous belong to GRB 061007 \citep{ryk09}, GRB 080810 \citep{pag09}, and finally the possible case of the ultra-luminous GRB 080607 \citep{per11}, which peaked so early that the rise (and possibly peak) was not detected in spite of extremely rapid follow-up. Then, between 0.5 and a few days in the $z=1$ frame, the afterglow of GRB\,161023A is seen to become one of the most luminous afterglows detected so far.
   
At millimetre wavelengths it is also one of the most luminous events ever recorded (see Fig.~\ref{Figmm}). In the observer frame, it is not among the brightest, where GRB\,030329 is still the record holder \citep{she03,res05}, followed by the recent GRB\,171205A \citep{deu17}, GRB\,100621A \citep{gre13}, GRB\,111005A \citep{mic16}, GRB\,100418A \citep[][ de Ugarte Postigo et al. in prep]{deu12a} and GRB\,110715A \citep{san17}, all of which happened at redshifts $z<1$. However, it is close to the group of most luminous millimetre GRB afterglows, which lie close to a peak luminosity of $10^{33}$~erg~s$^{-1}$~Hz$^{-1}$. Within the most luminous events we find GRB\,980329 \citep{smi99}, GRB\,080129 \citep{gre09}, GRB\,090313 \citep{mel10} and GRB\,050904 \citep{deu12a}, with GRB\,161023A being the fifth most luminous event for which we have millimetre data.

\subsection{Host galaxy}

Although we do not have direct observations of the host-galaxy emission lines, we are able to study the galaxy from within its interior using the afterglow light as a beacon that shines from the explosion site of the GRB. In this section we present different types of absorption line analysis that give us insight into the dynamics, abundances, dust depletion, molecular content, and structure of the host galaxy.

\subsubsection{Constraints from photometric upper limits}

The {\it Spitzer} non-detection at the GRB location can be used to constrain the properties of the host galaxy. The measured limit of $>25.3$~mag corresponds to an absolute magnitude limit of $>-20.0$~mag at a rest wavelength of 1 $\mu$m. For a continuous star formation history, this non-detection places a limit on the mass of the host to be $<4 \times 10^9 M_\odot$ (alternatively, the empirical method of \citealt{per16} provides a consistent upper limit).
This limit is about the mass of the Large Magellanic Cloud, or about 0.05~$L_*$ at the redshift of the burst. 
Assuming minimal dust ($< 0.5$ mag) and a >10 Myr stellar population, the FORS2 $r^\prime$-band non-detection limits the star formation rate to $<2 M_\odot$ yr$^{-1}$, although it could be substantially higher in the presence of dust extinction.

      \begin{figure*}
   \centering
   \includegraphics[width=13cm]{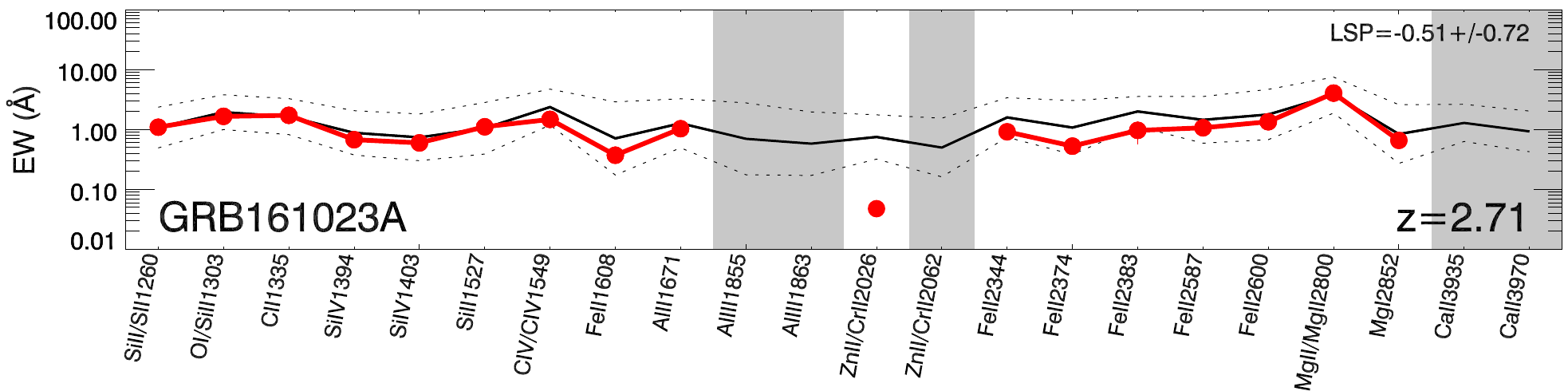}
      \caption{Equivalent width diagram obtained for the spectral features of the GRB\,161023A afterglow with X-shooter. It shows the strength of the lines measured for our GRB (in red) as compared to the average and standard deviation of strengths of a sample of GRB afterglows (in black), as described by \cite{deu12b}. The gray areas mask features that were not measured because they fell on telluric bands.
              }
         \label{FigLSP}
   \end{figure*}
   
\subsubsection{Line strength analysis}

We perform an analysis of the strong features detected at the redshift of the GRB in the X-shooter spectrum following the prescription of \citet{deu12b}, obtaining a value for the Line Strength Parameter of $\mathrm{LSP}=-0.51\pm0.72$. This number is indicative of the strength of absorption features and implies that the lines in the spectrum of GRB\,161023A are slightly weaker than the average of the sample presented in that paper (an average value would have returned an LSP equal to zero). We create a line-strength diagram, shown in Fig.~\ref{FigLSP}, which reveals a very low equivalent width (EW) in the \ion{Zn}{II}/\ion{Cr}{II} features, slightly low \ion{Fe}{II} and slightly low ionisation (\ion{C}{IV}/\ion{C}{II}, \ion{Si}{IV}/\ion{Si}{II}). Overall the line-strength parameter is below average, although the very weak \ion{Zn}{II}/\ion{Cr}{II} feature is responsible for lowering it and for increasing the uncertainty in the value of the LSP.

\begin{table*}
\caption{Components and column densities (N) with their corresponding $b$-parameters in the GRB system. Theif velocity ($v$) is definded with respect to the highest redshift component, which we define as having $v=0$\,km\,s$^{-1}$.  Transitions in brackets are detected but were not used for the fit due to either blending with other lines or large errors due to sky emission lines. }             
\label{tab:voigt}     
\centering      
\scriptsize
\begin{tabular}{l c c c c c c c c c c}       \\ \hline
 &  & \multicolumn{2}{c}{\textbf{Component I}}& \multicolumn{2}{c}{\textbf{Component II}} & \multicolumn{2}{c}{\textbf{Component III}} & \multicolumn{2}{c}{\textbf{Component IV}} & \textbf{total}  \\   \hline
    \textbf{Ion}   &\textbf{Transitions}  & \textbf{v~~b} &\textbf{log(N/cm$^{-2}$) }  & \textbf{v~~b}&\textbf{log(N/cm$^{-2}$) }  &\textbf{v~~b}&\textbf{log(N/cm$^{-2}$) } &\textbf{v~~b}&\textbf{log(N/cm$^{-2}$) } & \textbf{log(N/cm$^{-2}$) }  \\ 
       &\scriptsize \textbf{ \AA}		&\scriptsize \textbf{$\mathbf{(km~s^{-1})}$}&     &\scriptsize \textbf{$\mathbf{(km~s^{-1})}$}&      &\scriptsize \textbf{$\mathbf{(km~s^{-1})}$}&    & \scriptsize \textbf{$\mathbf{(km~s^{-1})}$}&  &  \\
\hline\hline  
HI		& \scriptsize Ly$\alpha$ to Ly9 &---	&---	&---	&---	&---	&--- &---	&---	&20.97$\pm$0.01	\\
 \ion{S}{II} 	& \scriptsize 1250, 1253, (1259)  &  0~~22 &14.69$\pm$0.02	& --76~~15 	& 14.33$\pm$0.04 & --- & --- & --188~~28		& 14.20$\pm$0.05 	& 14.94$\pm$0.02\\ %
 \ion{Si}{II} 	& \scriptsize 1260, 1304, 1526, 1808& 0~~22 & 14.54$\pm$0.02   	&  --76~~15 	& 14.77$\pm$0.03 & --- & ---	& --188~~28	& 13.97$\pm$0.02		&15.04$\pm$0.02 \\ %
     \ion{O}{I} 	& \scriptsize 	1302			& 0~~15 & $<$17.2	&  --76~~15 & $<$16.0 	& --- & --- &  --188~~30	& 14.58$\pm$0.01	&$<$17.2\\
 \ion{C}{II} 	& \scriptsize  1334		& 0~~22 & 15.30$\pm$0.10  	&  --76~~15 	& 15.80$\pm$0.10 	& --142~~30 & 13.78$\pm$0.02 &  --188~~28		& 14.56$\pm$0.01		& 15.94 $\pm$0.04 \\ %
 \ion{Fe}{II} 	&\scriptsize    1608, 2260, 2344, 2374	& 0~~22 & 14.43$\pm$0.02   	&  	  --76~~15	& 14.18$\pm$0.02& --- & --- 	&  	 --188~~28&  13.56$\pm$0.06	&14.66$\pm$0.01\\ 
  	&\scriptsize    (2382, 2586, 2600)	&   &    	&  	  	& &  &  	&  & 	&\\ %
 \ion{Al}{II} 	& \scriptsize  	1670	& 0~~22 & 13.23$\pm$0.01	& --76~~15 	& 14.03$\pm$0.05 & --142~~30 & 12.99$\pm$0.01	&  --188~~28	&12.81$\pm$0.02 	& 14.14$\pm$0.02\\ %
 \ion{Zn}{II} 	& \scriptsize   2026	& 0~~22 &12.01$\pm$0.09  	&  --76~~15	& 11.84$\pm$0.12	&  ---	& ---  & --188~~30 & 11.96$\pm$0.10 & 12.42$\pm$0.06	\\ %
 \ion{Mg}{II} 	& \scriptsize  2796, 2803 		& 0~~21 & 13.80$\pm$0.1  			&  --95~~43 	& 14.85$\pm$0.27 & --- & --- 	& 	 --188~~32	& 13.86$\pm$0.04  	& 15.08$\pm$0.20\\ %
 \ion{Mg}{I} 	& \scriptsize 	2026, 2852			& 0~~50 & 12.24$\pm$0.06	&  --95~~25& 12.17$\pm$0.06 	& --- & --- &  --188~~32	& 12.42$\pm$0.03	&12.76$\pm$0.03\\ \hline
 \ion{Fe}{II}1s& \scriptsize  	2333, 2612, 2626			& 0~~13 & 13.32$\pm$0.02	&  --76	~~25& 13.26$\pm$0.03 	& --- & --- &  --188~~20	& 12.92$\pm$0.04	& 13.68$\pm$0.08\\
 \ion{Fe}{II}3s& \scriptsize  2338, 2359, (2631)			& 0~~13 & 12.63$\pm$0.08 	&  --76~~25	& 12.80$\pm$0.06 & --- & --- 	&  --188~~20	& 12.49$\pm$0.12		& 13.13$\pm$0.05\\
 \ion{Fe}{II}4s& \scriptsize 2345, 2629			& 0~~13 & 12.90$\pm$0.03 	&  --76~~25	& 12.48$\pm$0.10 	&  ---	& ---	& --- & --- 	& 13.03$\pm$0.02\\
\ion{Fe}{II} 5s& \scriptsize  1637, 1702 	& 0~~13 & 13.19$\pm$0.04 &  --76~~25&13.13$\pm$0.06 &---	& ---	& --188~~20	&12.89$\pm$0.09 	&13.56$\pm$0.03\\
 \ion{Ni}{II}2s& \scriptsize   2217, 2316				& 0~~13 & 13.17$\pm$0.01 &  --76~~25	& 13.03$\pm$0.02 & --- & ---  	&  --188~~20	&12.59 $\pm$0.04	& 13.46$\pm$0.01\\
 \ion{Si}{II*} & \scriptsize	1309, 1533		& 0~~27 & 14.08$\pm$0.01 &  --76~~35	& 14.04$\pm$0.01	& --142~~30 & 13.22$\pm$0.05 &  --188~~38	& 13.95$\pm$0.01	& 14.50$\pm$0.01\\
 \ion{C}{II*} & \scriptsize	1335.6, 1335.7 		& 0~~27 & 13.87$\pm$0.01 &  --76~~35	& 14.38$\pm$0.01	& --- & --- &  --188~~38	& 14.48$\pm$0.01 & 14.79$\pm$0.01\\
 \ion{O}{I*}& \scriptsize 1304 & 0~~33 & 14.24$\pm$0.01 &  --63~~27	& 14.09$\pm$0.02	& --- & --- &  --176~~30	& 14.50$\pm$0.01 & 14.79$\pm$0.01\\
  \ion{O}{I**}& \scriptsize 1306 & 0~~50 & 14.52$\pm$0.06 &  --63~~15	& 14.40$\pm$0.06	& --102~~16 & 14.53$\pm$0.03&  --176~~30	& 14.46$\pm$0.03 & 15.23$\pm$0.04\\
   &  & 51~~33 & 14.70$\pm$0.04 &  	& 	&  &  &  	&  & \\
\hline\hline
\end{tabular}
\vspace{1mm}

\begin{tabular}{l c c c c c c c c c c}       
 &  & \multicolumn{2}{c}{\textbf{Component I}}& \multicolumn{2}{c}{\textbf{Component II}} & \multicolumn{2}{c}{\textbf{Component III}} & \multicolumn{2}{c}{\textbf{Component IV}} & \textbf{total}  \\   \hline
    \textbf{Ion}   &\textbf{Transitions}  & \textbf{v~~b} &\textbf{log(N/cm$^{-2}$)  }  & \textbf{v~~b}&\textbf{log(N/cm$^{-2}$)  }  &\textbf{v~~b}&\textbf{log(N/cm$^{-2}$)  } &\textbf{v~~b}&\textbf{log(N/cm$^{-2}$)  } & \textbf{log(N/cm$^{-2}$)  }  \\ 
       &\scriptsize \textbf{ \AA}		&\scriptsize \textbf{$\mathbf{(km~s^{-1})}$}&    &\scriptsize \textbf{$\mathbf{(km~s^{-1})}$}&    &\scriptsize \textbf{$\mathbf{(km~s^{-1})}$}&     & \scriptsize \textbf{$\mathbf{(km~s^{-1})}$}&   &   \\
\hline\hline  
 \ion{C}{IV} 	& \scriptsize1548, 1550 & +6~~29 &13.51$\pm$0.03 & --83~~43 	& 13.19$\pm$0.27 & --114~~27		& 15.42$\pm$0.08	& --188~~28  & 13.62$\pm$0.03  &15.43$\pm$0.07\\%
  \ion{Si}{IV} 	& \scriptsize1402 & +6~~29 &13.17$\pm$0.02 & --83~~43 	& 13.65$\pm$0.02 & --114~~27		& 14.44$\pm$0.02	& --188~~28  & 13.33$\pm$0.02  &14.55$\pm$0.01\\
   \ion{N}{V} 	& \scriptsize 1238, 1242  & --- & --- &--83~~43 &13.47$\pm$0.03 & --114~~27	& 13.54$\pm$0.03 & --188~~28		& 13.08$\pm$0.05	  &13.88$\pm$0.02\\%
    \ion{S}{IV} 	& \scriptsize 1062 & --- & --- &--83~~43 &14.42$\pm$0.02 & --114~~27 	& 14.07$\pm$0.05 & --188~~28		& 13.68$\pm$0.06	  &14.63$\pm$0.02\\
     \ion{S}{VI} 	& \scriptsize 933, 944 & --- & --- & --83~~43 &13.52$\pm$0.05 & --114~~27	& 13.97$\pm$0.03 & --188~~28		& 13.16$\pm$0.09	 &14.15$\pm$0.03\\%
      \ion{O}{VI} 	& \scriptsize 1031, (1037) & +6~~29 &13.77$\pm$0.03 & --83~~43 	& 14.22$\pm$0.02 & --114~~27		& 14.33$\pm$0.02	& --188~~28  & 13.94$\pm$0.02  &14.72$\pm$0.15\\%
       \ion{P}{V} 	& \scriptsize 1117, 1128 & --- & --- & --28~~25 &13.54$\pm$0.01 & --103~~25	& 13.13$\pm$0.02 & ---		& ---  &13.68$\pm$0.01\\ \hline \hline

\end{tabular}
\end{table*}

\subsubsection{Column densities and abundances in the host galaxy}\label{sect:abundances}
We fit the absorption lines of the host complex with Voigt profiles using the FITLYMAN environment in Midas \citep{fon95}. We use as zero velocity the highest redshift component, which we find to be at $z=2.71064\pm0.00003$. All transitions commonly observed in GRB afterglow spectra from Ly$\alpha$ to \ion{Mg}{I} are detected in our X-shooter data. The \ion{Ca}{II} doublet lies on the red end of the atmospheric band between {\it J} and {\it H} band and cannot be fitted. \ion{Mn}{II} transitions of $\lambda$ 2594 and 2606 \AA{} are detected but lie in a region contaminated by skylines beyond 9600 \AA{} and were not fitted. \ion{O}{I} is clearly saturated and has only one transition available for fitting, hence only an upper limit can be given, assuming a $b$-parameter of 15 km\,s$^{-1}$. Corresponding fine-structure transitions of Fe, Ni, Si, and C are measured, as well as metastable transitions of Ni and Fe up to \ion{Fe}{II} 5s, although \ion{Fe}{II} 2s could not be fitted due to issues with skylines and blending with \ion{Fe}{II} 3s. We further detect the strong high-ionisation lines of \ion{C}{IV}, \ion{Si}{IV},  and \ion{N}{V} as well as several transitions of \ion{O}{VI}, \ion{S}{IV},  \ion{S}{VI} and possibly \ion{P}{V} in the Ly$\alpha$ forest. The detection of \ion{P}{V} is not secure as it seems somewhat redshifted compared to the other high-ionisation lines, however, both transitions of \ion{P}{V} do show absorption at the same (redshifted) velocity. We hence urge caution concerning the detection of this line. 

\begin{figure}
   \centering
   \includegraphics[width=8cm]{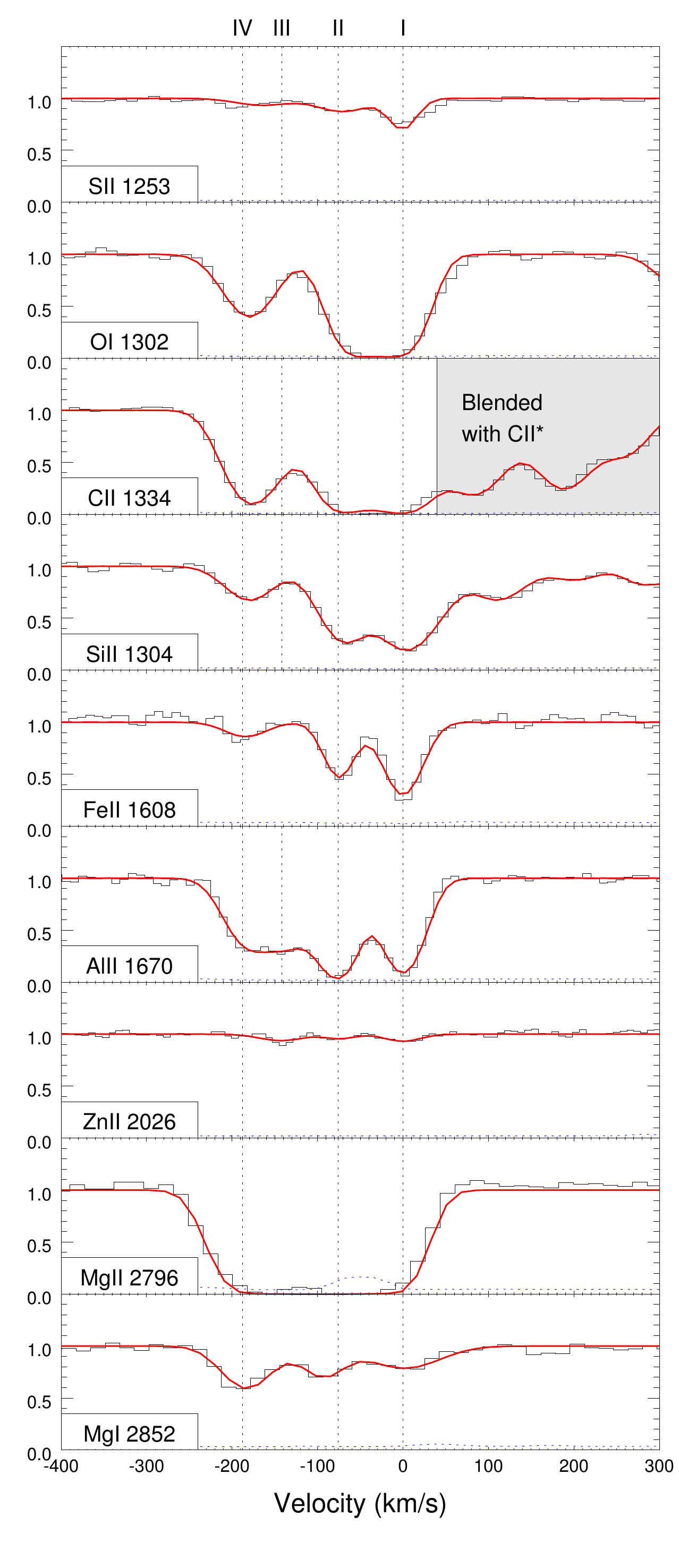}
      \caption{Voigt profile fits of low-ionisation features (red) at the redshift of the host galaxy overplotted on the normalised spectra. The error spectrum is shown with a dotted blue line. The different velocity components are marked with roman numbers at the top and vertical dotted lines.
              }
         \label{Fig:voigt}
   \end{figure}

\begin{figure}
   \centering
   \includegraphics[width=8cm]{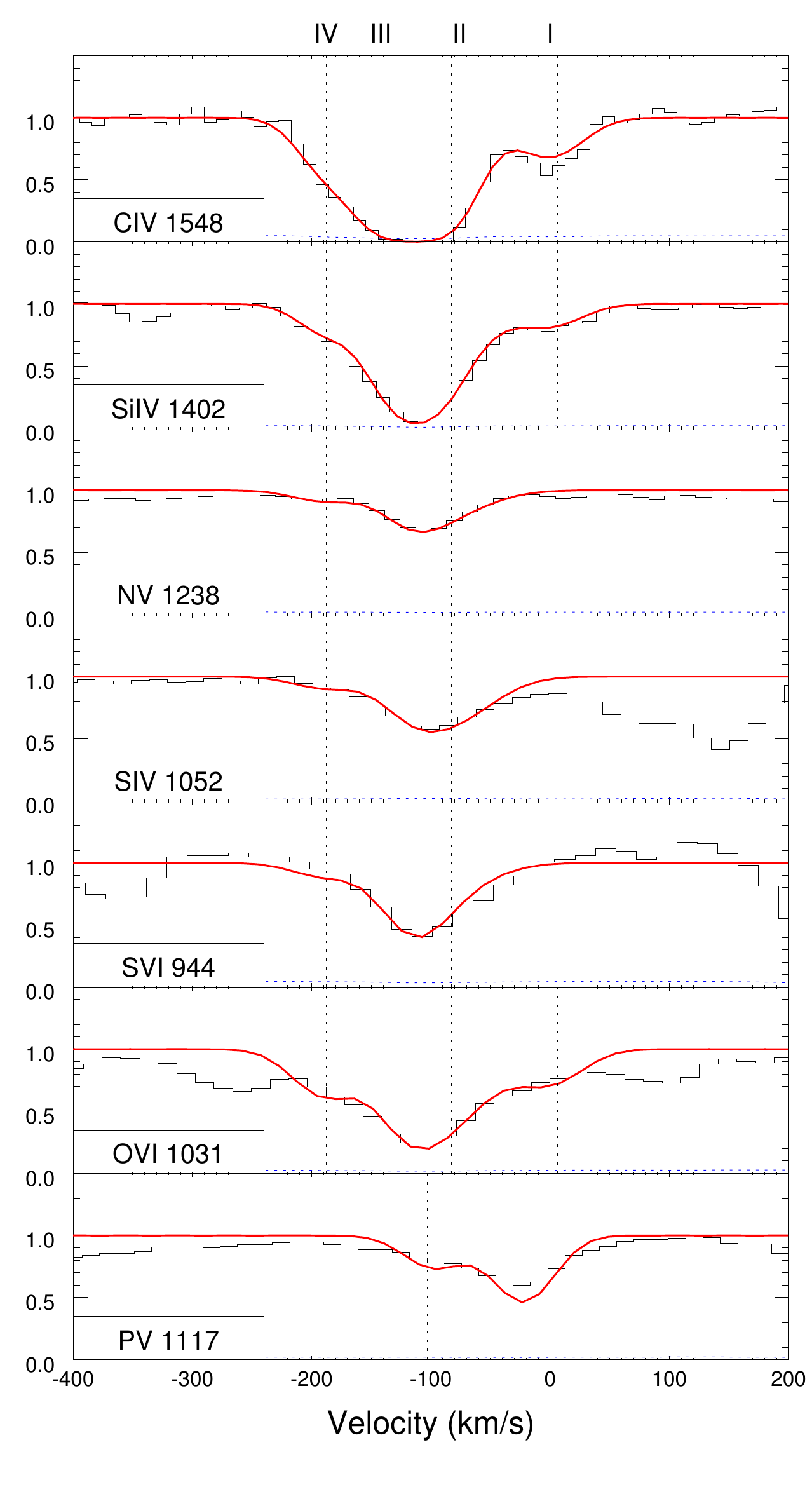}
      \caption{Voigt profile fits of high-ionisation features at the redshift of the host galaxy, using the same plotting scheme as in Fig.~\ref{Fig:voigt}.
              }
        \label{Fig:voigthigh}
   \end{figure}

\begin{figure}
   \centering
   \includegraphics[width=8cm]{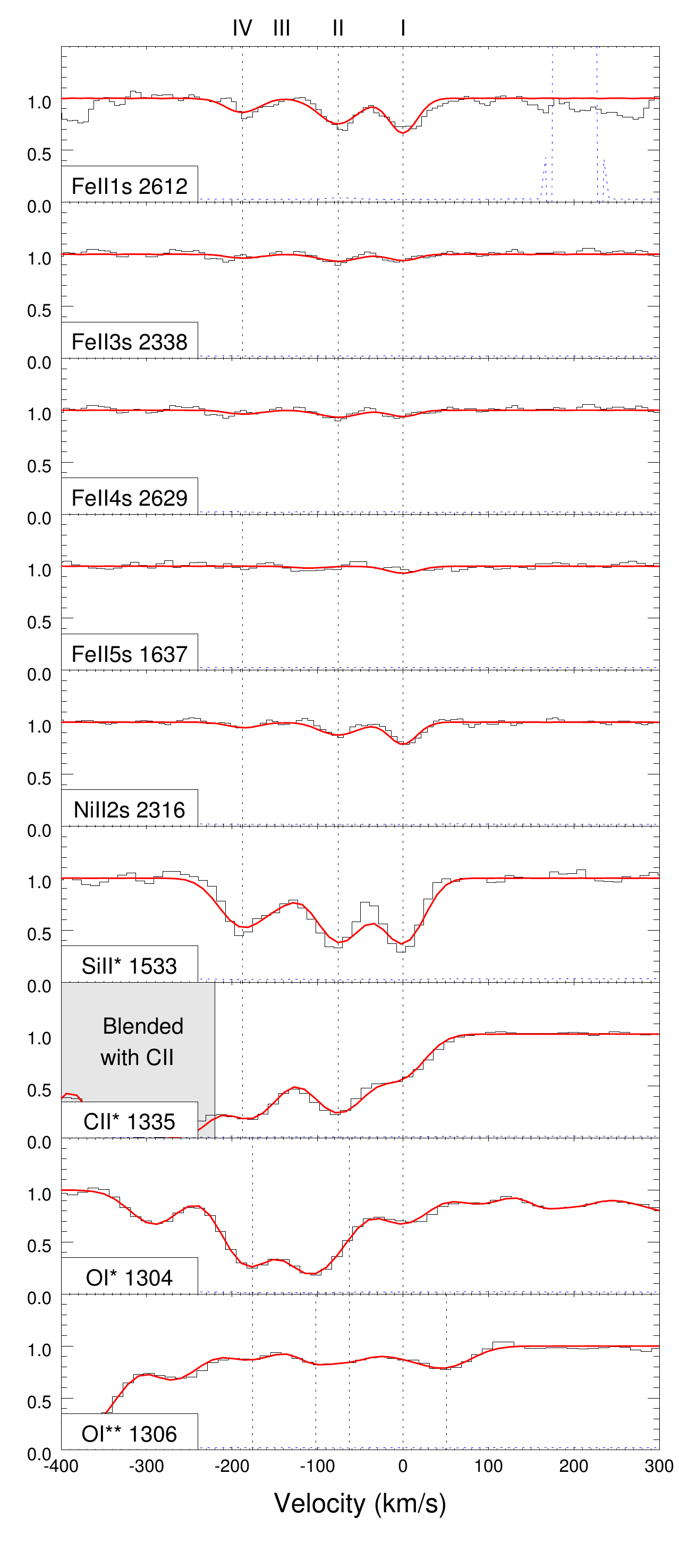}
      \caption{Voigt profile fits of fine-structure and metastable features at the redshift of the host galaxy, using the same plotting scheme as in Fig.~\ref{Fig:voigt}.
              }
         \label{Fig:voigtfine}
   \end{figure}

Whenever possible we fitted at least two transitions of the same elementary species and ionisation state to make sure that one of them is not contaminated by a transition from an intervening system, an atmospheric line, or the Ly$\alpha$ forest in case of the high-ionisation transitions. The majority of neutral and single ionised lines show three velocity components. \ion{Al}{II} and \ion{Si}{II*}$\lambda$1309 have an additional component between component I and III, also for \ion{O}{I**} we need an additional component to properly fit the absorption complex. \ion{Si}{II}$\lambda$1526 looks like an extended absorption system; however, the additional components red and blue of the four main components are \ion{C}{IV} doublets from an extended intervening system at $z=2.65$, as is some absorption within the complex that would correspond to a component II, but which in reality is \ion{C}{IV} $\lambda$ 1548 from the intervening system (see Sect. \ref{sec:int}). To determine the $b$-parameter, we fitted the unblended and unsaturated transitions of \ion{Fe}{II} 1608 and \ion{Si}{II*} 1309 and used the fitted $b$-parameters for all single ionised and fine-structure/metastable transitions respectively, with exceptions for lines with additional components. Somewhat different $b$-parameters were derived for the neutral and high-ionisation species using the strongest, non-saturated absorption line for each ionisation or fine-structure state. The fitted column densities for the individual components of all resonant and fine-structure/metastable transitions and the common $b$-parameters are listed in Table~\ref{tab:voigt}. In Figs.~\ref{Fig:voigt}, \ref{Fig:voigthigh}, and \ref{Fig:voigtfine} we plot some of the transitions and the corresponding fits.

The abundances derived for the high-ionisation lines as well as HI and the value for the metallicity are in good agreement with \cite{hei18} who used the same data set but a different Voigt-profile fitting code. Using the total column density of \ion{S}{II} we derive a metallicity of\footnote{For all abundances [X/Y] is relative to the solar ratio and hence defined as $\log_{10}(\mathrm{N}(\mathrm{X})/\mathrm{N}(\mathrm{Y}))-\log_{10}(\mathrm{N}(\mathrm{X})_\odot/\mathrm{N}(\mathrm{Y})_\odot$).} [S/H]~$=-1.15\pm0.04$ for the host galaxy. A similar metallicity is derived from Zn, [Zn/H]~$=-1.11\pm0.07$. This is a rather average metallicity for a GRB host at this redshift \citep{tho13,spa14,cuc15}. We tried to disentangle different components in the Ly absorption series which we detect down to Ly 9 in order to derive separate metallicities for the three different components. However, all Lyman transitions are either contaminated by other Lyman forest lines or transitions from intervening systems in the red and blue wing or completely saturated, making a separation of three possible sub-components impossible. Hence we can only derive a metallicity for the entire host system including all absorption components. 

\subsubsection{Dust depletion in the host}

\begin{table*}
\caption{Abundances, dust depletion and corrected metallicities using different elements \citep[according to][]{dec16}. Errors are propagated from the column density measurements and the errors on the depletion values in \cite{dec16}. No depletion values are given for Al and C. $\mathrm{[Zn/Fe]}=0.70\pm0.08$, $0.52\pm0.11$, $0.60\pm0.14$, and $1.34\pm0.13$ for the total host absorption and component I, II and IV respectively.}            
\label{tab:depletion}     
\centering       
\scriptsize
\begin{tabular}{c | c c c c c c c c | c c}       
\hline\hline                
Element&$\delta_X$ host & $\delta_X$ (I) & $\delta_X$ (II) & $\delta_X$ (IV) &$\mathcal{{DT\!\!M}}$ host & $\mathcal{{DT\!\!M}}$ (I) & $\mathcal{{DT\!\!M}}$ (II)& $\mathcal{{DT\!\!M}}$ (III) & [X/H]&[X/H]$_\mathrm{tot}$\\ \hline
O       &   --0.13$\pm$0.12 & --0.10$\pm$0.11  & --0.11$\pm$0.12    & --0.22$\pm$0.16    & --- & ---- & ---- & --- & $<$--0.46 & $<$--0.36 \\
Zn     &  --0.19$\pm$0.03 & --0.14$\pm$0.04 & --0.16$\pm$0.04    & --0.36$\pm$0.05    & --- & ---- & ---- & --- & --1.11$\pm$0.07 & --0.92$\pm$0.08 \\
S      &  --0.24$\pm$0.09 & --0.19$\pm$0.06 & --0.21$\pm$0.07    & --0.42$\pm$0.12    & --- & ---- & ---- & --- & --1.15$\pm$0.04 & --0.91$\pm$0.08 \\
Si     & --0.47$\pm$0.09 & --0.36$\pm$0.08 &  --0.42$\pm$0.10   & --0.87$\pm$0.12    &  0.68$\pm$0.11 &0.57$\pm$0.11 & 0.62$\pm$0.14 & 0.88$\pm$0.17 & --1.44$\pm$0.04 & --0.97$\pm$0.08  \\
Mg    &  --0.46$\pm$0.22 & --0.35$\pm$0.09  &  --0.40$\pm$0.10   & --0.84$\pm$0.12    & 0.66$\pm$0.13 & 0.56$\pm$0.114 & 0.61$\pm$0.14 &  0.88$\pm$0.18 & --1.49$\pm$0.20 & --1.03$\pm$    0.22 \\
Fe    &  --0.89$\pm$0.11 & --0.67$\pm$0.15 &  --0.77$\pm$0.18   & --1.70$\pm$0.18    & 0.89$\pm$0.23 & 0.80$\pm$0.27 & 0.85$\pm$0.35 & 1.00$\pm$0.34 & --1.81$\pm$0.04 & --0.92$\pm$0.12 \\
Al     &  --- &  &     &     & --- & ---- & ---- & --- & --1.28$\pm$0.04 & --- \\
C      &  --- &  &     &     & --- & ---- & ---- & --- & --1.46$\pm$0.07 & --- \\
\hline                                
\end{tabular}
\end{table*}

Above we determined the metallicity of the host from two of the elements least depleted by dust, although there might be some issues with sulphur that can show depletion up to 1 dex in very dusty systems \citep{jen09} but can under most circumstances be treated as a mildly depleted element together with zinc. Indeed, if we determine the metallicity from other elements we obtain lower values (see Table~\ref{tab:depletion}). For oxygen we obtain a limit of  [O/H]~$<-0.46$. Oxygen is also considered to be undepleted, however, the transition is usually saturated, and there is no other transition available at the observable wavelengths, hence a determination of metallicity is often difficult, which is also the case for GRB\,161023A. Carbon is an equally difficult case with only one, often saturated transition at $\lambda$1335, which is furthermore blended with two of its fine-structure transitions. 

\begin{figure}
   \includegraphics[width=\hsize]{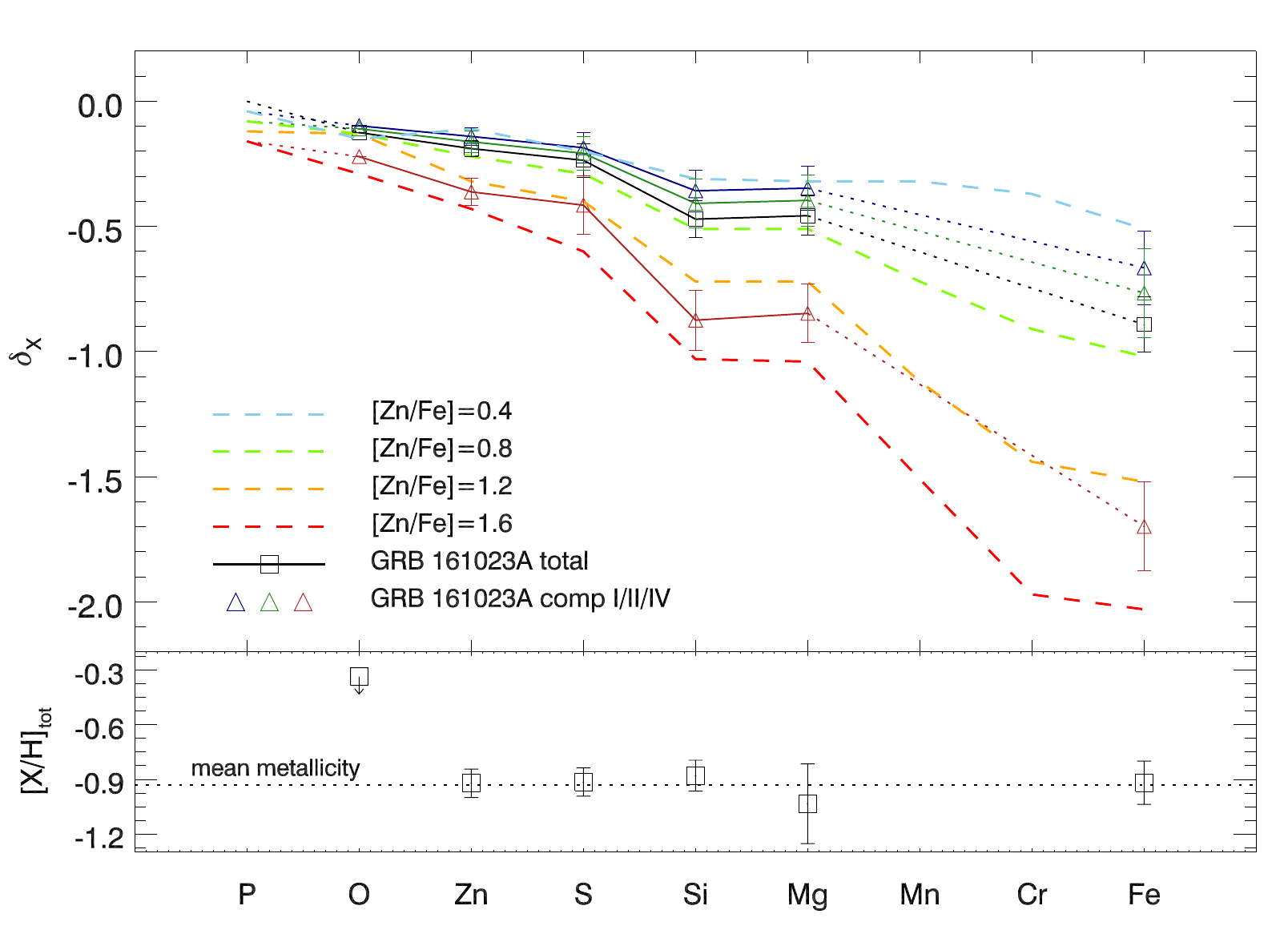}
      \caption{Dust-depletion sequence of the host of GRB\,161023A and the three main components resolved. The relative abundances are [Zn/Fe]~$=0.7$ for the total host system and 0.52, 0.6 and 1.34 for components I, II and IV respectively. As comparison, we show dust-depletion patterns from fitting DLA absorption systems \citep[see][]{dec16}. In the bottom panel we plot the dust-corrected metallicities (see also Table~\ref{tab:depletion}).}
         \label{fig:depletion}
   \end{figure}

If the different metallicities are due to dust depletion only, we should be able to verify this by comparing them to common dust-depletion patterns. \cite{dec16,dec17} recently extended the analysis of \cite{jen09} for relative abundances along Galactic sight lines to the (higher) abundances measured in damped Lyman-$\alpha$ system (DLA) sight lines and determined the depletion $\delta_X$ and the nucleosynthetic over-/under-abundances $\alpha_X$ of different elements relative to [Zn/Fe]. The authors furthermore provide new depletion patterns, substituting the commonly used patterns for the warm/cool disk and halo by \cite{sav96}. Since we detect Zn and Fe in all components, we can derive depletion values for both the combined host absorption and the three main components. In Fig. \ref{fig:depletion} we plot $\delta_X$ of different elements in order of increasing depletion and for the different absorption-line components. We determined $\delta_X$ from fits relative to [Zn/Fe] where $\delta_X$~=~A$_2$ + B$_2$ $\times$ [Zn/Fe] \citep[see][Eq. (5)]{dec16} and the coefficients for each element, A$_2$ and B$_2$ are given in \citet[][Table~3]{dec16}. The values obtained for $\delta_X$ in our spectra of GRB\,161023A are given in Table~\ref{tab:depletion} for the different components. Component IV (the bluest component) has the highest depletion with values close to Milky Way gas which is quite rare for quasi-stellar object (QSO) -DLAs \citep[see][]{dec16}. However, this is the component where zinc is only marginally detected, hence we caution against any strong conclusions being derived from it.

With the depletion values derived from [Zn/Fe], we can now determine the total abundances corrected for dust depletion and hence establish a corrected metallicity of the host [X/H]$_\mathrm{tot}$~=~$[$X/H$]$-$\delta_X$ \citep[see][Eqs. (1) and (2)]{dec17}, which we also list in Table~\ref{tab:depletion} and plot in the bottom panel of Fig. \ref{fig:depletion}.  If nucleosynthesis effects played a major role, we would expect deviations of some corrected metallicities, however, for GRB\,161023A all are consistent within errors. GRB\,161023A shows that possible nucleosynthetic over- or under-abundances are difficult to measure, even in GRB absorption systems at high signal-to-noise ratio and with many different elements detected. To derive a final metallicity for the host of GRB 161023A, we take the average of all depletion-corrected metallicities from lines with well-constrained column densities and available depletion patterns: Zn, S, Si, Mg, and Fe. Taking into account the errors for the individual corrected metallicities, we infer an average, dust-corrected metallicity of $-0.95\pm0.05$, which we adopt as the metallicity for the host of GRB 161023A.

The depletion values furthermore allow us to calculate the dust-to-metals (dtm) ratio normalised to the Galactic value ($\mathcal{{DT\!\!M}}$) where $\mathcal{{DT\!\!M}}=(1-10^{\delta_X}$)/dtm(Gal) and the Galactic ratio dtm(Gal) is 0.98 \citep{dec13}. The $\mathcal{{DT\!\!M}}$ values for different elements are listed in  Table~\ref{tab:depletion}. The $\mathcal{{DT\!\!M}}$ values for Fe, Si, and Mg, elements that can be depleted onto dust grains, are in good agreement and well within the spread of values for other DLAs at a metallicity of $\sim-1.0$ \citep[see Fig. 16 in][]{dec16}. The exception here is component IV, which has a high depletion factor as mentioned above and has a $\mathcal{{DT\!\!M}}$ at the upper end of the distribution for DLAs at that metallicity. We note, however, that we are not able to resolve the metallicity for different components. Finally, from the $\mathcal{{DT\!\!M}}$ we can also derive an overall extinction of the sight lines in the host \citep[see][Eq. 8]{dec16} and obtain an estimate of $A_V=0.04$ mag, not far from the value obtained from the SED fit (Sect. \ref{Sect:lum}).

\subsubsection{Limits on the detection of molecules from ALMA spectroscopy}

The ALMA observing windows of Band 3 were tuned to cover the range in which we would expect to observe several molecular transitions at the redshift of the GRB. In particular, we covered the CO(2-3) 
and the HCO+(3-4) transitions. Although there is no detection of any absorption features, these observations can be used to determine a limit on the column density of each of the molecules along the line of sight toward the GRB within its host galaxy. The column density of a specific molecule with a transition from level J to J+1 (absorption) is given by the relation (assuming local thermodynamic equilibrium conditions, based on the formulation of \citealt{wik95,wik97})
\begin{equation}
N=\frac{8\pi}{c^3}\frac{\nu^3}{A_{J+1,J}g_{J+1}}\frac{Q(T_x){\rm exp}(E_J/kT_x)}{1-{\rm exp}(-h\nu/kT_x)}\int\tau_{\nu}dV
\end{equation}
where $\nu$ is the observed frequency, $A_{J+1,J}$ is the Einstein coefficient of the transition, $g_{J+1}$ is 2(J+1)+1, $Q(T_x)$ is the partition function, and $T_x$ is the rotational temperature, which, as a lower limit, we can assume to be that of the cosmic microwave background (CMB) at the redshift of the burst ($T_x=2.725(1+z)=10.1$ K). The energy of the lower level is $E_J$, $\int\tau_{\nu}dV$ is the observed optical depth integrated over the line for a given transition, where $V$ is the velocity relative to the line rest frame, $h$ is the Plank constant, and $k$ is the Boltzmann constant. In our case we can get a limit on the value of the integral by using Eq. (5) from \citet{wik97},
\begin{equation}
\int\tau_{\nu}dV\simeq3\sigma_{\tau}\sqrt{\Delta V \delta V},
\end{equation}
 $\Delta V$ being the velocity width of the line, $\delta V$ the spectral resolution and $\sigma_{\tau}$ the r.m.s. of the opacity, which is equivalent to 
\begin{equation}\sigma_{\tau}=-{\rm ln}(1-\sigma_{norm}),\end{equation}
where $\sigma_{norm}$ is the r.m.s. of the continuum-normalised spectrum. Assuming a $\delta v=13$ km s$^{-1}$, similar to the $b$-parameters measured for the absorption lines in the X-shooter spectrum (see Sect.~ \ref{sect:abundances}), we get a limit value of 3.9 km s$^{-1}$ for the integral for any of the three transitions that we are considering.

\begin{table*}
\caption{Parameters for the molecular transitions studied in the spectrum of GRB\,161023A. Limits are 3-$\sigma$.}            
\label{table:par}     
\centering                       
\begin{tabular}{c c c c c c c c c}       
\hline\hline
Transition & $\nu_{obs}$ & A$_{J+1,J}$        & $g_{J+1}$ & $E_J$ & $\int\tau_{\nu}dv$  & $T_X$   & $Q(T_X)$& log(N/cm$^{-2}$)    \\
           & GHz         & s$^{-1}$           &           & (K)   & km s$^{-1}$         & (K)     &         &                     \\
\hline 
CO(2-3)    & 93.2316     & $2.5\times10^{-6}$ & 7         & 33.2  & 3.9                 & 10.1    & 4.02    & $<15.7$             \\
           &             &                    &           &       &                     & 20.0    & 7.56    & $<15.5$             \\
           &             &                    &           &       &                     & 30.0    & 11.14   & $<15.6$             \\
HCO+(3-4)  & 96.1807     & $2.5\times10^{-3}$ & 9         & 42.8  & 4.1                 & 10.1    & 5.08    & $<13.2$             \\
           &             &                    &           &       &                     & 20.0    & 9.68    & $<12.8$             \\
           &             &                    &           &       &                     & 30.0    & 14.32   & $<12.8$             \\
\hline                                
\end{tabular}
\end{table*}

   \begin{figure}
   \includegraphics[width=\hsize]{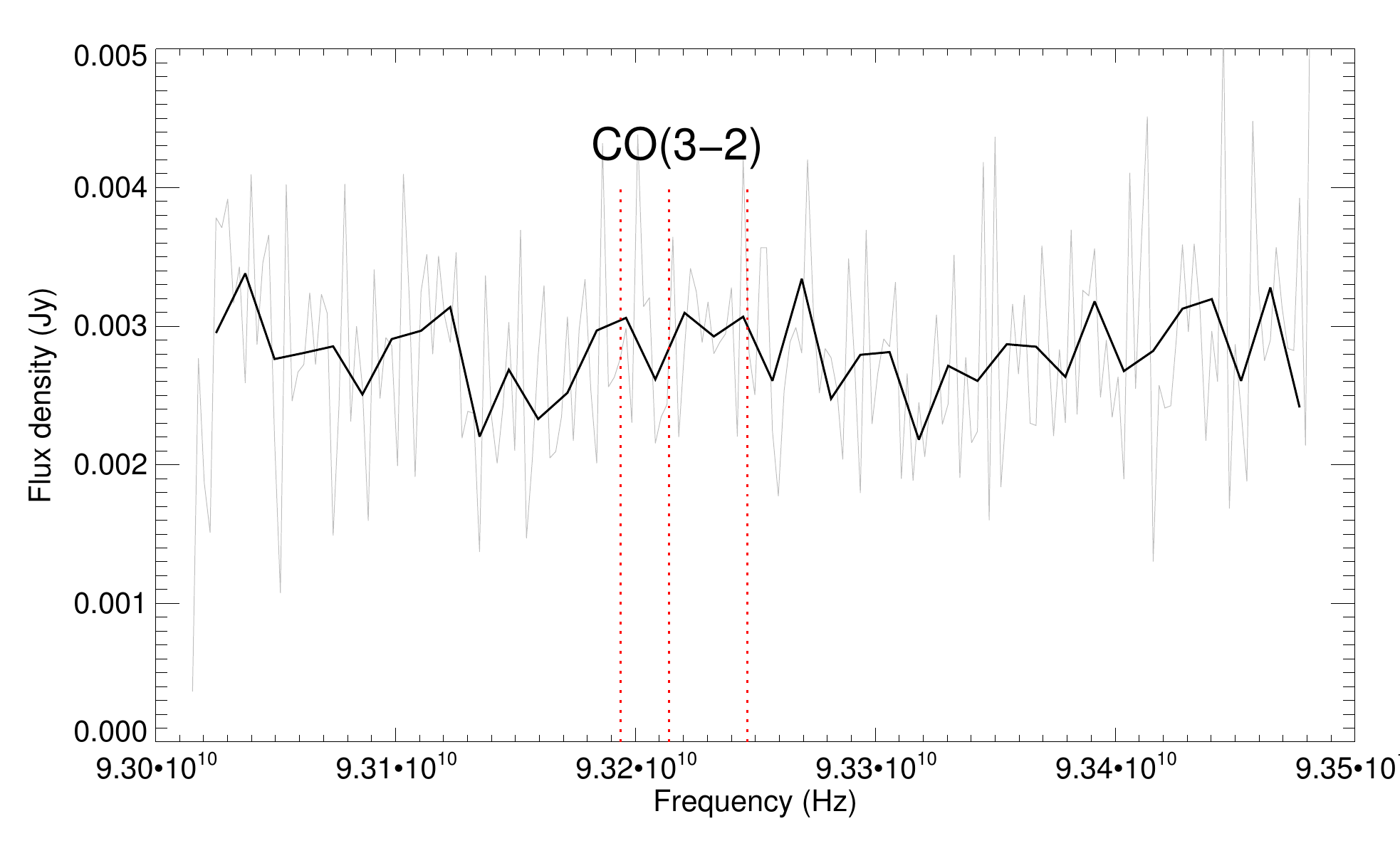}
   \includegraphics[width=\hsize]{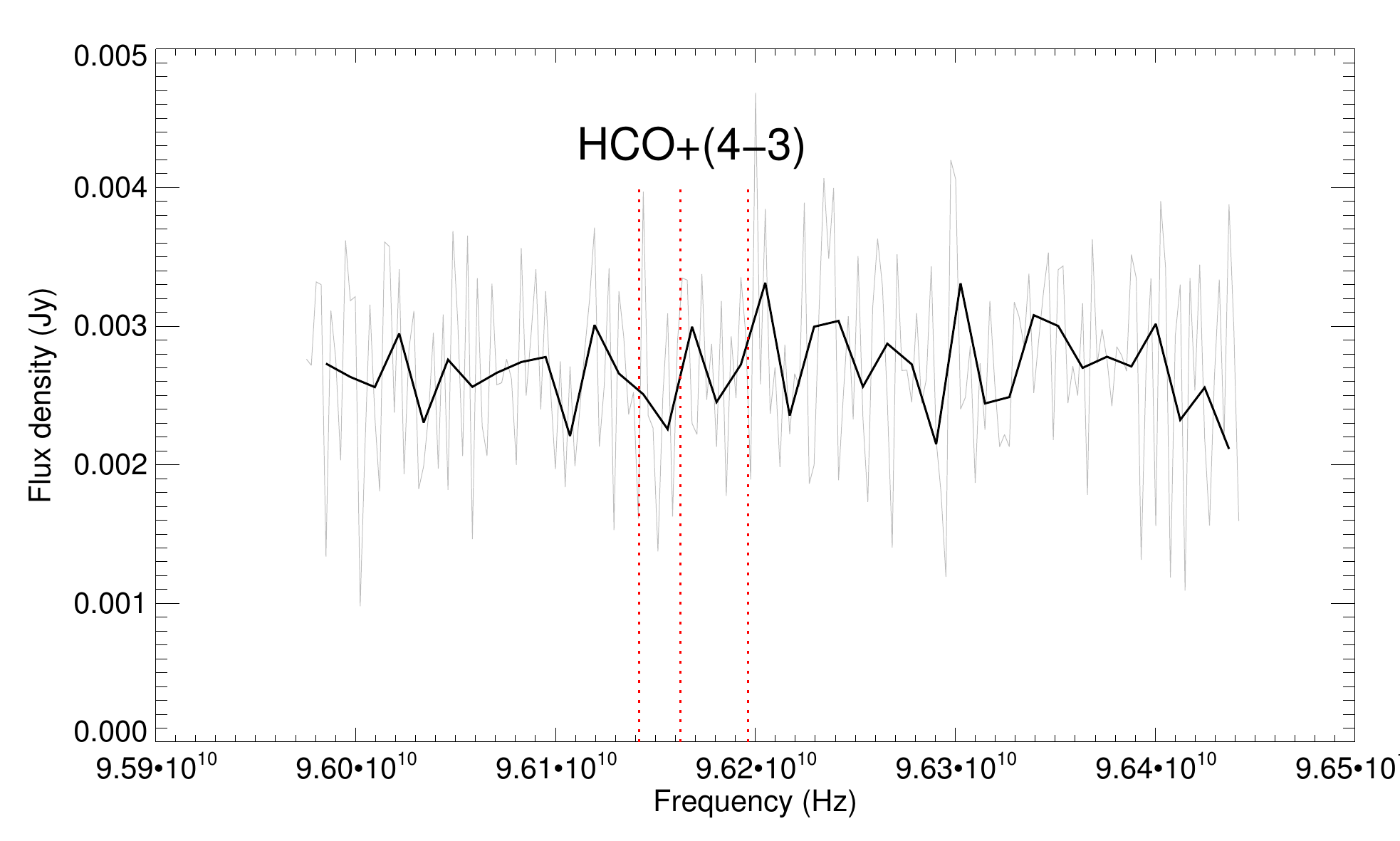}
      \caption{Regions of the ALMA spectra covering the CO(3-2) (top) and HCO+(4-3) (bottom) features. We do not see any significant absorption in any of the cases, allowing us only to place limits on the molecular content of the line of sight. The grey spectrum has a resolution of 2 km/s whereas the black line is for a binned resolution of 10 km/s. The red vertical lines mark the expected location of features at the redshift of the three velocity components seen in the optical spectra.
              }
         \label{FigALMAspec}
   \end{figure}

In this case, our attempt to detect absorption features in the line of sight of the GRB yielded only non-detections, from which we derive 
the abundance of molecules in the GRB host galaxy. Here, the host-galaxy environment is nearly dust-free, and has a lower-than-average metal content, in spite of the large amount of metal features that have been detected. In the future, millimetre and submillimetre absorption spectroscopy will have better chances of success if the search is focused towards bright bursts (to increase the signal-to-noise ratio) and/or highly-extinguished lines of sight, with plenty of dust and a high column density of metals, generally known as optically dark bursts.

\subsubsection{Limits on the detection of molecules from X-shooter spectroscopy}

   \begin{figure*}
   \centering
   \includegraphics[width=18cm]{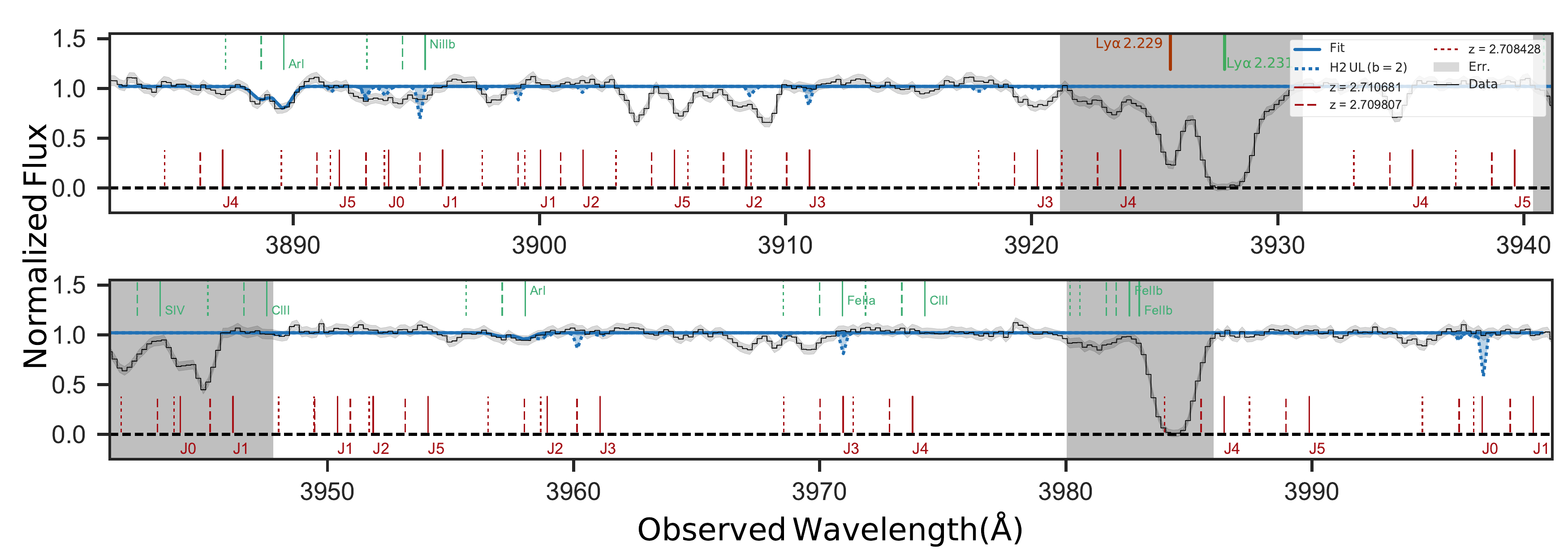}
      \caption{X-shooter spectrum from 3862.0\AA\, to 4001.6\AA. In green and
      red we indicate ion lines as well as the Lyman-Werner absorption bands of
      molecular hydrogen for each of the three absorption components given in Table
      \ref{tab:voigt}. The blue line shows the model with $N_{\mathrm{J0-5}}(\mathrm{H2}) <
      10^{15.2}\mathrm{cm}^{-2}$, including absorption from \ion{Ar}{i} in the GRB host
      galaxy. The grey shaded areas were excluded from the fit.}
         \label{fig:h2abs}
   \end{figure*}

We use the method developed in Bolmer et al. (in prep.) to search for absorption
of molecular hydrogen and carbon monoxide at the redshift of the individual
absorption components given in Table \ref{tab:voigt}. We use the {\tt PyMC}
Markov Chain Monte Carlo (MCMC) python package\footnote{https://pymc-devs.github.io/pymc/} to 
fit parts of the normalised X-shooter spectrum with the Lyman-Werner absorption
bands of molecular hydrogen or the CO bandheads, including any number of blending ion
lines for each of the individual absorption components, with the column density and $b$
as a free parameter (the Voigt profiles are convolved with the instrumental resolution).
In the case of molecular hydrogen, we here use the rotational levels
from $J=0$ to $J=5$ with a common $b$ parameter but individual column densities.
Although the Lyman and Werner bands fall between 910 \AA\, and 1152 \AA\, (rest frame),
we limit the fitted part of the spectrum to the range between 1040.7 \AA\, and 1078.4 \AA,
corresponding to 3862.0 \AA\, to 4001.6 \AA\ in the observer frame,
due to the sparseness of intervening systems and absorption lines in that region (see
Fig. \ref{fig:xs_spec}). Additionally we exclude several parts of the spectrum as
indicated by the grey shaded areas in Fig. \ref{fig:h2abs}. The spectrum shows evidence
for absorption from \ion{Ar}{i}, which was thus included in the fit.
We find no evidence for absorption from molecular hydrogen with an upper limit of
$N_{\mathrm{J0-5}}(\mathrm{H_2}) < 10^{15.2}\,\mathrm{cm}^{-2}\,(3\sigma)$, when assuming $b=2.0$ km $s^{-1}$
(smaller $b$ values lead to higher upper limits for the column density and therefore give
the most conservative result). 

   \begin{figure}
   \includegraphics[width=\hsize]{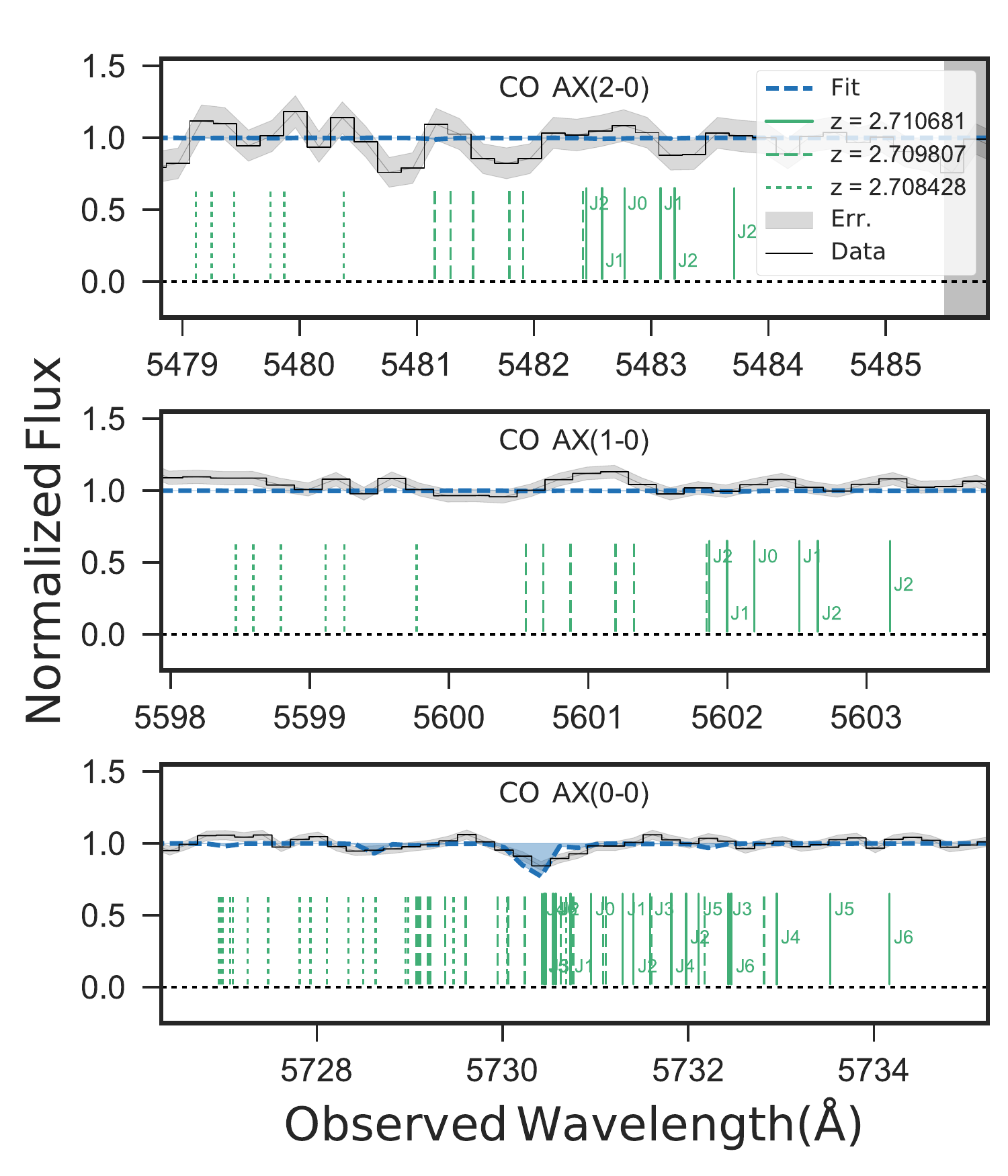}
      \caption{Three parts of the X-shooter spectrum covering CO AX(0-0), AX(1-0),
      and AX(2-0) lines, showing no indication for absorption from carbon monoxide down
      to $N_{\mathrm{J0-6}}(\mathrm{CO}) < 10^{14.5}\,\mathrm{cm}^{-2}$.}
         \label{fig:coabs}
   \end{figure}

We also find no evidence for absorption by carbon monoxide with an upper limit of
$N_{\mathrm{J0-6}}(\mathrm{CO}) < 10^{14.5}\,\mathrm{cm}^{-2}\,(3\sigma)$ when simultaneously
fitting the rotational levels from $J = 0$ to $J = 6$ of CO AX(0-0), AX(1-0), and AX(2-0)
when assuming $b=2.0$ km/s. The result is shown in Fig. \ref{fig:coabs}.

\subsubsection{Distance scales from the fine-structure lines}
\label{sec:finestruc}

Excited states in the surrounding medium of the GRB are routinely detected in afterglow spectra (see, e.g., \citealt{har13} and references therein). There is a general consensus that these features are produced by indirect UV pumping by the afterglow, that is, through the population of higher levels followed by the depopulation into the states responsible for the absorption features. This was demonstrated both by the detection of variability of fine-structure lines in multi-epoch spectroscopy \citep{vre07,del09}, and through the column-density ratios of different excited levels when multiple spectra were not available \citep{led09,del09b}.

The latter scenario holds for GRB\,161023A, where no multi-epoch observations are available. In fact, the high column density of the first metastable level of {\ion{Fe}{ii}} (a4F9/2, or 5s in Table 3) with respect to the fine-structure levels of the ground state (1s to 4s) can hardly be explained with a level population distribution given by a Boltzmann function \citep{vre07}, meaning that collisional excitations can be rejected at very high significance. The lack of multi-epoch spectroscopy does not completely rule out the possibility that the exciting UV flux comes from regions with high star formation rates and not from the GRB alone. In fact, fine-structure emission lines are present in high-redshift Lyman-break galaxies, but usually with lower column densities than observed in GRBs  \citep[see][]{sha03}. Assuming that the UV flux comes from the GRB, we can estimate the GRB-absorber distance, comparing the observed column densities to those predicted by a time-dependent photo-excitation code for the time when the spectroscopic observations were acquired. The photo-excitation code is that used by \citet{vre07} and \citet{del09}, to which we refer the reader for more details. Our equations take into account the $(4\pi)^{-1/2}$ correction factor to the flux experienced by the absorbing gas described by \citet{vre11}. We assume that the species for which we are running the code are in the ground state before the UV flash from the GRB reaches the gas. The behaviour of the GRB emission before the X-shooter observation was estimated using the light curve and spectral indices derived in Sect.~\ref{sect:SED}, with no spectral variation assumed during the time interval between the burst and our observation. 

We here concentrate on the {\ion{Fe}{ii}} levels because the multiple excited lines allow for a more precise fitting of the data. The initial column densities of the ground states were computed from the observed column densities of all the levels of each ion, that is, we are assuming that the species are not excited at $t = 0$. The initial values for the three components in which excited levels are detected are the following:  $\log (N_\ion{Fe}{ii}/{\rm cm}^{-2}) = 14.49$,   $\log (N_\ion{Fe}{ii}/{\rm cm}^{-2}) = 14.28$, and  $\log (N_\ion{Fe}{ii}/{\rm cm}^{-2}) = 13.74$ for component I, II, and IV, respectively. The Doppler parameter $b$ was fixed to $20$ km s$^{-1}$, which is the average value for the transitions of {\ion{Fe}{ii}} in the three components. Allowing for a variation in the range $10-30$ km s$^{-1}$ does not change our results significantly. We find the following values for the GRB-absorber distance: $d_I=990 \pm 60$ pc, $d_{II} = 820\pm 90$ pc, and $d_{IV}=700\pm 140$ pc for the three components. While for components II and IV the reduced $\chi^2$ values (in the range $1-2$) give a good agreement between data and model, for component I this is not the case. We allow for a variation of the total column density of {\ion{Fe}{ii}} to explore the possibility that the limited resolution of X-shooter may not allow us to distinguish between two sub-components at the velocity of component I, one close to the GRB and responsible for the excited level and one further away, not related to these transitions. However, the best fit to our data is still obtained with the original column density. We then try to compare data and model allowing for larger error bars in the excited levels and we obtain a satisfactory fit. The best value of the GRB absorber does not vary, but the computed error is somewhat larger ($\sim140$ pc instead of $60$ pc). This is a possible indication that the errors on the excited levels in the strongest component I are somewhat underestimated by the Voigt fitting procedure, since we cannot fully exclude saturation in this component.

\subsubsection{Dynamical analysis of the different components}

All resonant and fine-structure/metastable transitions show the same velocity structure of three (for some stronger lines four) absorption components at 0, --76 (--142) and --188 km\,s$^{-1}$ with a slightly different velocity for the second component of \ion{Mg}{I} and \ion{Mg}{II}. Both resonant and fine-structure transitions have the red-most component as the strongest absorption component which we therefore define as $v=0$ km\,s$^{-1}$. The only exception is \ion{Mg}{I} which has as strongest component the blue-most component, which is also obvious in the blend of \ion{Zn}{II} 2026 and \ion{Mg}{I} 2026, where the strongest component in the complex is the bluest line of \ion{Mg}{I}. \ion{O}{I}, on the other hand, shows a normal pattern like the other single ionised and fine-structure transitions. Such an ``outlier'' behaviour for \ion{Mg}{I}, but not for example for \ion{O}{I}, was also found in the host of GRB\,100219A \citep{tho13}. This might be an indication for Mg, or at least neutral Mg, being associated with gas different from the bulk of the ISM, for example in the halo of the galaxy.

The relative strength of the fine-structure versus resonant components seems to be similar in all three components, implying that the impact of the GRB radiation reaches across all of the absorbing gas, which is not usually the case \citep[see e.g.][]{tho08, tho13, del14}. This is in line with the distance of the absorption components derived from the modelling of the fine-structure line strength (see Sect. \ref{sec:finestruc}). Despite large uncertainties, our analysis shows that the least-redshifted component is actually the one closest to the GRB. If the GRB were located somewhere inside the central parts of the host, we would expect a more ordered spatial distribution of absorbing clouds along the sight line in the host, as this has usually been observed in GRB hosts \citep[see e.g.][]{del10, del11, tho13}. However, here we see clouds at very different velocities within only a few 100pc. In addition, there is a large total velocity distribution of more than 250 km\,s$^{-1}$, too high for the velocity field expected for a galaxy with a mass as low as for GRB\,161023A, besides that, those galaxies often do not even show ordered velocity fields. This points to a more turbulent gas distribution than has been usually observed for GRB hosts.

All high-ionisation lines again show the same velocity structure (except for \ion{P}{V} as noted above), however, with relative strengths different for the resonant and fine-structure lines. The structure consists of a deep central absorption at --114 km\,s$^{-1}$ with additional components in the wings. We also note a small additional component redwards of the main absorption line at +6 km\,s$^{-1}$ for \ion{Si}{IV}, \ion{C}{IV} and possibly \ion{O}{VI}, although the latter could be contaminated by Ly forest absorption. The velocities of the main component and the additional component at +6 km\,s$^{-1}$ are almost consistent with component I and III of the resonant lines. We can only speculate whether this is a coincidence or whether it implies that the material is actually at the same position in the galaxy but with very different relative strengths. High- and low-ionisation transitions frequently do not trace each other \citep[see e.g.][]{tho08,tho13, har15}. \cite{hei18} made a complete study of high-ionisation lines from the X-shooter GRB sample, including data from this burst. They found two classes of high-ionisation systems: those at the bulk velocity of the host absorption and those blueshifted by a few 100 km\,s$^{-1}$, the latter being associated with systems with log(N$_{H}$/cm$^{-2}$)~$<21.5$. This is interpreted as the high-ionisation absorption being associated with the HII region of the GRB and photoionised by the event, but at low N$_H$ the absorption only occurs at the edge of the HII region. An earlier study claimed that high-ionisation absorptions at several 1000 km\,s$^{-1}$ from the GRB redshift \citep{fox08} were likely all associated with galaxies near the GRB host, along the sight line; in fact, the closest intervening system of GRB\,161023A is only 1200 km\,s$^{-1}$ from the GRB redshift (see Sect. \ref{sec:int}). Also, only \ion{C}{IV} and \ion{S}{IV} have been detected in these systems, but none of the absorption species with high ionisation energies such as \ion{N}{V}.

The total absorption complex spans a velocity of nearly 200 km\,s$^{-1}$, which is consistent with the value of the velocity field of a relatively large galaxy. \cite{ara18} established correlations between metallicity and velocity for GRB hosts based on earlier works on DLAs, for example by \cite{led06}. If the absorption can be associated entirely with rotation in the host, we can obtain a mass estimate of the host according to the Tully-Fischer relation at $z\sim2.6$ in \cite{ueb17} and get log(M*/M$_{\odot}$)$\sim$10.5, within the range of GRB hosts at that redshift \citep[see e.g.][]{per16}. As described above, the presence of fine-structure absorption at almost equal relative strength compared to the resonant absorption implies that the absorbing gas might be concentrated in only a part of the galaxy or that the GRB is located in the ``foreground'' part of its host, hence the velocity width would only imply a lower limit on the total mass of the host.

\subsection{Intervening systems}
\label{sec:int}

Additionally, beyond the GRB host-galaxy absorption, we observe nine different intervening absorption systems, many of them with several velocity components. Only one other GRB so far has a similarly high number of identified intervening systems, GRB 130606A \citep{har15}, which, however, was at a much higher redshift of $z=5.91$. In Table~\ref{table:int} we present the systems that we have identified and their detected absorption features as well as the different redshifts of the individual components, the table with the column densities of all intervening systems and their absorption lines can be found in the Appendix, Table~\ref{table:intN}. In Figs.~\ref{Fig:int1} -- \ref{Fig:int9} in the Appendix we plot the strongest lines for each intervening system to show the absorption profiles.

The intervening system with the highest redshift is only $\sim1200$ km\,s$^{-1}$ from the central redshift of the host galaxy. While this is probably too far for it to be associated with the host itself, it is likely a smaller satellite galaxy of the host or another companion in a galaxy group to which the host belongs, and not related to the outflow from the GRB as, for example suggested in \cite{fox08}. This is also supported by the fact that we detect \ion{Fe}{II} absorption in this system and not only high-ionisation lines. The next-highest intervening system at $z=2.6$ seems to have an extended halo of hydrogen and we can fit a velocity complex of nearly 500 km\,s$^{-1}$ in both Ly$\alpha$/Ly$\beta$ (Ly$\gamma$ is contaminated by another Ly forest system) and \ion{C}{IV} $\lambda\lambda$ 1548, 1550. The 1548 \AA{} line of the latter is contaminating the \ion{Si}{II} $\lambda$ 1526 absorption complex of the host galaxy but the 1550 \AA{} line is separated and can be fit with the corresponding components. The system could be even larger with a width of $\sim1000$ km\,s$^{-1}$ evident in Ly$\alpha$, but since Ly$\beta$ seems to be contaminated we cannot verify its existence. Another large halo might be present in the system at $z=2.5$, with a velocity width of $\sim450$ km\,s$^{-1}$ and four to five different components in Ly$\alpha$ and Ly$\beta$. A very interesting system is the one at $z=1.24$, with six components in both \ion{Mg}{II} lines (and 4 in \ion{Mg}{I} and \ion{Fe}{II}) spanning a width of almost 500 km\,s$^{-1}$. Velocity widths above $\sim400$ km\,s$^{-1}$ are very unusual for intervening absorbers both in the sight lines towards GRBs and QSO-DLAs \citep[see e.g.][]{ara18, ber16}. 

Only two intervening systems are sub-DLAs or Lyman limit systems (LLS), the one at $z=2.69$ (sub-DLA, with log log(N$_{H}$/cm$^{-2})=19.2$) and the one at $z=2.40$ (LLS, log(N$_{H}$/cm$^{-2})=18.6$), the other intervening systems have log log(N$_{H}$/cm$^{-2})<17$. Most of our intervening systems show absorption of high-ionisation lines of \ion{C}{IV} and \ion{Si}{IV}, usually associated with the haloes of galaxies. We detect \ion{Mg}{II}, \ion{Mg}{I} absorption only at the redshift of the GRB and for the two lowest redshift systems, at $z=1.14$ and 1.24. In spite of the large number of intervening systems detected in the line of sight toward GRB\,161023A, only the one at $z=1.24$ can be considered a strong Mg absorber with a rest frame EW of $1.73\pm0.02$ \AA{} for the \ion{Mg}{II} $\lambda2796$ line. Strong intervening Mg absorbers are commonly found along GRB and QSO sight lines \citep[see e.g,][]{ver09,cuc13,Christensen}. 

\begin{table}
\caption{Intervening systems along the line of sight to GRB\,161023A.}            
\label{table:int}     
\centering                       
\begin{tabular}{c c }       
\hline\hline                
Redshift & Features \\  
\hline              
2.6956 & \ion{C}{IV}, \ion{Fe}{II} \\
2.6952 & Ly-series (Ly-$\alpha$--Ly-7), \ion{C}{IV}, \ion{Fe}{II}\\ 
2.6949 & Ly-series (Ly-$\alpha$--Ly-7) \\
\hline 
2.6606 & Ly-series (Ly-$\alpha$--Ly-$\gamma$), \ion{Si}{IV}, \ion{C}{IV} \\ 
2.6594 & Ly-series (Ly-$\alpha$--Ly-$\gamma$), \ion{Si}{IV}, \ion{C}{IV} \\
2.6591 & Ly-series (Ly-$\alpha$--Ly-$\gamma$), \ion{C}{IV} \\
2.6584 & Ly-series (Ly-$\alpha$--Ly-$\gamma$), \ion{C}{IV}\\
2.6579 & Ly-series (Ly-$\alpha$--Ly-$\gamma$), \ion{C}{IV}\\
2.6572 & Ly-series (Ly-$\alpha$--Ly-5), \ion{C}{IV} \\ 
2.6551 & Ly-series (Ly-$\alpha$--Ly-$\gamma$),\ion{C}{IV} \\
2.6540 & Ly-series (Ly-$\alpha$--Ly-$\gamma$)  \\
\hline 
2.5275 & Ly-$\alpha$, Ly-$\beta$, \ion{C}{IV} \\  
2.5268 & Ly-$\alpha$, Ly-$\beta$,\ion{C}{IV} \\  
2.5257 & Ly-$\alpha$, Ly-$\beta$ \\
2.5240 & Ly-$\alpha$, Ly-$\beta$ \\
2.5227 & Ly-$\alpha$, Ly-$\beta$ \\
\hline 
2.4055 & Ly-$\alpha$, Ly-$\beta$, \ion{Si}{II}, \ion{Si}{IV}, \ion{C}{II}, \ion{C}{IV}, \ion{Al}{III} \\ 
2.4040 & Ly-$\alpha$, Ly-$\beta$, \ion{Si}{IV}, \ion{C}{II}, \ion{C}{IV}, \ion{Al}{III} \\ 
2.4029 & Ly-$\alpha$, Ly-$\beta$, \ion{Si}{IV}, \ion{C}{II}, \ion{C}{IV}, \ion{Al}{III} \\   
\hline 
2.3111  & \ion{C}{IV} \\
2.3103 & Ly-$\alpha$, \ion{C}{IV} \\ 
\hline
2.2313 & Ly-$\alpha$, \ion{C}{IV} \\  
2.2306 & Ly-$\alpha$, \ion{C}{IV} \\ 
\hline 
1.9267 & Ly-$\alpha$, \ion{Si}{IV}, \ion{C}{IV} \\ 
\hline 
1.2446 & \ion{Mg}{II} \\ 
1.2440 &\ion{Fe}{II},\ion{Mg}{II}, \ion{Mg}{I} \\ 
1.2434 &\ion{Fe}{II},\ion{Mg}{II}, \ion{Mg}{I} \\ 
1.2430 &\ion{Fe}{II}, \ion{Mg}{II},\ion{Mg}{I} \\ 
1.2425 &\ion{Fe}{II}, \ion{Mg}{II}, \ion{Mg}{I} \\ 
1.2413 & \ion{Mg}{II} \\   
\hline 
1.1387 & \ion{Mg}{II} \\
1.1382 & \ion{Mg}{II} \\
1.1376 & \ion{Fe}{II},\ion{Mg}{II} \\
\hline                                
\end{tabular}
\end{table}




\section{Discussion and conclusions}

\begin{table*}
\caption{Measurements of different parameters of GRBs for which molecular searches have been performed. The top part of the table shows events for which no H$_2$ was detected \citep{led09}. The second part of the table shows events for which there were successful H$_2$ detections \citep{pro09,del14,kru13,fri15}. The metallicities are based on Zn, except for those marked with $^a$ which use O as reference or those marked with $^b$ which use S. Extinctions earlier than 2010 are obtained from \citet{kan10} and \citet{per11} for the case of GRB\,080607. At the bottom we also put the values for GRB\,161023A with the mean metallicity derived in this work. The molecular fraction is defined as $f(\mathrm{H}_2)=2\mathrm{N}(\mathrm{H}_2)/(2\mathrm{N}(\mathrm{H}_2)+\mathrm{N}(\mathrm{H}))$.}            
\label{table:mol}     
\centering        
\scriptsize{
\begin{tabular}{l c c c c c c c c c}       
\hline\hline                
GRB &  Redshift & log(N$_H$/cm$^{-2}$) & log(N$_{H_2}$/cm$^{-2}$) & log(N$_{CO}$/cm$^{-2}$) & log($f$(H$_2$)) & N$_{CO}$/N$_{H_2}$ & [X/H]         & [Zn/Fe]       & $A_V$ (mag)  \\  
\hline
050730  & 3.969 & $22.10\pm0.10$       & $<13.8$                  & ---                     & $<-8.0$       & ---                & $-2.18\pm0.11^b$ & ---           & $0.10\pm0.02$ \\
050820A & 2.615 & $21.05\pm0.10$       & $<14.1$                  & ---                     & $<-6.7$       & ---                & $-0.39\pm0.10$ & $0.80\pm0.03$ & $0.07\pm0.01$             \\
050922C & 2.200 & $21.55\pm0.10$       & $<14.6$                  & ---                     & $<-6.7$       & ---                & $-1.82\pm0.11^a$ & ---           & $<0.03$             \\
060607A & 3.075 & $16.95\pm0.03$       & $<13.5$                  & ---                     & $<-3.1$       & ---                & ---            & ---           & $<0.24$             \\
071031  & 2.692 & $22.15\pm0.05$       & $<14.1$                  & ---                     & $<-7.8$       & ---                & $-1.73\pm0.05$ & $0.04\pm0.02$ & $0.14\pm0.13$      \\
080310  & 2.427 & $18.70\pm0.10$       & $<14.3$                  & ---                     & $<-4.2$       & ---                & $\leq-1.91\pm0.13^b$ & ---       & $0.19\pm0.05$             \\
080413A & 2.435 & $21.85\pm0.15$       & $<15.8$                  & ---                     & $<-5.7$       & ---                & $-1.60\pm0.16$ & $0.13\pm0.07$ &  $<0.51$            \\
\hline
080607  & 3.036 & $22.70\pm0.15$       & $21.2\pm0.15$            & $16.5\pm0.3$            & $<0.13$($\sim-1.2$) & $2\times10^{-5}$ & $>-0.2$    & ---            & $3.3\pm0.4$    \\
120327A & 2.813 & $22.01\pm0.09$       & $16.5\pm1.2$             & ---                     & $-5.2\pm1.2$  & ---                & $-1.17\pm0.11$    & $0.56\pm0.14$  & $<0.03$        \\
120815A & 2.358 & $21.95\pm0.10$       & $20.54\pm0.15$           & ---                     & $-1.14\pm0.15$& ---                & $-1.15\pm0.12$ & $1.01\pm0.10$  & $<0.15$        \\
121024A & 2.302 & $21.88\pm0.10$       & $19.85\pm0.15$           & $<14.4$                 & $-1.4$        & $<3\times10^{-6}$  & $\sim-0.3$     & $0.85\pm0.04$  & $0.09\pm0.06$  \\
\hline
161023A & 2.710 & $20.97\pm0.01$       & $<15.2$                  & $<14.5$                 & $<-5.0$       & ---                & $-0.95\pm0.05$ & $0.7\pm0.04$   & $0.09\pm0.03$  \\
\hline                                
\end{tabular}}
\end{table*}

GRB\,161023A was an exceptionally luminous GRB. Although being at a redshift of $z=2.710$, its afterglow reached a peak magnitude of $r^{\prime}=12.6$ mag (equivalent to an absolute magnitude of M$_U=-34.43\pm0.13$ mag). It was also energetic in the high-energy regime, although not extraordinary, emitting E$_{{\rm iso,\gamma}}=(6.3\pm0.9)\times10^{53}$ erg. The burst is consistent with the E$_p$-E$_{{\rm iso,\gamma}}$ correlation \citep{ama02}.
Its afterglow was amongst the few most luminous ever detected, both in optical and millimetre wavelengths.
As such, this burst is a perfect example of how GRB afterglows can be exceptional beacons to study lines of sight across a big fraction of the Universe's history, at the same time as they allow us to study their own host galaxy from the inside.

Our earliest observations, using the Watcher telescope, started just 40 seconds after the GRB onset. The very early optical light curve is characterised by optical flickering simultaneous to the gamma-ray emission. After 80 s, the emission rapidly rises, peaking at about 240 s and henceforth decaying steadily. This peak may represent a reverse-shock flash transitioning into a classical forward-shock afterglow, but a more detailed analysis is beyond the scope of this paper. This afterglow emission shows a constant decay until 2.81 days after the burst, when it steepens due to what we interpret as a jet break. From this break time we estimate a relatively narrow jet half opening angle of $\theta_j =3.74\pm0.09$ deg, well below the average value of 6.5 deg. This narrow jet contributed to the unusual observed luminosity of the event.

We construct an SED at the time of the ALMA observations, 2.1 days after the burst, combining millimetre data from ALMA, NIR and optical data from GROND, and X-ray data from {\it Swift}/XRT. We find that the Band 3 data are below the characteristic frequency $\nu_m$ whereas the Band 7 data are already above $\nu_m$ but still below $\nu_c$. The optical data seem to be in the same regime as the X-ray data, both beyond $\nu_c$, although the cooling break is probably not far from the NIR data at the time of the ALMA observation. 

Taking advantage of the brightness of GRB\,161023A we performed early spectroscopy at UV, optical and NIR wavelengths with X-shooter at ESO VLT and, for the first time, also in millimetre wavelengths with ALMA, in search of possible absorption from the CO and HCO+ molecules. 

The X-shooter spectrum shows a plethora of absorption features. We fit the absorption features using Voigt profiles to determine the column densities of all species at the redshift of the GRB, which shows three different velocity components within a range of 200 km s$^{-1}$. This was done independently for low- and high-ionisation lines, which showed a slightly different velocity structure, as has been seen before \citep{hei18}, and also for fine-structure and metastable levels, which are expected to be indirectly excited by the UV emission of the GRB.

Using these column densities we derive a metallicity of [S/H]~$=-1.15\pm0.04$ and [Zn/H]~$=-1.11\pm0.07$ for the host galaxy, which is an average metallicity for a GRB host at this redshift. Thanks to the good signal-to-noise of the spectrum and the abundance of resolved features of different species we can study the dust-depletion patterns of the three velocity components observed within the host galaxy of the GRB using recently established correlations based on [Zn/Fe] for QSO-DLAs by \cite{dec16,dec17}. We measure only a mild depletion for components I and III and the average of the host with [Zn/Fe] $=0.5-0.7$. Component IV, which is the one closest to the GRB, might have a higher depletion, similar to MW-disc-like extinction, but we caution against over-interpreting our findings as Zn absorption is very weak in this component. The dust-corrected average metallicity of the host of GRB 161023A is $-0.94\pm0.08$. The dust-to-metals ratio derived from the depletion analysis is average for a DLA at that redshift and metallicity \citep{dec16}. The estimate of the total extinction of $A_V=0.04$ mag is not far from the value derived from SED fitting. 

We assume that the fine-structure and metastable levels detected for \ion{Fe}{II} are excited by indirect UV pumping, as has been shown for other GRBs in the past. Combining the column densities of these species and the information on spectral and decay slopes of the afterglow to predict how the afterglow would have excited them, we estimate the physical distance to the three systems detected in the host galaxy: $d_I=990 \pm 60$ pc, $d_{II} = 820\pm 90$ pc, and $d_{IV}=700\pm 140$ pc. Hence, the closest system to the GRB afterglow is in this case the one that is blueshifted the furthest in the spectrum.

\begin{figure}
   \includegraphics[width=\hsize]{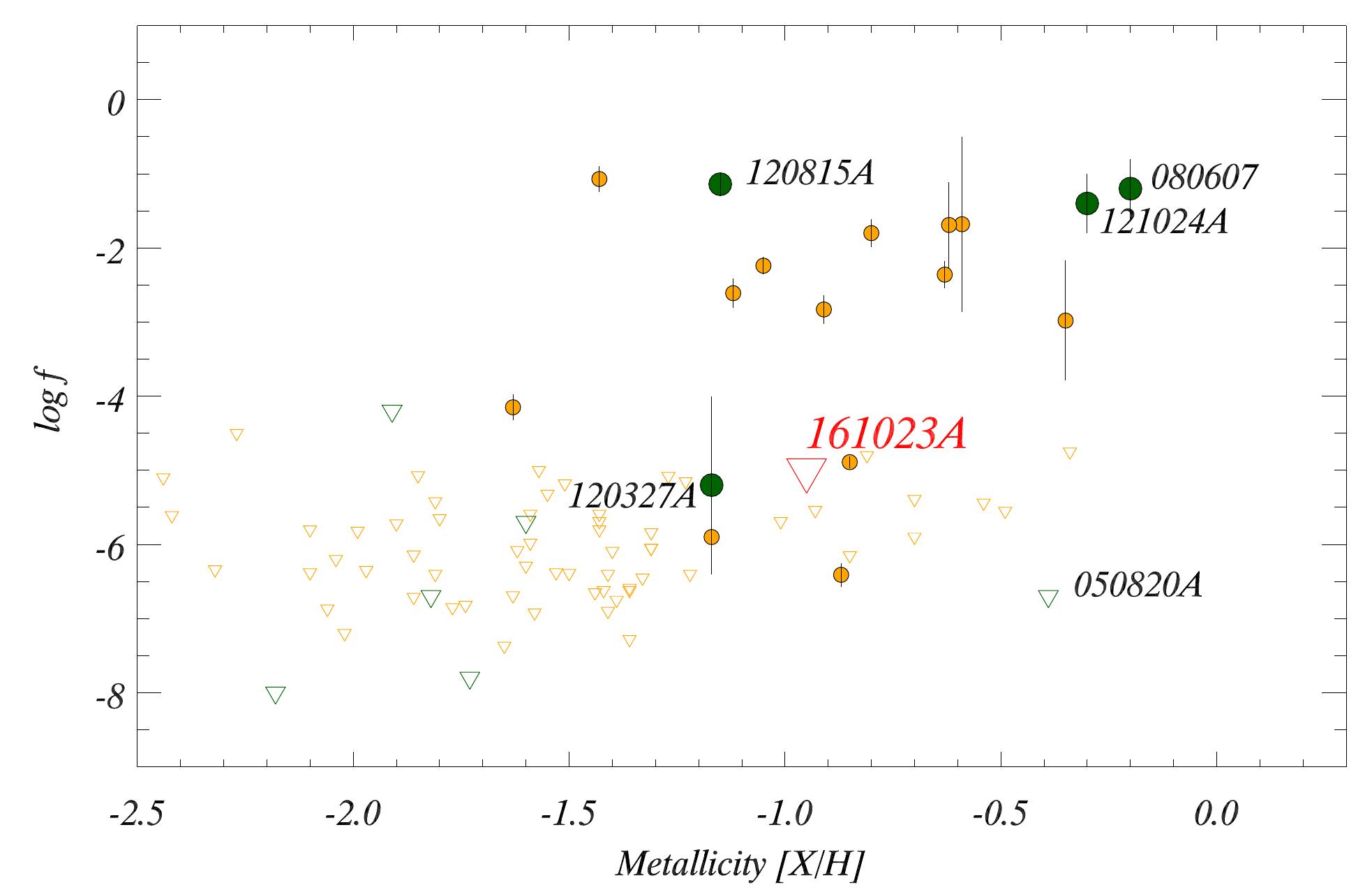}
      \caption{Molecular fraction of GRB (in green, Table~\ref{table:mol}) and quasar DLAs \citep[in yellow, from][]{not08} compared to their metallicity. In both cases the molecular fraction is higher at higher metallicity.}
         \label{fig:molfrac}
   \end{figure}

The ALMA observations did not yield a detection of molecules but allow us to impose limits on the column density of CO and HCO+ along the GRB line of sight, and can be used as a guide for future observations. We determine a limit for the CO molecule from the CO(2-3) transition of log(N$_{CO}$/cm$^{-2}$)~$<15.7$ and for HCO+ of log(N$_{HCO+}$/cm$^{-2}$)~$<13.2$. On the other hand, the redshift of the GRB is ideal for the search for molecular features in the ultraviolet range. Based on a search for the rotational Lyman-Werner bands of hydrogen molecules in the Lyman-$\alpha$ forest we impose a detection limit of log(N$_{H_2}$/cm$^{-2}$)~$<15.2$. We also look for carbon monoxide in the X-shooter spectrum, determining a limit of log(N$_{CO}$/cm$^{-2}$)~$<14.5$. These values can be compared to the ones shown in Table~\ref{table:mol} for a sample of GRB afterglow spectra in which a search for molecules was performed. The limit obtained for N$_{H_2}$ in GRB\,161023A would have been good enough to allow the detection of molecular hydrogen features for similar column densities as those previously measured in some GRBs. Furthermore, for a case like GRB\,080607, the most extreme line of sight regarding molecular content, ALMA would have easily detected CO features. The host galaxy of GRB\,161023A has a molecular fraction smaller than 10$^{-5}$. This is much lower than cases like GRBs\,080607, 120815A or 121024A, which are all in the range between 0.03 and 0.1, but consistent with the measurement of GRB\,120327A at $10^{-5.2}$, and comparable with the sample of \cite{led09}, which showed no detections down to limits that reached 10$^{-8}$. In Fig.~\ref{fig:molfrac} we compare the molecular fraction and the metallicity of GRBs and quasar DLAs. We notice that although there does not seem to be a direct correlation between the molecular fraction and the metallicity, detections of H$_2$ in both GRB and quasar DLA spectra have only been achieved for higher metallicities: in the case of GRBs, only for [X/H]~$>-1.2$. Furthermore, there seems to be a clustering of molecular fractions between $-3<$~log($f$)~$<-0.5$ with a gap of more than an order of magnitude until the following detection. There does not seem to be a correlation either for the molecular fraction with extinction. It is known, however, that high-extinction lines of sight normally imply high column densities of hydrogen and consequently will make the detection of H$_2$ easier. Furthermore, the presence of molecules such as CO are known to be correlated with dust clouds, implying that their detection will be more likely in highly extinguished lines of sight. As a unique significant example, the extreme sight line toward GRB\,080607 had both, close-to-solar metallicity and high extinction. 

We note that both ultraviolet and millimetre methods complement each other well at this redshift. However, at lower redshifts ($z\lesssim2.2$) the UV method stops being useful as the features are no longer visible from ground-based observatories and at higher redshifts ($z\gtrsim4.0$) the Lyman forest becomes more populated and less transparent and the analysis becomes unpractical. On the other hand, millimetre observations do not have this redshift constraint. Moreover, most of these molecules are correlated with the presence of dust, implying that the best chance of measuring them is along dusty sight lines. However, this same dust strongly attenuates ultraviolet radiation, making optical studies much more complicated than millimetre ones, as these are not affected by extinction. Hence, we expect that in the future, searches for molecules at millimetre wavelengths will be more successful for very bright afterglows and for those with very high optical extinction due to dust (dark bursts).

The sight line of GRB\,161023A is amongst the ones with the largest number of intervening systems detected up to now \citep{fyn09,har15,sel18}. We find nine independent absorption systems ranging from $z=1.138$ to $z=2.695$, most of them with multiple velocity components, for a total of 34 different components. Only one system has a detection of strong \ion{Mg}{II}. We performed Voigt profile fits to the components for which a fit was feasible. The ALMA observations allowed us to serendipitously tentatively identify four emitters at 350 GHz within 7$^{\prime\prime}$ of the GRB. Future studies may tell if any of these systems are related to the intervening systems detected in the spectrum.

\begin{acknowledgements}
      
      AdUP and CT acknowledge support from Ram\'on y Cajal fellowships RyC-2012-09975 and RyC-2012-09984 and the Spanish Ministry of Economy and Competitiveness through projects AYA2014-58381-P and AYA2017-89384-P, AdUP furthermore from the BBVA foundation. DAK acknowledges support from the Spanish research project AYA 2014-58381-P, and from Juan de la Cierva Incorporaci\'on fellowship IJCI-2015-26153.
      FEB acknowledges support from CONICYT-Chile (Basal-CATA PFB-06/2007) and the Ministry of Economy, Development, and Tourism's Millennium Science Initiative through grant IC120009, awarded to The Millennium Institute of Astrophysics, MAS.
      Part of the funding for GROND (both hardware as well as personnel) was generously granted from the Leibniz-Prize to Prof. G. Hasinger "(DFG grant HA 1850/28-1).JB acknowledges support through the Sofja Kovalevskaja Award to P. Schady from the Alexander von Humboldt Foundation of Germany.
      MJM acknowledges the support of the National Science Centre, Poland through the POLONEZ grant 2015/19/P/ST9/04010; this project has received funding from the European Union's Horizon 2020 research and innovation programme under the Marie Sk{\l}odowska-Curie grant agreement No. 665778. AG acknowledges the financial support from the Slovenian Research Agency (research core funding No. P1-0031 and project grant No. J1-8136) and networking support by the COST Action GWverse CA16104. 
      
\end{acknowledgements}

\bibliographystyle{aa}
\bibliography{GRB161023A_ALMAXS}

\begin{appendix}

\section{Photometry and spectroscopy tables}

\longtab[1]{
\begin{longtable}{c c c c c}   
\caption{Optical/NIR observations. Magnitudes (in the AB system) are as observed without correction for Galactic extinction, whereas the flux densities have been corrected for it. 
\label{tab:phot}}\\
\hline\hline                 
T-T$_0$ 			& Telescope/Instrument 	& Band			& Magnitude 	& Flux density\\    
(days)     			&                        &     		      	&			& ($\mu$Jy) 	\\
\hline                        
\endfirsthead
\caption{continued.}\\
\hline\hline 
T-T$_0$ 			& Telescope/Instrument 	& Band			& Magnitude 	& Flux density\\     
(days)     			&                       &     		      		&			& ($\mu$Jy) \\
\hline                        	
\endhead
\hline
\endfoot
        0.46868 & 2.0mFaulkes   & $g^\prime$     & 19.08$\pm$0.07 &   94.02$\pm$  6.07  \\
        0.53520 & 1.0mZadko-Gingin & $g^\prime$  & $>18.8$ &   $<121.67$ \\
        0.56466 & 2.0mFaulkes   & $g^\prime$     & 19.30$\pm$0.07 &   76.77$\pm$  4.95  \\
        1.05378 & 2.2mMPG/GROND & $g^\prime$     & 19.82$\pm$0.04 &   47.42$\pm$  1.70  \\
        1.06005 & 2.2mMPG/GROND & $g^\prime$     & 19.80$\pm$0.02 &   48.43$\pm$  0.88  \\
        1.06869 & 2.2mMPG/GROND & $g^\prime$     & 19.79$\pm$0.02 &   48.67$\pm$  0.76  \\
        1.08363 & 2.2mMPG/GROND & $g^\prime$     & 19.84$\pm$0.02 &   46.85$\pm$  0.80  \\
        1.09443 & 2.2mMPG/GROND & $g^\prime$     & 19.85$\pm$0.02 &   46.45$\pm$  0.79  \\
        1.12178 & 2.2mMPG/GROND & $g^\prime$     & 19.86$\pm$0.02 &   45.66$\pm$  0.78  \\
        1.13536 & 2.2mMPG/GROND & $g^\prime$     & 19.90$\pm$0.02 &   44.13$\pm$  0.75  \\
        2.06320 & 2.2mMPG/GROND & $g^\prime$     & 20.82$\pm$0.02 &   19.01$\pm$  0.34  \\
        2.07639 & 2.2mMPG/GROND & $g^\prime$     & 20.85$\pm$0.02 &   18.45$\pm$  0.31  \\
        3.06313 & 2.2mMPG/GROND & $g^\prime$     & 21.37$\pm$0.03 &   11.43$\pm$  0.28  \\
        3.07624 & 2.2mMPG/GROND & $g^\prime$     & 21.36$\pm$0.02 &   11.55$\pm$  0.22  \\
        4.06666 & 2.2mMPG/GROND & $g^\prime$     & 21.99$\pm$0.03 &    6.44$\pm$  0.16  \\
        4.07855 & 2.2mMPG/GROND & $g^\prime$     & 22.00$\pm$0.02 &    6.41$\pm$  0.14  \\
        5.06072 & 2.2mMPG/GROND & $g^\prime$     & 22.48$\pm$0.04 &    4.11$\pm$  0.14  \\
        6.08295 & 2.2mMPG/GROND & $g^\prime$     & 22.83$\pm$0.05 &    2.96$\pm$  0.13  \\
       14.09933 & 2.2mMPG/GROND & $g^\prime$     & $>24.43$       & $<   0.68$          \\
\hline
        0.00051 & 0.4mWatcher   & $r^\prime$     & 16.03$\pm$0.03 &  1514.96$\pm$ 37.55  \\
        0.00058 & 0.4mWatcher   & $r^\prime$     & 14.82$\pm$0.03 &  4583.53$\pm$116.47  \\
        0.00066 & 0.4mWatcher   & $r^\prime$     & 16.33$\pm$0.04 &  1146.04$\pm$ 41.65  \\
        0.00074 & 0.4mWatcher   & $r^\prime$     & 15.94$\pm$0.03 &  1645.89$\pm$ 53.41  \\
        0.00085 & 0.4mWatcher   & $r^\prime$     & 16.23$\pm$0.03 &  1253.14$\pm$ 36.06  \\
        0.00092 & 0.4mWatcher   & $r^\prime$     & 15.91$\pm$0.03 &  1688.88$\pm$ 49.81  \\
        0.00100 & 0.4mWatcher   & $r^\prime$     & 15.48$\pm$0.02 &  2502.65$\pm$ 56.28  \\
        0.00108 & 0.4mWatcher   & $r^\prime$     & 15.21$\pm$0.02 &  3224.04$\pm$ 69.26  \\
        0.00115 & 0.4mWatcher   & $r^\prime$     & 15.15$\pm$0.03 &  3397.82$\pm$ 86.34  \\
        0.00123 & 0.4mWatcher   & $r^\prime$     & 14.59$\pm$0.03 &  5712.15$\pm$134.76  \\
        0.00134 & 0.4mWatcher   & $r^\prime$     & 14.23$\pm$0.02 &  7892.23$\pm$181.74  \\
        0.00142 & 0.4mWatcher   & $r^\prime$     & 13.76$\pm$0.02 & 12189.89$\pm$274.12  \\
        0.00150 & 0.4mWatcher   & $r^\prime$     & 13.60$\pm$0.02 & 14164.45$\pm$311.22  \\
        0.00157 & 0.4mWatcher   & $r^\prime$     & 13.53$\pm$0.02 & 15135.61$\pm$348.54  \\
        0.00165 & 0.4mWatcher   & $r^\prime$     & 13.47$\pm$0.02 & 15966.14$\pm$350.81  \\
        0.00173 & 0.4mWatcher   & $r^\prime$     & 13.25$\pm$0.02 & 19498.44$\pm$428.42  \\
        0.00184 & 0.4mWatcher   & $r^\prime$     & 13.12$\pm$0.03 & 22120.75$\pm$562.10  \\
        0.00191 & 0.4mWatcher   & $r^\prime$     & 13.05$\pm$0.03 & 23593.91$\pm$556.63  \\
        0.00199 & 0.4mWatcher   & $r^\prime$     & 12.97$\pm$0.02 & 25374.64$\pm$570.60  \\
        0.00207 & 0.4mWatcher   & $r^\prime$     & 13.03$\pm$0.02 & 23900.11$\pm$550.37  \\
        0.00214 & 0.4mWatcher   & $r^\prime$     & 12.89$\pm$0.03 & 27164.38$\pm$640.87  \\
        0.00222 & 0.4mWatcher   & $r^\prime$     & 12.95$\pm$0.02 & 25680.29$\pm$577.48  \\
        0.00251 & 0.4mWatcher   & $r^\prime$     & 12.92$\pm$0.02 & 26448.43$\pm$544.74  \\
        0.00259 & 0.4mWatcher   & $r^\prime$     & 12.77$\pm$0.03 & 30338.90$\pm$751.95  \\
        0.00267 & 0.4mWatcher   & $r^\prime$     & 12.87$\pm$0.02 & 27771.53$\pm$596.63  \\
        0.00274 & 0.4mWatcher   & $r^\prime$     & 12.86$\pm$0.02 & 27925.43$\pm$613.57  \\
        0.00286 & 0.4mWatcher   & $r^\prime$     & 12.86$\pm$0.02 & 28028.50$\pm$630.28  \\
        0.00293 & 0.4mWatcher   & $r^\prime$     & 12.91$\pm$0.02 & 26693.15$\pm$573.47  \\
        0.00301 & 0.4mWatcher   & $r^\prime$     & 12.86$\pm$0.02 & 28028.50$\pm$615.84  \\
        0.00308 & 0.4mWatcher   & $r^\prime$     & 12.85$\pm$0.03 & 28313.91$\pm$684.59  \\
        0.00316 & 0.4mWatcher   & $r^\prime$     & 12.89$\pm$0.02 & 27164.38$\pm$596.85  \\
        0.00324 & 0.4mWatcher   & $r^\prime$     & 12.84$\pm$0.02 & 28654.96$\pm$615.61  \\
        0.00335 & 0.4mWatcher   & $r^\prime$     & 12.91$\pm$0.02 & 26767.01$\pm$601.92  \\
        0.00343 & 0.4mWatcher   & $r^\prime$     & 12.94$\pm$0.02 & 26013.56$\pm$558.87  \\
        0.00351 & 0.4mWatcher   & $r^\prime$     & 12.96$\pm$0.03 & 25609.43$\pm$650.75  \\
        0.00358 & 0.4mWatcher   & $r^\prime$     & 12.92$\pm$0.02 & 26497.19$\pm$582.19  \\
        0.00366 & 0.4mWatcher   & $r^\prime$     & 12.87$\pm$0.02 & 27643.93$\pm$593.89  \\
        0.00373 & 0.4mWatcher   & $r^\prime$     & 12.94$\pm$0.02 & 26037.53$\pm$559.38  \\
        0.00384 & 0.4mWatcher   & $r^\prime$     & 12.92$\pm$0.02 & 26594.99$\pm$612.43  \\
        0.00392 & 0.4mWatcher   & $r^\prime$     & 12.82$\pm$0.02 & 29026.85$\pm$623.60  \\
        0.00400 & 0.4mWatcher   & $r^\prime$     & 12.98$\pm$0.02 & 25142.00$\pm$465.47  \\
        0.00407 & 0.4mWatcher   & $r^\prime$     & 12.96$\pm$0.03 & 25585.85$\pm$618.62  \\
        0.00415 & 0.4mWatcher   & $r^\prime$     & 13.01$\pm$0.02 & 24299.64$\pm$546.43  \\
        0.00423 & 0.4mWatcher   & $r^\prime$     & 13.01$\pm$0.03 & 24322.03$\pm$588.07  \\
        0.00434 & 0.4mWatcher   & $r^\prime$     & 13.01$\pm$0.02 & 24479.35$\pm$550.47  \\
        0.00442 & 0.4mWatcher   & $r^\prime$     & 13.09$\pm$0.02 & 22740.49$\pm$499.65  \\
        0.00450 & 0.4mWatcher   & $r^\prime$     & 13.11$\pm$0.02 & 22181.96$\pm$487.38  \\
        0.00457 & 0.4mWatcher   & $r^\prime$     & 13.09$\pm$0.03 & 22636.01$\pm$534.03  \\
        0.00465 & 0.4mWatcher   & $r^\prime$     & 13.06$\pm$0.02 & 23334.57$\pm$524.73  \\
        0.00472 & 0.4mWatcher   & $r^\prime$     & 13.06$\pm$0.02 & 23334.57$\pm$480.61  \\
        0.00484 & 0.4mWatcher   & $r^\prime$     & 13.06$\pm$0.02 & 23227.36$\pm$510.35  \\
        0.00491 & 0.4mWatcher   & $r^\prime$     & 13.13$\pm$0.02 & 21797.16$\pm$490.16  \\
        0.00499 & 0.4mWatcher   & $r^\prime$     & 13.20$\pm$0.02 & 20398.58$\pm$448.19  \\
        0.00507 & 0.4mWatcher   & $r^\prime$     & 13.19$\pm$0.02 & 20739.57$\pm$466.37  \\
        0.00515 & 0.4mWatcher   & $r^\prime$     & 13.16$\pm$0.02 & 21144.62$\pm$464.59  \\
        0.00522 & 0.4mWatcher   & $r^\prime$     & 13.18$\pm$0.02 & 20854.50$\pm$458.21  \\
        0.00534 & 0.4mWatcher   & $r^\prime$     & 13.24$\pm$0.03 & 19696.99$\pm$476.24  \\
        0.00541 & 0.4mWatcher   & $r^\prime$     & 13.31$\pm$0.03 & 18501.20$\pm$436.48  \\
        0.00549 & 0.4mWatcher   & $r^\prime$     & 13.33$\pm$0.02 & 18230.55$\pm$400.56  \\
        0.00557 & 0.4mWatcher   & $r^\prime$     & 13.28$\pm$0.02 & 19072.16$\pm$428.88  \\
        0.00565 & 0.4mWatcher   & $r^\prime$     & 13.35$\pm$0.02 & 17881.33$\pm$402.10  \\
        0.00572 & 0.4mWatcher   & $r^\prime$     & 13.42$\pm$0.02 & 16734.00$\pm$367.68  \\
        0.00584 & 0.4mWatcher   & $r^\prime$     & 13.32$\pm$0.02 & 18281.00$\pm$411.09  \\
        0.00591 & 0.4mWatcher   & $r^\prime$     & 13.42$\pm$0.02 & 16672.47$\pm$374.92  \\
        0.00599 & 0.4mWatcher   & $r^\prime$     & 13.43$\pm$0.03 & 16519.61$\pm$389.73  \\
        0.00606 & 0.4mWatcher   & $r^\prime$     & 13.51$\pm$0.02 & 15360.31$\pm$337.49  \\
        0.00614 & 0.4mWatcher   & $r^\prime$     & 13.39$\pm$0.02 & 17218.68$\pm$387.20  \\
        0.00622 & 0.4mWatcher   & $r^\prime$     & 13.47$\pm$0.02 & 15922.08$\pm$349.84  \\
        0.00633 & 0.4mWatcher   & $r^\prime$     & 13.50$\pm$0.02 & 15488.16$\pm$356.66  \\
        0.00641 & 0.4mWatcher   & $r^\prime$     & 13.45$\pm$0.02 & 16338.04$\pm$358.98  \\
        0.00648 & 0.4mWatcher   & $r^\prime$     & 13.50$\pm$0.02 & 15502.43$\pm$356.99  \\
        0.00656 & 0.4mWatcher   & $r^\prime$     & 13.49$\pm$0.02 & 15703.62$\pm$361.62  \\
        0.00664 & 0.4mWatcher   & $r^\prime$     & 13.53$\pm$0.02 & 15107.75$\pm$339.73  \\
        0.00671 & 0.4mWatcher   & $r^\prime$     & 13.68$\pm$0.02 & 13121.99$\pm$295.08  \\
        0.00683 & 0.4mWatcher   & $r^\prime$     & 13.64$\pm$0.02 & 13677.28$\pm$307.56  \\
        0.00690 & 0.4mWatcher   & $r^\prime$     & 13.52$\pm$0.02 & 15191.47$\pm$341.61  \\
        0.00698 & 0.4mWatcher   & $r^\prime$     & 13.69$\pm$0.02 & 13097.85$\pm$294.53  \\
        0.00705 & 0.4mWatcher   & $r^\prime$     & 13.59$\pm$0.02 & 14269.21$\pm$320.87  \\
        0.00713 & 0.4mWatcher   & $r^\prime$     & 13.69$\pm$0.03 & 12989.72$\pm$314.07  \\
        0.00721 & 0.4mWatcher   & $r^\prime$     & 13.76$\pm$0.02 & 12223.62$\pm$274.88  \\
        0.00732 & 0.4mWatcher   & $r^\prime$     & 13.81$\pm$0.03 & 11619.83$\pm$288.00  \\
        0.00809 & 0.4mWatcher   & $r^\prime$     & 13.91$\pm$0.02 & 10616.95$\pm$218.67  \\
        0.00884 & 0.4mWatcher   & $r^\prime$     & 14.08$\pm$0.02 &  9086.57$\pm$187.15  \\
        0.00959 & 0.4mWatcher   & $r^\prime$     & 14.29$\pm$0.03 &  7509.31$\pm$177.16  \\
        0.01034 & 0.4mWatcher   & $r^\prime$     & 14.35$\pm$0.02 &  7079.46$\pm$155.55  \\
        0.04896 & 0.25mTAROT & Clear($r^\prime$)     & 15.62$\pm$0.20 & 2203.94$\pm$449.64  \\
        0.06458 & 0.25mTAROT & Clear($r^\prime$)     & 15.99$\pm$0.13 & 1567.47$\pm$202.68  \\
        0.08771 & 0.25mTAROT & Clear($r^\prime$)     & 16.78$\pm$0.36 &  757.18$\pm$300.26  \\
        0.10739 & 8.2mVLT/FORS2 & $R_{\rm special}$($r^\prime$)  & 16.82$\pm$0.04 &  729.79$\pm$ 26.89  \\        
        0.10851 & 8.2mVLT/X-shooter & $r^\prime$     & 16.83$\pm$0.05 &  723$\pm$ 33  \\
        0.10853 & 8.2mVLT/FORS2 & $R_{\rm special}$($r^\prime$)  & 16.84$\pm$0.04 &  716.47$\pm$ 26.40  \\
        0.11851 & 8.2mVLT/FORS2 & $R_{\rm special}$($r^\prime$)  & 17.03$\pm$0.05 &  601.45$\pm$ 27.71  \\
        0.14799 & 8.2mVLT/FORS2 & $R_{\rm special}$($r^\prime$)  & 17.44$\pm$0.05 &  412.29$\pm$ 18.99  \\
        0.44539 & 2.0mFaulkes   & $r^\prime$     & 18.57$\pm$0.03 &  145.61$\pm$  4.02  \\
        0.46184 & 2.0mFaulkes   & $r^\prime$     & 18.52$\pm$0.03 &  152.48$\pm$  4.21  \\
        0.55816 & 2.0mFaulkes   & $r^\prime$     & 18.69$\pm$0.03 &  130.38$\pm$  3.60  \\
        0.57271 & 1.0mZadko-Gingin & Clear($r^\prime$)     & 18.69$\pm$0.21 &  130.38$\pm$ 27.83  \\
        0.80740 & 2.0mFaulkes   & $r^\prime$     & 19.00$\pm$0.05 &   97.99$\pm$  4.51  \\
        0.81434 & 0.4mWatcher   & $r^\prime$     & 19.16$\pm$0.09 &    84.57$\pm$  7.34  \\
        0.84346 & 0.4mWatcher   & $r^\prime$     & 19.25$\pm$0.10 &    77.84$\pm$  7.68  \\
        0.87875 & 0.4mWatcher   & $r^\prime$     & 19.34$\pm$0.12 &    71.65$\pm$  8.70  \\

        1.05378 & 2.2mMPG/GROND & $r^\prime$     & 19.37$\pm$0.02 &   69.95$\pm$  1.32  \\
        1.05421 & 8.2mVLT/FORS2 & $R_{\rm special}$($r^\prime$)  & 19.39$\pm$0.05 &   68.42$\pm$  3.15  \\
        1.06005 & 2.2mMPG/GROND & $r^\prime$     & 19.36$\pm$0.02 &   70.11$\pm$  1.16  \\
        1.06869 & 2.2mMPG/GROND & $r^\prime$     & 19.38$\pm$0.02 &   69.08$\pm$  1.05  \\
        1.08363 & 2.2mMPG/GROND & $r^\prime$     & 19.42$\pm$0.02 &   66.74$\pm$  1.22  \\
        1.09443 & 2.2mMPG/GROND & $r^\prime$     & 19.43$\pm$0.02 &   66.06$\pm$  1.00  \\
        1.12178 & 2.2mMPG/GROND & $r^\prime$     & 19.43$\pm$0.02 &   65.76$\pm$  1.00  \\
        1.13536 & 2.2mMPG/GROND & $r^\prime$     & 19.48$\pm$0.02 &   62.92$\pm$  0.98  \\
        1.81227 & 2.0mFaulkes   & $r^\prime$     & 20.08$\pm$0.07 &   36.24$\pm$  2.34  \\
        2.06320 & 2.2mMPG/GROND & $r^\prime$     & 20.38$\pm$0.02 &   27.51$\pm$  0.55  \\
        2.07639 & 2.2mMPG/GROND & $r^\prime$     & 20.39$\pm$0.02 &   27.21$\pm$  0.55  \\
        3.06313 & 2.2mMPG/GROND & $r^\prime$     & 20.91$\pm$0.02 &   16.86$\pm$  0.34  \\
        3.07624 & 2.2mMPG/GROND & $r^\prime$     & 20.92$\pm$0.02 &   16.72$\pm$  0.32  \\
        3.54572 & 2.0mFaulkes   & $r^\prime$     & 21.11$\pm$0.13 &   14.03$\pm$  1.81  \\
        4.06666 & 2.2mMPG/GROND & $r^\prime$     & 21.53$\pm$0.02 &    9.53$\pm$  0.20  \\
        4.07855 & 2.2mMPG/GROND & $r^\prime$     & 21.57$\pm$0.03 &    9.21$\pm$  0.22  \\
        5.06072 & 2.2mMPG/GROND & $r^\prime$     & 22.02$\pm$0.03 &    6.08$\pm$  0.16  \\
        5.46725 & 2.0mFaulkes   & $r^\prime$     & 22.26$\pm$0.18 &    4.87$\pm$  0.90  \\
        6.08295 & 2.2mMPG/GROND & $r^\prime$     & 22.37$\pm$0.05 &    4.40$\pm$  0.21  \\
       14.09933 & 2.2mMPG/GROND & {\it r}     & 24.45$\pm$0.20 &    0.65$\pm$  0.13  \\
        519.431 & 8.2mVLT/FORS2 & $R_{\rm special}$($r^\prime$)  & $>26.5$ &   $<0.10$  \\
\hline
        0.45296 & 2.0mFaulkes   & $i^\prime$     & 18.38$\pm$0.03 &  170.29$\pm$  4.71  \\
        0.46904 & 2.0mFaulkes   & $i^\prime$     & 18.35$\pm$0.03 &  175.07$\pm$  4.84  \\
        0.55810 & 2.0mFaulkes   & $i^\prime$     & 18.50$\pm$0.03 &  152.48$\pm$  4.21  \\
        0.80747 & 2.0mFaulkes   & $i^\prime$     & 18.77$\pm$0.06 &  118.90$\pm$  6.57  \\
        1.05378 & 2.2mMPG/GROND & $i^\prime$     & 19.16$\pm$0.05 &   82.89$\pm$  3.98  \\
        1.06005 & 2.2mMPG/GROND & $i^\prime$     & 19.14$\pm$0.03 &   84.43$\pm$  1.98  \\
        1.06869 & 2.2mMPG/GROND & $i^\prime$     & 19.16$\pm$0.02 &   82.76$\pm$  1.60  \\
        1.08363 & 2.2mMPG/GROND & $i^\prime$     & 19.20$\pm$0.03 &   80.15$\pm$  2.16  \\
        1.09443 & 2.2mMPG/GROND & $i^\prime$     & 19.22$\pm$0.02 &   78.76$\pm$  1.65  \\
        1.12178 & 2.2mMPG/GROND & $i^\prime$     & 19.24$\pm$0.02 &   76.97$\pm$  1.57  \\
        1.13536 & 2.2mMPG/GROND & $i^\prime$     & 19.30$\pm$0.02 &   72.93$\pm$  1.57  \\
        2.06320 & 2.2mMPG/GROND & $i^\prime$     & 20.17$\pm$0.03 &   32.67$\pm$  0.83  \\
        2.07639 & 2.2mMPG/GROND & $i^\prime$     & 20.16$\pm$0.03 &   33.12$\pm$  0.78  \\
        3.06313 & 2.2mMPG/GROND & $i^\prime$     & 20.69$\pm$0.04 &   20.31$\pm$  0.69  \\
        3.07624 & 2.2mMPG/GROND & $i^\prime$     & 20.76$\pm$0.03 &   18.99$\pm$  0.60  \\
        3.55208 & 2.0mFaulkes   & $i^\prime$     & 20.90$\pm$0.10 &   16.72$\pm$  1.54  \\
        4.06666 & 2.2mMPG/GROND & $i^\prime$     & 21.35$\pm$0.06 &   11.08$\pm$  0.59  \\
        4.07855 & 2.2mMPG/GROND & $i^\prime$     & 21.37$\pm$0.04 &   10.83$\pm$  0.36  \\
        5.06072 & 2.2mMPG/GROND & $i^\prime$     & 21.79$\pm$0.05 &    7.35$\pm$  0.33  \\
        6.08295 & 2.2mMPG/GROND & $i^\prime$     & 22.17$\pm$0.09 &    5.17$\pm$  0.42  \\
       14.09933 & 2.2mMPG/GROND & $i^\prime$     & $>23.72$       & $<   1.25$          \\
\hline
        0.47510 & 2.0mFaulkes   & $z^\prime$     & 18.19$\pm$0.08 &  200.08$\pm$ 14.76  \\
        0.56468 & 2.0mFaulkes   & $z^\prime$     & 18.35$\pm$0.10 &  172.66$\pm$ 15.93  \\
        1.05378 & 2.2mMPG/GROND & $z^\prime$     & 19.08$\pm$0.08 &   87.77$\pm$  6.99  \\
        1.06005 & 2.2mMPG/GROND & $z^\prime$     & 19.02$\pm$0.04 &   93.27$\pm$  3.36  \\
        1.06869 & 2.2mMPG/GROND & $z^\prime$     & 18.98$\pm$0.03 &   96.83$\pm$  3.07  \\
        1.08363 & 2.2mMPG/GROND & $z^\prime$     & 19.08$\pm$0.04 &   87.77$\pm$  3.21  \\
        1.09443 & 2.2mMPG/GROND & $z^\prime$     & 19.04$\pm$0.03 &   91.73$\pm$  2.91  \\
        1.12178 & 2.2mMPG/GROND & $z^\prime$     & 19.03$\pm$0.03 &   91.95$\pm$  2.80  \\
        1.13536 & 2.2mMPG/GROND & $z^\prime$     & 19.09$\pm$0.03 &   87.65$\pm$  2.78  \\
        2.06977 & 2.2mMPG/GROND & $z^\prime$     & 19.93$\pm$0.03 &   40.41$\pm$  1.28  \\
        3.06975 & 2.2mMPG/GROND & $z^\prime$     & 20.44$\pm$0.04 &   25.08$\pm$  0.88  \\
        4.07206 & 2.2mMPG/GROND & $z^\prime$     & 21.13$\pm$0.04 &   13.30$\pm$  0.52  \\
        5.06072 & 2.2mMPG/GROND & $z^\prime$     & 21.61$\pm$0.05 &    8.57$\pm$  0.43  \\
        6.08295 & 2.2mMPG/GROND & $z^\prime$     & 22.38$\pm$0.15 &    4.22$\pm$  0.64  \\
       14.09933 & 2.2mMPG/GROND & $z^\prime$     & $>22.93$       & $<   2.55$          \\
\hline
        1.05761 & 2.2mMPG/GROND & {\it J}     & 18.73$\pm$0.05 &  119.29$\pm$  6.05  \\
        1.07086 & 2.2mMPG/GROND & {\it J}     & 18.68$\pm$0.05 &  125.77$\pm$  6.03  \\
        1.08404 & 2.2mMPG/GROND & {\it J}     & 18.82$\pm$0.07 &  109.80$\pm$  7.70  \\
        1.09446 & 2.2mMPG/GROND & {\it J}     & 18.75$\pm$0.05 &  117.54$\pm$  5.64  \\
        1.12181 & 2.2mMPG/GROND & {\it J}     & 18.82$\pm$0.04 &  109.90$\pm$  4.46  \\
        1.13539 & 2.2mMPG/GROND & {\it J}     & 18.82$\pm$0.05 &  110.40$\pm$  5.09  \\
        2.06982 & 2.2mMPG/GROND & {\it J}     & 19.60$\pm$0.06 &   53.73$\pm$  3.22  \\
        3.06981 & 2.2mMPG/GROND & {\it J}     & 20.16$\pm$0.10 &   32.14$\pm$  2.99  \\
        4.07313 & 2.2mMPG/GROND & {\it J}     & 20.43$\pm$0.11 &   25.04$\pm$  2.61  \\
        5.06076 & 2.2mMPG/GROND & {\it J}     & 20.97$\pm$0.18 &   15.21$\pm$  2.80  \\
        6.08299 & 2.2mMPG/GROND & {\it J}     & $>21.16$       & $<  12.71$          \\
       14.09937 & 2.2mMPG/GROND & {\it J}     & $>21.66$       & $<   8.07$          \\
\hline
        1.05761 & 2.2mMPG/GROND & {\it H}     & 18.50$\pm$0.07 &  147.00$\pm$  9.08  \\
        1.07086 & 2.2mMPG/GROND & {\it H}     & 18.45$\pm$0.06 &  153.64$\pm$  8.35  \\
        1.08404 & 2.2mMPG/GROND & {\it H}     & 18.52$\pm$0.08 &  143.78$\pm$ 10.87  \\
        1.09446 & 2.2mMPG/GROND & {\it H}     & 18.42$\pm$0.05 &  158.53$\pm$  8.03  \\
        1.12181 & 2.2mMPG/GROND & {\it H}     & 18.60$\pm$0.05 &  133.63$\pm$  6.77  \\
        1.13539 & 2.2mMPG/GROND & {\it H}     & 18.60$\pm$0.06 &  133.63$\pm$  7.51  \\
        2.06982 & 2.2mMPG/GROND & {\it H}     & 19.44$\pm$0.09 &   61.56$\pm$  5.11  \\
        3.06981 & 2.2mMPG/GROND & {\it H}     & 19.96$\pm$0.12 &   38.36$\pm$  4.60  \\
        4.07313 & 2.2mMPG/GROND & {\it H}     & $>20.17$       & $<  31.53$          \\
        5.06076 & 2.2mMPG/GROND & {\it H}     & $>20.53$       & $<  22.67$          \\
        6.08299 & 2.2mMPG/GROND & {\it H}     & $>20.56$       & $<  21.96$          \\
       14.09937 & 2.2mMPG/GROND & {\it H}     & $>21.13$       & $<  12.96$          \\
\hline
        1.05761 & 2.2mMPG/GROND & $K_S$     & 18.18$\pm$0.07 &  195.43$\pm$ 13.51  \\
        1.07086 & 2.2mMPG/GROND & $K_S$     & 18.23$\pm$0.09 &  186.29$\pm$ 15.46  \\
        1.08404 & 2.2mMPG/GROND & $K_S$     & 18.34$\pm$0.18 &  168.88$\pm$ 30.49  \\
        1.09446 & 2.2mMPG/GROND & $K_S$     & 18.22$\pm$0.10 &  188.44$\pm$ 17.55  \\
        1.13539 & 2.2mMPG/GROND & $K_S$     & 18.30$\pm$0.10 &  175.87$\pm$ 17.03  \\
        2.06982 & 2.2mMPG/GROND & $K_S$     & 19.04$\pm$0.13 &   89.04$\pm$ 11.76  \\
        3.06981 & 2.2mMPG/GROND & $K_S$     & 19.50$\pm$0.20 &   58.07$\pm$ 11.84  \\
        4.07313 & 2.2mMPG/GROND & $K_S$     & $>19.73$       & $<  47.03$          \\
        5.06076 & 2.2mMPG/GROND & $K_S$     & $>20.11$       & $<  33.23$          \\
        6.08299 & 2.2mMPG/GROND & $K_S$     & $>19.93$       & $<  39.17$          \\
       14.09937 & 2.2mMPG/GROND & $K_S$     & $>19.99$       & $<  37.12$          \\
\hline
        291.792 & 0.8m{\it Spitzer}/IRAC & 3.6~$\mu$m    & $>25.3$ & $ < 0.28$                 \\
\hline
\end{longtable}
}

\longtab[2]{
\begin{longtable}{cccc}
\caption{Equivalent widths of the spectral features measured in the X-shooter spectrum. This list does not include features within the Lyman-$\alpha$ forest.}
\label{tab:ew}\\
\hline\hline
Wavelength&   Feature                   & $z$       & EW                   \\
 ({\AA})  &                             &           & ({\AA})              \\
\hline
\endfirsthead
\caption{continued.}\\
\hline\hline
Wavelength&   Feature                   & $z$       & EW                    \\
 ({\AA})  &                             &           & ({\AA})               \\
\hline
\endhead
\hline
\endfoot
   4531.1 & \ion{C}{IV}$\lambda$1548.20   & 1.92669  &   0.356$\pm$ 0.035   \\
   4538.6 & \ion{C}{IV}$\lambda$1550.77   & 1.92668  &   0.286$\pm$ 0.029   \\
   4544.8 & \ion{C}{II}$\lambda$1334.53   & 2.40550  &   0.310$\pm$ 0.026   \\
   4595.1 & \ion{N}{V}$\lambda$1238.82    & 2.70937  &   0.277$\pm$ 0.017   \\
   4609.9 & \ion{N}{V}$\lambda$1242.80    & 2.70928  &   0.147$\pm$ 0.014   \\
   4640.2 & \ion{S}{II}$\lambda$1250.58   & 2.70970  &   0.134$\pm$ 0.015   \\
   4651.7 & \ion{S}{II}$\lambda$1253.52   & 2.71090  &   0.369$\pm$ 0.023   \\
   4675.9 & \ion{S}{II}$\lambda$1259.52   & 2.70970  &   4.063$\pm$ 0.019   \\
          & \ion{Si}{II}$\lambda$1260.42  & 2.70970  &                      \\
   4691.8 & \ion{Si}{II*}$\lambda$1264.74 &  2.70970 &   4.107$\pm$ 0.017   \\
   4744.6 & \ion{Si}{IV}$\lambda$1393.76  &  2.40420 &   0.391$\pm$ 0.021   \\
   4774.5 & \ion{Si}{IV}$\lambda$1402.77  &  2.40366 &   0.120$\pm$ 0.017   \\
   4831.1 & \ion{O}{I}$\lambda$1302.17    &  2.71008 &   2.832$\pm$ 0.023    \\
   4840.4 & \ion{Si}{II}$\lambda$1304.37  &  2.70970 &   3.316$\pm$ 0.032    \\
          & \ion{O}{I*}$\lambda$1304.86    &  2.70970 &                       \\
   4857.0 & \ion{Si}{II*}$\lambda$1309.28 &  2.70964 &   1.224$\pm$ 0.022    \\
   4887.4 & \ion{Ni}{II}$\lambda$1317.22  &  2.71041 &   0.076$\pm$ 0.014    \\
   4951.7 & \ion{C}{II}$\lambda$1334.53   &  2.70970 &   6.403$\pm$ 0.017    \\
          & \ion{C}{II*}$\lambda$1335.66  &  2.70970 &                       \\
          & \ion{C}{II*}$\lambda$1335.71  &  2.70970 &                       \\
   5002.5 & \ion{C}{IV}$\lambda$1548.20   &  2.23116 &   0.125$\pm$ 0.015    \\
   5010.6 & \ion{C}{IV}$\lambda$1550.77   &  2.23107 &   0.092$\pm$ 0.014    \\
   5083.8 & \ion{Ni}{II}$\lambda$1370.13  &  2.71044 &   0.061$\pm$ 0.014    \\
   5094.1 & \ion{C}{IV}$\lambda$1548.20   &  2.29032 &   0.136$\pm$ 0.015    \\
   5101.7 & \ion{Si}{IV}$\lambda$1393.76  &  2.66039 &   0.293$\pm$ 0.016    \\
   5125.2 & \ion{C}{IV}$\lambda$1548.20   &  2.31041 &   0.090$\pm$ 0.014    \\
   5134.2 & \ion{C}{IV}$\lambda$1550.77   &  2.31076 &   0.221$\pm$ 0.018    \\
          & \ion{Si}{IV}$\lambda$1402.77  &  2.66004 &                       \\
   5169.7 & \ion{Si}{IV}$\lambda$1393.76  &  2.70921 &   2.505$\pm$ 0.020    \\
   
   5203.2 & \ion{Si}{II}$\lambda$1526.71  &  2.40540 &   2.221$\pm$ 0.026    \\
          & \ion{Si}{IV}$\lambda$1402.77  &  2.70920 &                       \\
   5258.7 & \ion{Fe}{II}$\lambda$2344.21  &  1.24325 &   0.478$\pm$ 0.018    \\
   5270.2 & \ion{C}{IV}$\lambda$1548.20   &  2.40407 &   0.634$\pm$ 0.019    \\
   5278.7 & \ion{C}{IV}$\lambda$1550.77   &  2.40389 &   0.321$\pm$ 0.020    \\
   5326.5 & \ion{Fe}{II}$\lambda$2374.46  &  1.24324 &   0.149$\pm$ 0.015    \\
   5461.3 & \ion{C}{IV}$\lambda$1548.20   &  2.52752 &   0.129$\pm$ 0.018    \\
   5470.6 & \ion{C}{IV}$\lambda$1550.77   &  2.52767 &   0.079$\pm$ 0.018    \\
   5558.5 & \ion{Fe}{II}$\lambda$2600.17  &  1.13774 &   0.158$\pm$ 0.025    \\
\hline
   5664.4 & \ion{Si}{II}$\lambda$1526.71  &  2.70970 &   4.119$\pm$ 0.032    \\
          & \ion{C}{IV}$\lambda$1548.20   & 2.6590   &                      \\                    
          & \ion{C}{IV}$\lambda$1550.77   & 2.6590   &                      \\                     
   5688.5 & \ion{Si}{II*}$\lambda$1533.43 &  2.70966 &   2.084$\pm$ 0.029   \\
   5721.0 & \ion{C}{IV}$\lambda$1548.20   &  2.69526 &   0.171$\pm$ 0.022  \\
   5730.5 & \ion{C}{IV}$\lambda$1550.77   &  2.69527 &   0.085$\pm$ 0.017  \\
   5742.6 & \ion{C}{IV}$\lambda$1548.20   &  2.70920 &   3.004$\pm$ 0.044  \\
   5752.1 & \ion{C}{IV}$\lambda$1550.77   &  2.70919 &   2.475$\pm$ 0.040  \\
   5802.4 & \ion{Fe}{II}$\lambda$2586.65  &  1.24321 &   0.371$\pm$ 0.032  \\
   5832.9 & \ion{Fe}{II}$\lambda$2600.17  &  1.24326 &   0.972$\pm$ 0.034  \\
   5892.2 & \ion{Na}{I}$\lambda$5891.58   &  0.00010 &   0.140$\pm$ 0.029  \\
   5967.6 & \ion{Fe}{II}$\lambda$1608.45  &  2.71018 &   1.375$\pm$ 0.037  \\
   5978.2 & \ion{Mg}{II}$\lambda$2796.35  &  1.13784 &   0.877$\pm$ 0.032  \\
   5993.4 & \ion{Mg}{II}$\lambda$2803.53  &  1.13780 &   0.485$\pm$ 0.030  \\
   6019.1 & \ion{Al}{III}$\lambda$1862.79 &  2.23125 &   0.042$\pm$ 0.015  \\
   6042.1 & \ion{Fe}{II2s}$\lambda$1629.16 &  2.70870 &   0.051$\pm$ 0.019  \\
   6044.7 & \ion{Fe}{II2s}$\lambda$1629.16 &  2.71032 &   0.116$\pm$ 0.023  \\
   6072.1 & \ion{Fe}{II5s}$\lambda$1637.40&  2.70838 &   0.047$\pm$ 0.014  \\
   6198.0 & \ion{Al}{II}$\lambda$1670.79  &  2.70964 &   3.834$\pm$ 0.030  \\
   6273.1 & \ion{Mg}{II}$\lambda$2796.35  &  1.24330 &   4.012$\pm$ 0.038  \\
   6288.9 & \ion{Mg}{II}$\lambda$2803.53  &  1.24322 &   3.055$\pm$ 0.035  \\
   6311.5 & \ion{Fe}{II5s}$\lambda$1702.04 &  2.70820 &   0.061$\pm$ 0.016  \\
   6315.0 & \ion{Fe}{II5s}$\lambda$1702.04 &  2.71025 &   0.232$\pm$ 0.021  \\
   6343.7 & \ion{Ni}{II}$\lambda$1709.60 &  2.71063 &   0.049$\pm$ 0.016  \\
   6399.3 & \ion{Mg}{I}$\lambda$2852.96   &  1.24304 &   0.497$\pm$ 0.023  \\
   6704.8 & \ion{Si}{II}$\lambda$1808.01  &  2.70836 &   0.030$\pm$ 0.013  \\
   6708.5 & \ion{Si}{II}$\lambda$1808.01  &  2.71042 &   0.281$\pm$ 0.020  \\
   7514.8 & \ion{Zn}{II}$\lambda$2026.14  &  2.7084 &   0.066$\pm$ 0.013  \\
          & \ion{Cr}{II}$\lambda$2026.27   & 2.7084 &                      \\   
          & \ion{Mg}{I}$\lambda$2026.48   & 2.7084 &                      \\  
   7516.3 & \ion{Zn}{II}$\lambda$2026.14  &  2.7097 &   0.045$\pm$ 0.014  \\
          & \ion{Cr}{II}$\lambda$2026.27   & 2.7097 &                      \\   
          & \ion{Mg}{I}$\lambda$2026.48   & 2.7097 &                      \\   
   7518.3 & \ion{Zn}{II}$\lambda$2026.14  &  2.7105 &   0.078$\pm$ 0.015  \\
          & \ion{Cr}{II}$\lambda$2026.27   & 2.7105 &                      \\   
          & \ion{Mg}{I}$\lambda$2026.48   & 2.7105 &                      \\    
   7983.6 & \ion{Fe}{II}$\lambda$2344.21  &  2.40567 &   0.088$\pm$ 0.014  \\
   8037.6 & \ion{Ni}{II*}$\lambda$2166.23  &  2.71041 &   0.329$\pm$ 0.019  \\
   8114.7 & \ion{Fe}{II}$\lambda$2382.76  &  2.40558 &   0.232$\pm$ 0.012  \\
   8388.0 & \ion{Fe}{II}$\lambda$2260.78  &  2.71020 &   0.251$\pm$ 0.023  \\
   8591.3 & \ion{Ni}{II*}$\lambda$2316.70 &  2.70841 &   0.061$\pm$ 0.012  \\
   8596.0 & \ion{Ni}{II*}$\lambda$2316.70 &  2.71043 &   0.396$\pm$ 0.017  \\
   8656.0 & \ion{Fe}{II*}$\lambda$2333.52 &  2.70944 &   0.709$\pm$ 0.032  \\
   8661.7 & \ion{Fe}{II}$\lambda$2344.21  &  2.70970 &   0.040$\pm$ 0.013  \\
          & \ion{Fe}{II4s}$\lambda$2345.00&  2.70970 &                     \\
          
   8672.6 & \ion{Fe}{II3s}$\lambda$2338.72  &  2.70824 &   0.075$\pm$ 0.018  \\
   8676.0 & \ion{Fe}{II3s}$\lambda$2338.72  &  2.70972 &   0.088$\pm$ 0.017  \\
   8678.3 & \ion{Fe}{II3s}$\lambda$2338.72  &  2.71067 &   0.087$\pm$ 0.017  \\
          
   8714.2 & \ion{Fe}{II*}$\lambda$2349.02&  2.70973 &   0.723$\pm$ 0.026  \\
   8756.4 & \ion{Fe}{II3s}$\lambda$2359.83&  2.71060 &   0.243$\pm$ 0.023  \\
   8772.4 & \ion{Fe}{II*}$\lambda$2365.55 &  2.70838 &   0.057$\pm$ 0.011  \\
   8776.5 & \ion{Fe}{II*}$\lambda$2365.55 &  2.71014 &   0.319$\pm$ 0.024  \\
   8809.6 & \ion{Fe}{II}$\lambda$2374.46  &  2.71016 &   1.957$\pm$ 0.022  \\
   8836.3 & \ion{Fe}{II*}$\lambda$2381.49  &  2.71039 &   0.925$\pm$ 0.076  \\
   8840.6 & \ion{Fe}{II}$\lambda$2382.76  &  2.71024 &   0.925$\pm$ 0.076  \\
   8855.1 & \ion{Fe}{II}$\lambda$2600.17  &  2.40560 &   0.181$\pm$ 0.014  \\
   8860.9 & \ion{Fe}{II*}$\lambda$2389.36 &  2.70850 &   0.095$\pm$ 0.015  \\
   8865.3 & \ion{Fe}{II*}$\lambda$2389.36 &  2.71033 &   0.439$\pm$ 0.022  \\
   8891.5 & \ion{Fe}{II*}$\lambda$2396.36 &  2.71043 &   1.137$\pm$ 0.362  \\
   8901.8 & \ion{Fe}{II*}$\lambda$2399.98 &  2.70830 &   0.347$\pm$ 0.076  \\
   8932.0 & \ion{Fe}{II3s}$\lambda$2407.39 &  2.71025 &   0.228$\pm$ 0.018  \\
   8946.2 & \ion{Fe}{II3s}$\lambda$2411.25 &  2.70970 &   0.702$\pm$ 0.004  \\
          & \ion{Fe}{II4s}$\lambda$2411.80 &  2.70970 &                     \\
          & \ion{Fe}{II3s}$\lambda$2414.60 &  2.70970 &                     \\
   9596.2 & \ion{Fe}{II}$\lambda$2586.65  &  2.70991 &   3.973$\pm$ 0.043  \\
   9607.1 & \ion{Fe}{II}$\lambda$2600.17  &  2.69480 &   0.427$\pm$ 0.020  \\
   9639.0 & \ion{Fe}{II*}$\lambda$2599.15 &  2.70970 &   0.714$\pm$ 0.025  \\
          & \ion{Fe}{II}$\lambda$2600.17  &  2.70970 &                     \\
   9694.0 & \ion{Fe}{II*}$\lambda$2612.65  &  2.71042 &   0.690$\pm$ 0.035  \\
   9745.1 & \ion{Fe}{II*}$\lambda$2626.45  &  2.71038 &   0.489$\pm$ 0.050  \\
   9753.1 & \ion{Fe}{II3s}$\lambda$2629.08 &  2.70971 &   0.054$\pm$ 0.018  \\
   9755.7 & \ion{Fe}{II4s}$\lambda$2629.08 &  2.71070 &   0.226$\pm$ 0.023  \\
   9759.8 & \ion{Fe}{II3s}$\lambda$2631.83 &  2.70836 &   0.282$\pm$ 0.025  \\
   9766.1 & \ion{Fe}{II3s}$\lambda$2631.83 &  2.71076 &   1.189$\pm$ 0.056  \\
\hline
  10372.3 & \ion{Mg}{II}$\lambda$2796.35  &  2.70921 &   8.881$\pm$ 0.216  \\
  10398.5 & \ion{Mg}{II}$\lambda$2803.50  &  2.70921 &   6.208$\pm$ 0.133  \\
  10582.1 & \ion{Mg}{I}$\lambda$2852.96   &  2.70918 &   2.437$\pm$ 0.108  \\
\hline
\end{longtable}
}

\section{Voigt profile fit of intervening systems}

In this Appendix we present the results of the Voigt profile fits to the intervening systems.

   \begin{figure}[ht]
   \centering
   \includegraphics[width=8cm]{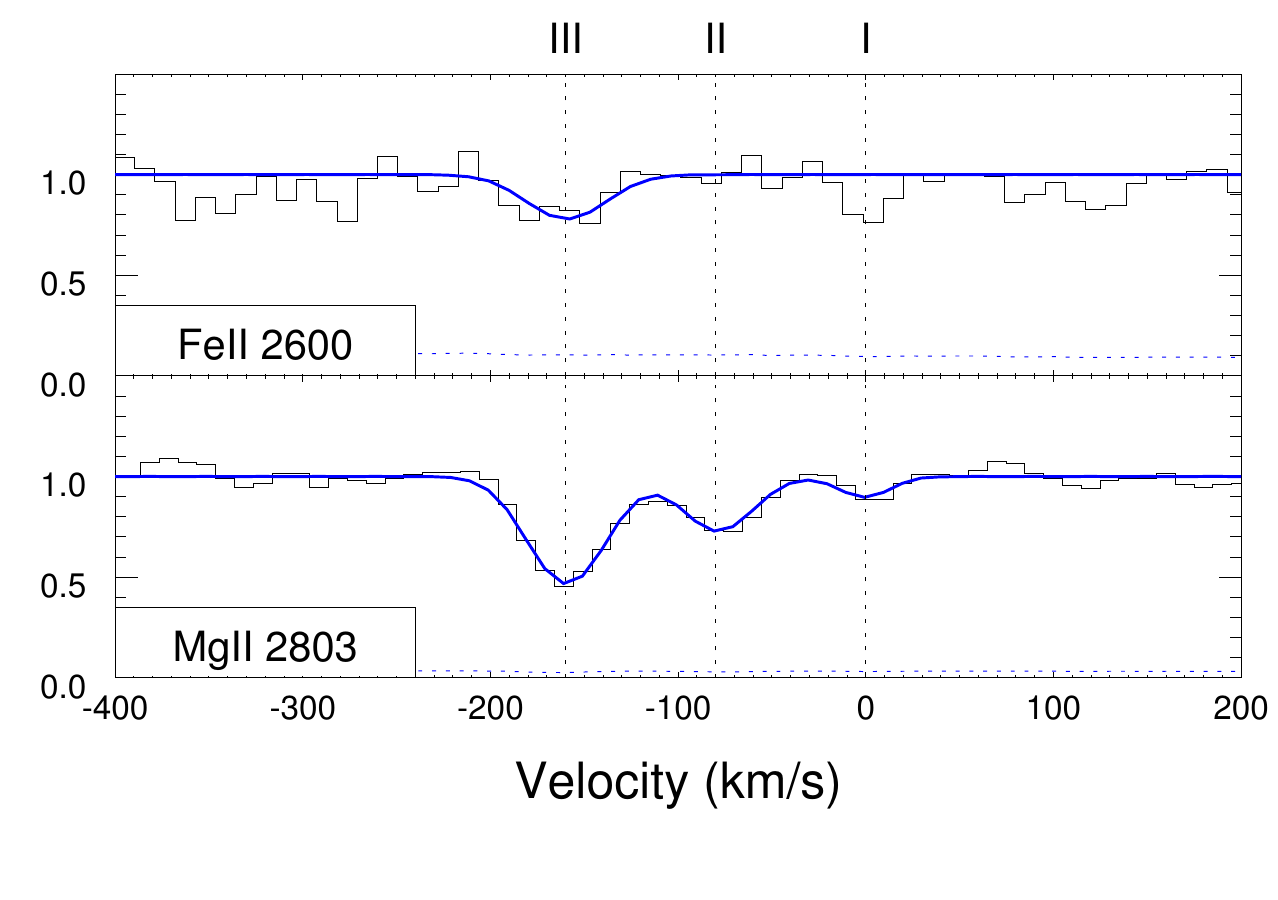}
      \caption{Intervening system at $z=1.138$. This and the following plots use the same scheme as in Fig.~\ref{Fig:voigt}, with the only difference that the fits of the intervening systems are shown in blue.}
         \label{Fig:int1}
   \end{figure}

   \begin{figure}[ht]
   \centering
   \includegraphics[width=8cm]{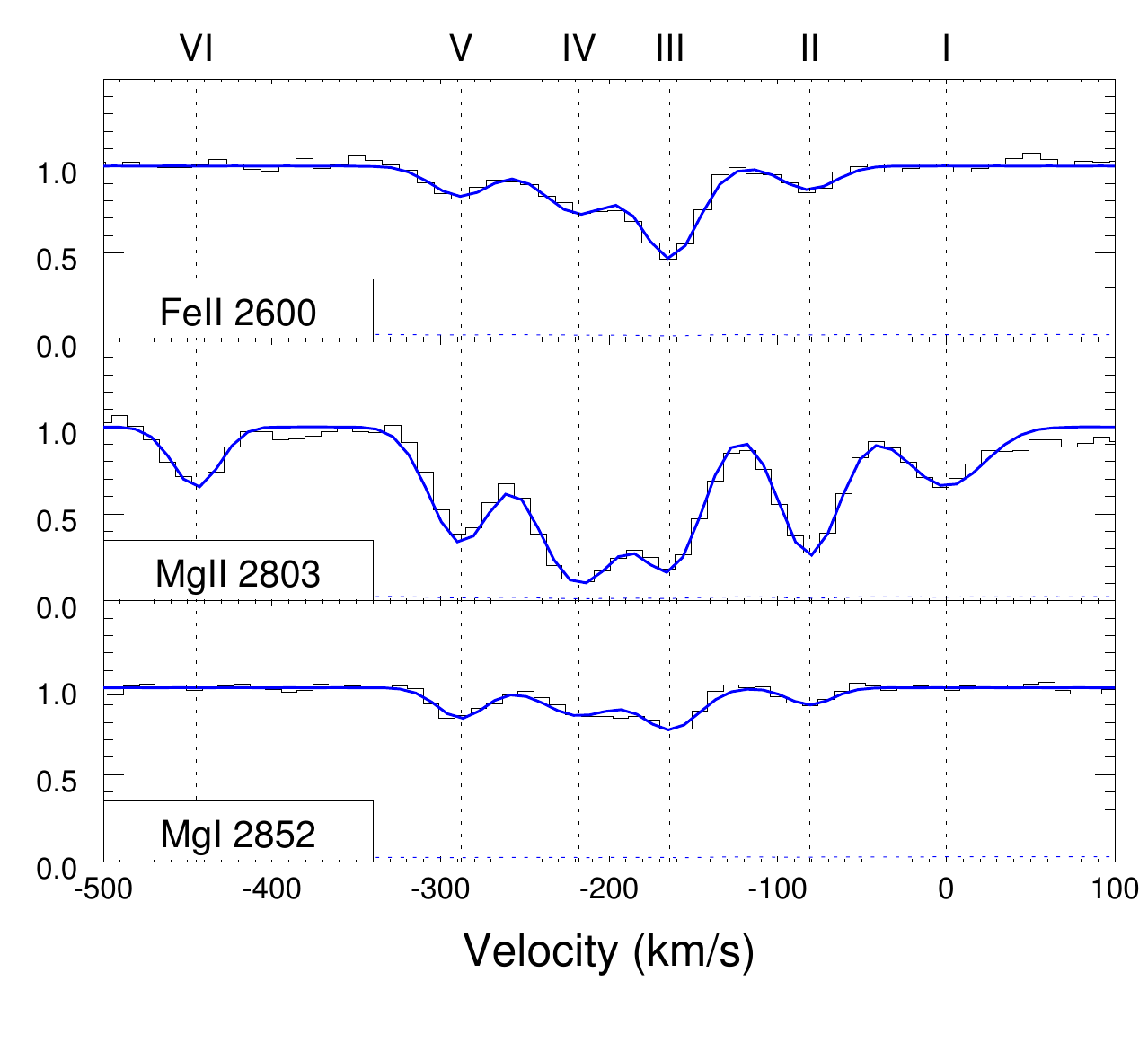}
      \caption{Intervening system at $z=1.243$.}
         \label{Fig:int2}
   \end{figure}
   
      \begin{figure}[ht]
   \centering
   \includegraphics[width=8cm]{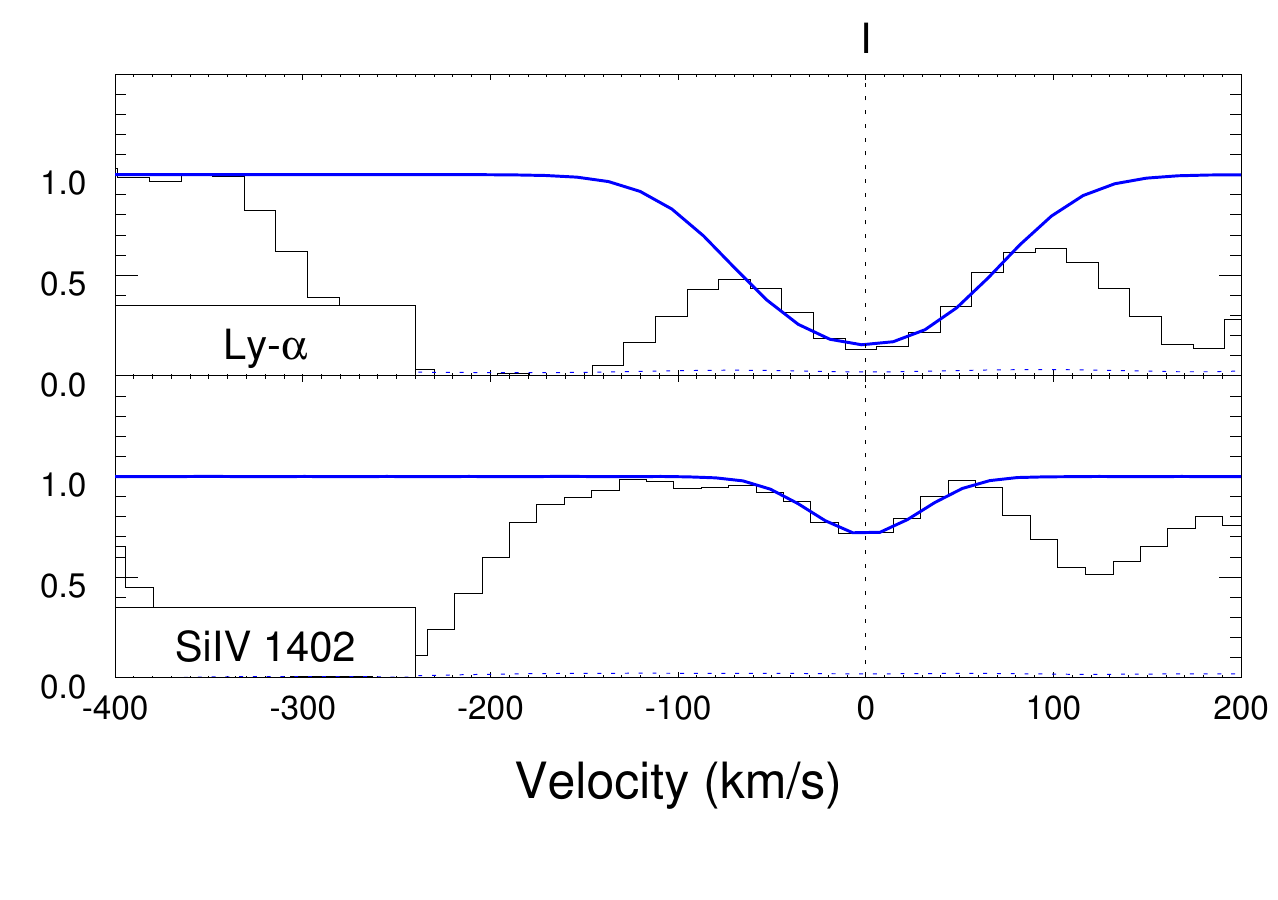}
      \caption{Intervening system at $z=1.9267$.}
         \label{Fig:int3}
   \end{figure}
   
      \begin{figure}[ht]
   \centering
   \includegraphics[width=8cm]{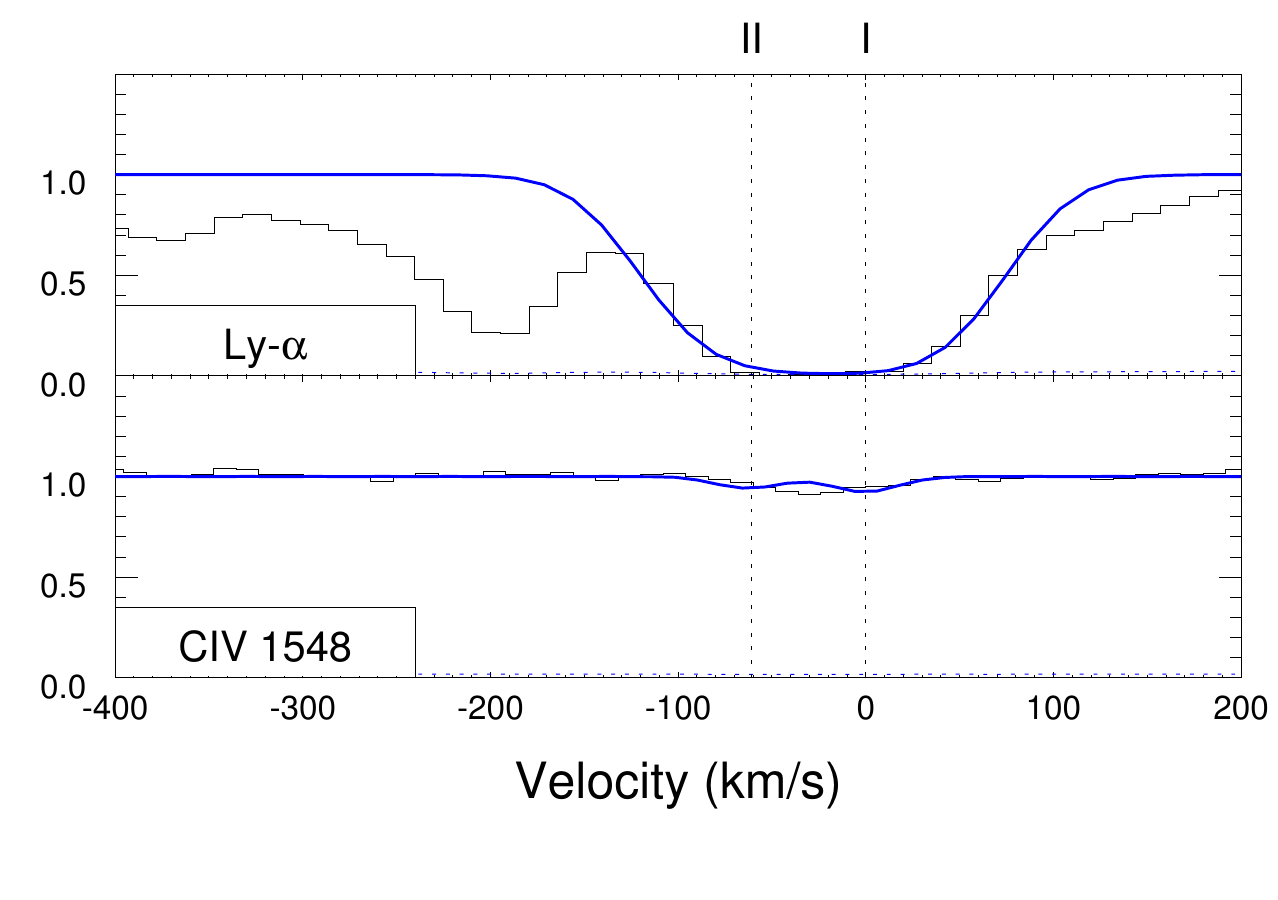}
      \caption{Intervening system at $z=2.231$.}
         \label{Fig:int4}
   \end{figure}
   
      \begin{figure}[ht]
   \centering
   \includegraphics[width=8cm]{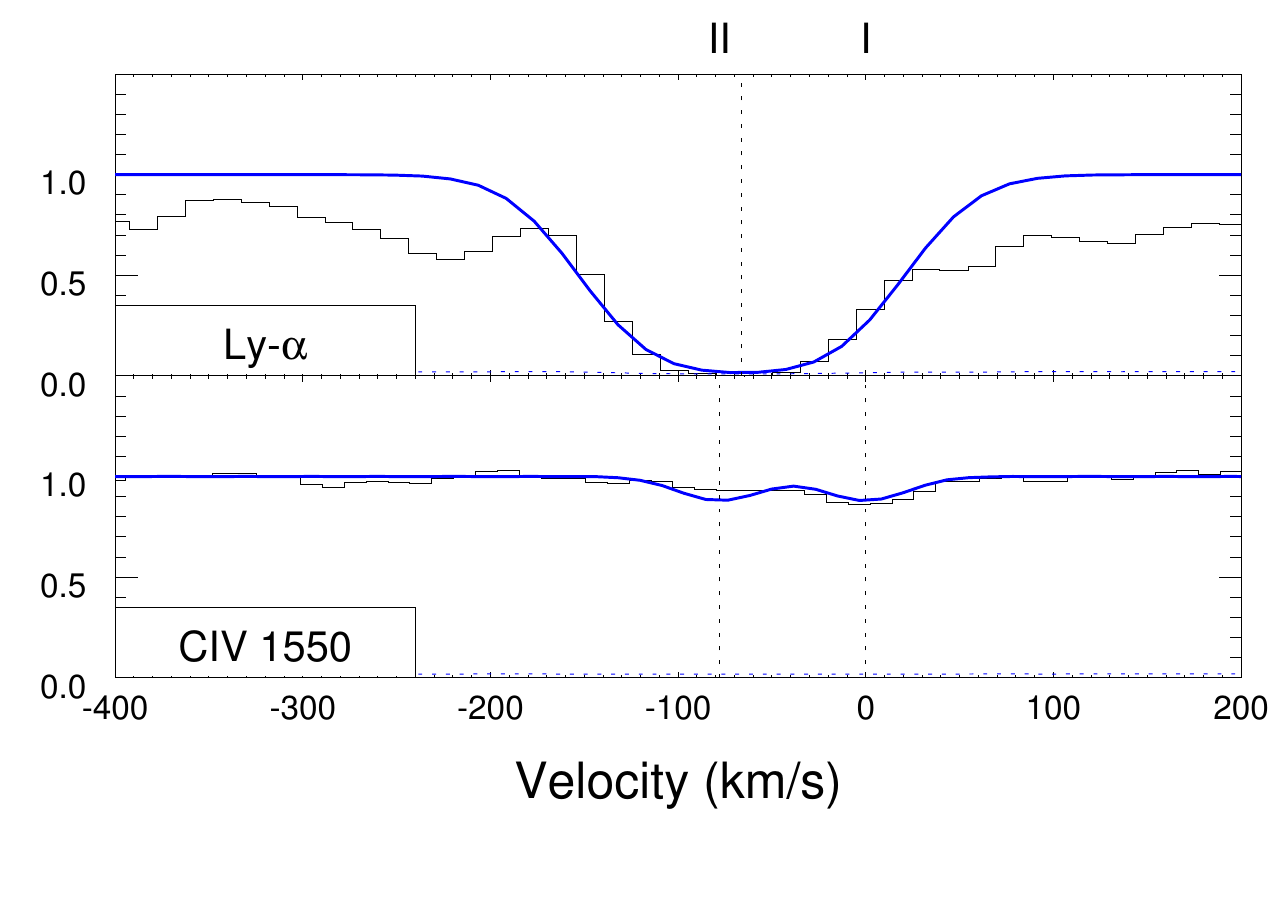}
      \caption{Intervening system at $z=2.311$.}
         \label{Fig:int5}
   \end{figure}
   
      \begin{figure}[ht]
   \centering
   \includegraphics[width=8cm]{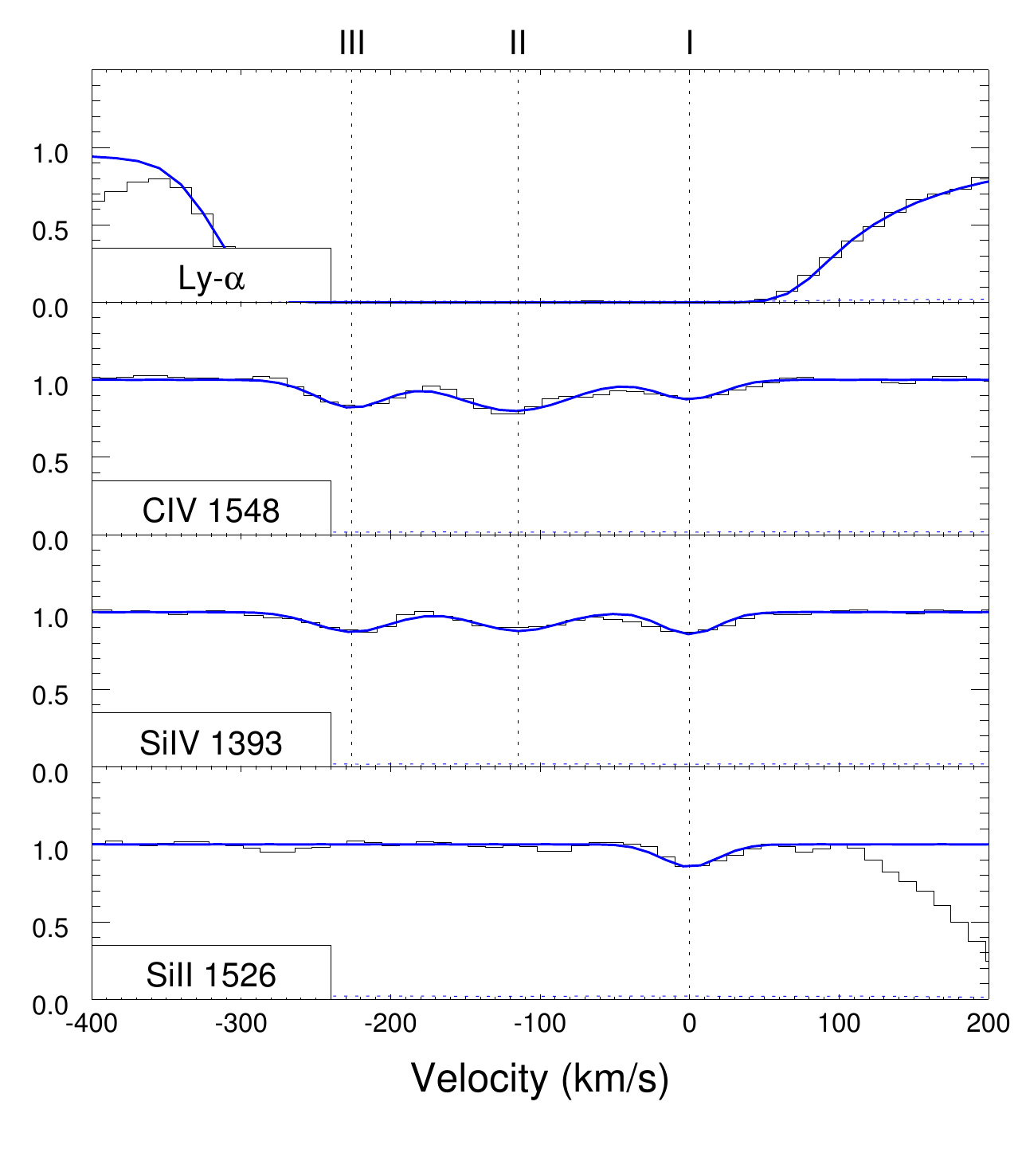}
      \caption{Intervening system at $z=2.404$.}
         \label{Fig:int6}
   \end{figure}

      \begin{figure}[ht]
   \centering
   \includegraphics[width=8cm]{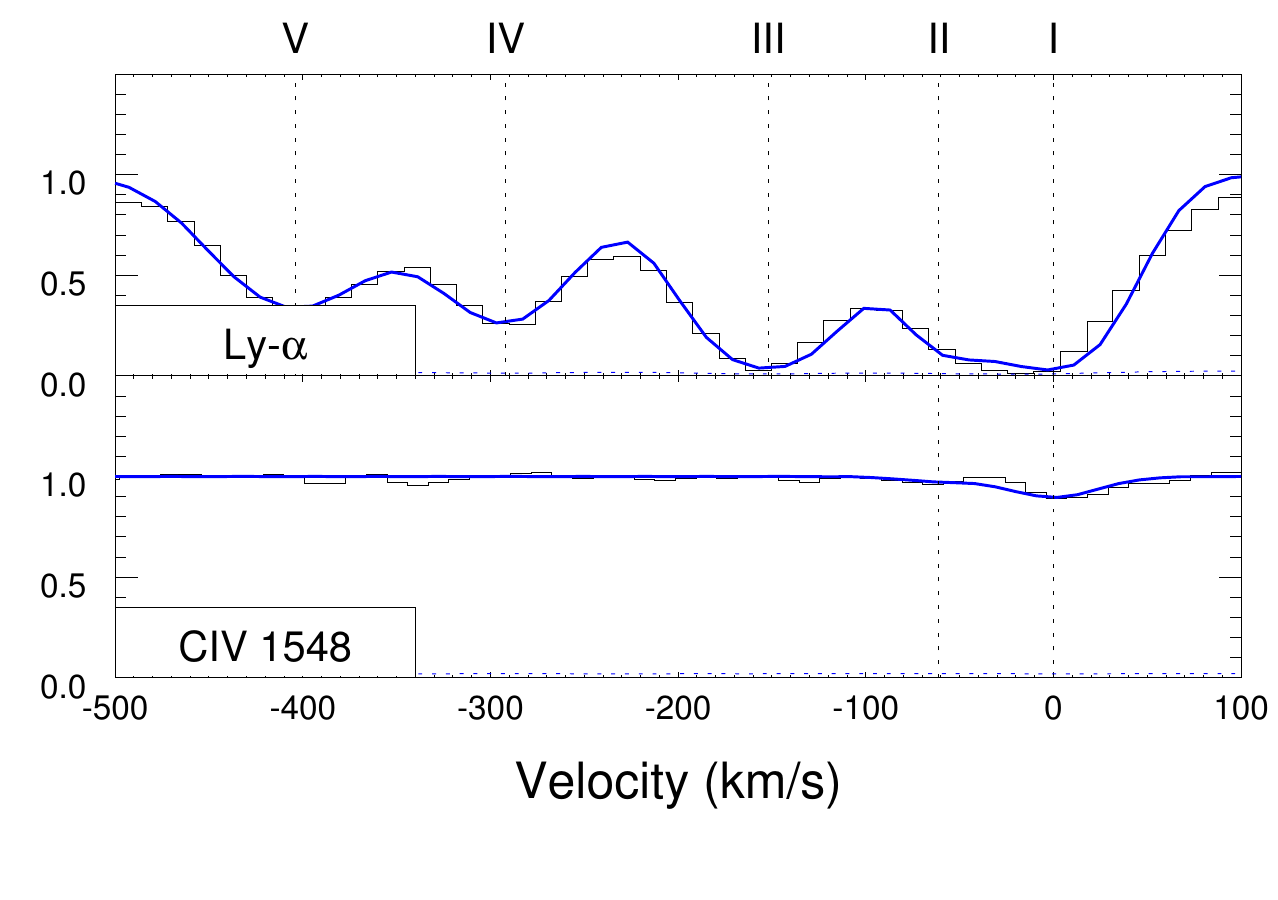}
      \caption{Intervening system at $z=2.525$.}
         \label{Fig:int7}
   \end{figure}
   
      \begin{figure}[ht]
   \centering
   \includegraphics[width=8cm]{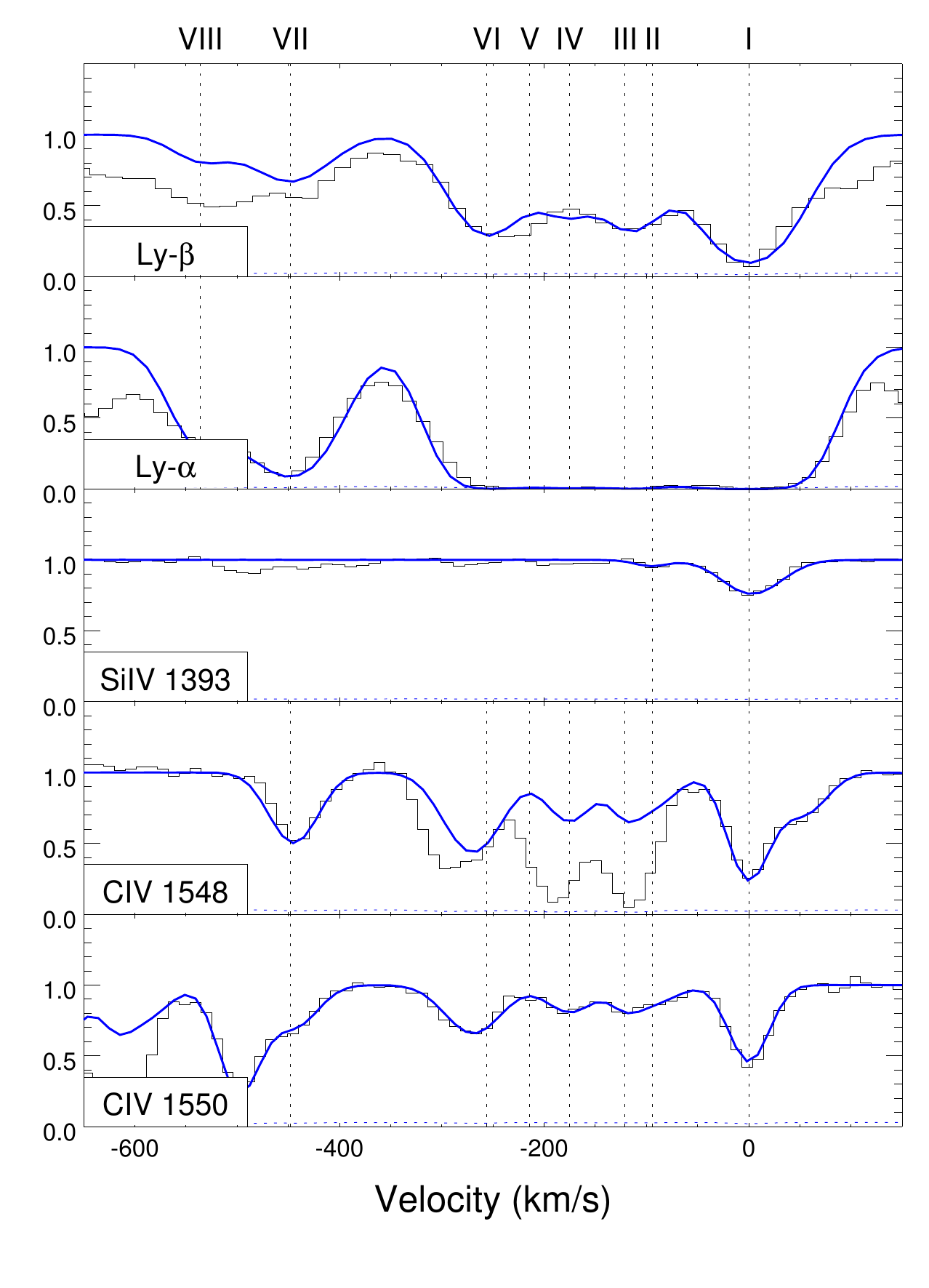}
      \caption{Intervening system at $z=2.659$. The additional absorption in \ion{C}{IV} $\lambda$ 1548 belongs to \ion{Si}{II} $\lambda$1526 in the GRB host.}
         \label{Fig:int8}
   \end{figure}
   
      \begin{figure}[ht]
   \centering
   \includegraphics[width=8cm]{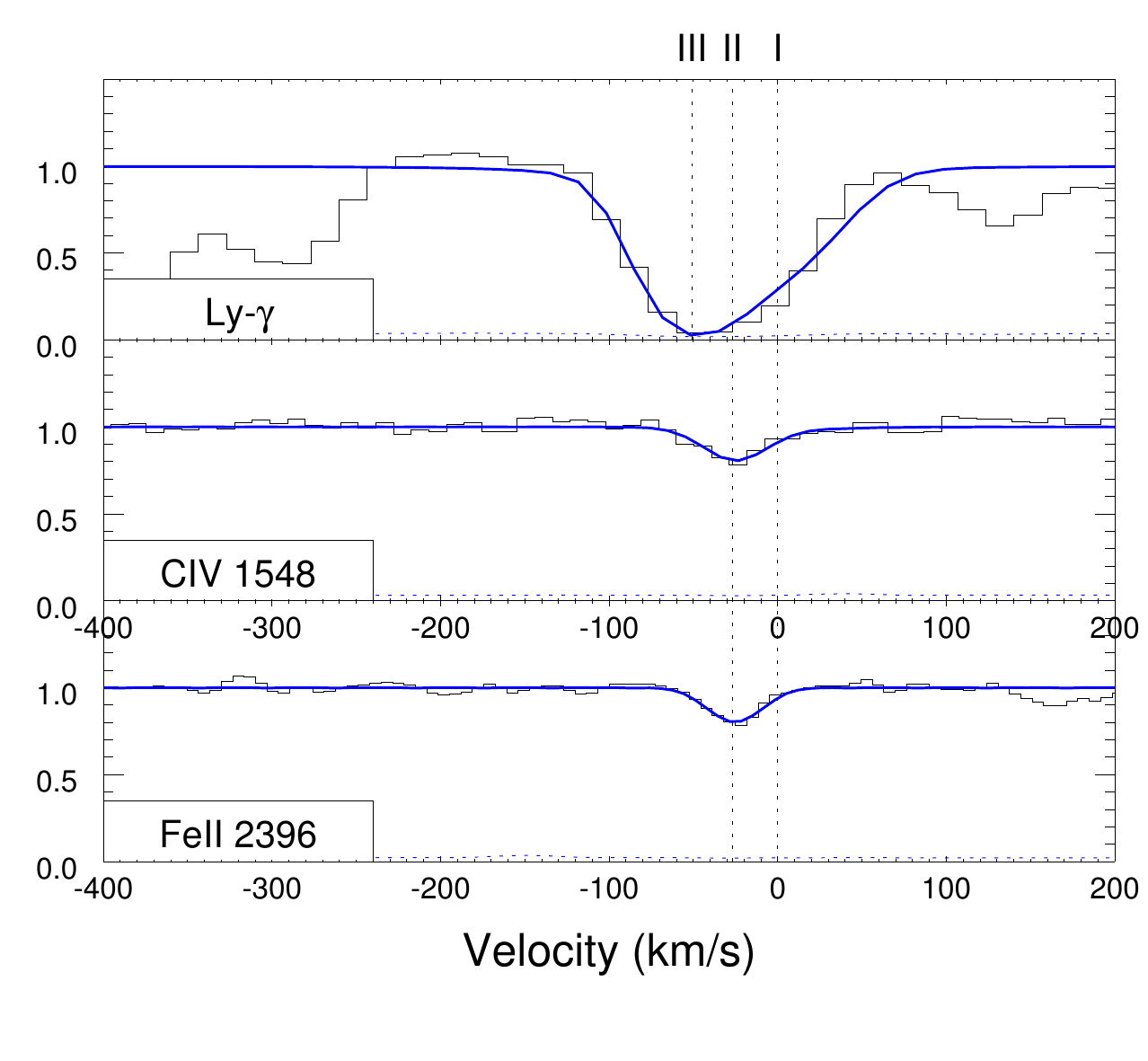}
      \caption{Intervening system at $z=2.695$.}
         \label{Fig:int9}
   \end{figure}

\begin{table*}
\caption{Column densities (N) and b-parameters from Voigt profile fits to the intervening systems along the line of sight to GRB\,161023A ordered by redshift. Different systems in the table are separated by double lines. For the fitting of the Lyman series we do not give b-parameters. We note that the fitted transitions are different in each intervening system. The \ion{C}{II} line of the $z=2.404$ system as well as the \ion{C}{IV} transitions of the $z=1.9267$ system were not fitted as they are located in the red wing of Ly$\alpha$ of the GRB system. }
\label{table:intN}     
\centering                       
\begin{tabular}{c c c c c c c c}       
\hline\hline                
Redshift & log(N/cm$^{-2}$)  & log(N/cm$^{-2}$) & b & log(N/cm$^{-2}$)  & b & log(N/cm$^{-2}$)  & b \\  
         &                   &                  & km\,s$^{-1}$ &        & km\,s$^{-1}$ &        & km\,s$^{-1}$ \\  
\hline \hline                 
 & HI  & \ion{C}{IV} & & \ion{Fe}{II}&    \\  
  & \scriptsize{Ly$\alpha$ - Ly7} & \scriptsize{1548, 1550} & &\scriptsize{2600} &  \\   \hline
 2.6955709   &  ---  & 12.28$\pm$0.55     &51 & 5.96$\pm$0.08 & 88 \\  
 2.6952579   & 15.00$\pm$0.02     &  13.08$\pm$0.07   &18 & 12.68$\pm$0.03& 13 \\
 2.6949573   & 19.19$\pm$0.02     &  ---     & --- & --- & ---  \\ 
\hline \hline
 &  HI   & CIV & & \ion{Si}{IV} & & & \\ 
  & \scriptsize{Ly$\alpha$-L$\gamma$}  & \scriptsize{1548, 1550} & & \scriptsize{1393, 1402} &  & & \\ \hline
 2.6605325  &  15.04$\pm$0.01 &  14.00$\pm$0.1     & 17 &13.01$\pm$0.02 & 32 & \\
 2.6593809   &  14.08$\pm$0.04 &  13.30$\pm$0.07    & 26  & 12.84$\pm$0.17 & 5 & \\
 2.6590521   &  14.32$\pm$0.03 &  13.14$\pm$0.09    & 10 & & & \\
 2.6583936   &  14.40$\pm$0.02 &  13.47$\pm$0.04     &24& & & \\
 2.6572421   &  14.68$\pm$0.01 &  13.84$\pm$0.02     & 31 & & & \\
 2.6551034   &  14.23$\pm$0.01 &  13.70$\pm$0.02     & 25 & & & \\
 2.6539929   &  13.80$\pm$0.00  &  ---     &  & & & \\
\hline \hline 
 &  HI   &  CIV & & & & & \\ 
  &  \scriptsize{Ly$\alpha$, Ly$\beta$} & \scriptsize{1548, 1550} & & & & & \\ \hline
 2.5275          &  14.34$\pm$0.02    & 12.92$\pm$0.02      & 24 & & & \\
 2.5268          &   13.97$\pm$0.04   & 12.18$\pm$0.21      & 15 & & & \\
 2.5257          &  14.34$\pm$0.01    & ---       &--- & & & \\
 2.5240          &  13.92$\pm$0.01    & ---       &---  & & & \\  
 2.5227          &  13.89$\pm$0.01    & ---       & --- & & & \\
\hline \hline
  & HI & \ion{C}{IV} & & \ion{Si}{IV} & & \ion{Si}{II} & \\ 
  &  \scriptsize{Ly$\alpha$, Ly$\beta$}& \scriptsize{1548, 1550} & & \scriptsize{1393, 1402} & &\scriptsize{1526}  & \\ \hline
2.4054        & 18.57$\pm$0.01  &12.71$\pm$0.03  &38 & 13.12$\pm$0.03  & 38 & 13.16$\pm$0.04 & 38 \\ 
2.4041       & 15.91$\pm$0.29&  12.66$\pm$0.03& 30 & 13.34$\pm$0.02 & 30&  --- & \\ 
2.4029        & 15.06$\pm$0.03& 12.65$\pm$0.03 & 26 & 13.20$\pm$0.02 & 26 &  --- & \\  
\hline \hline
      & HI & \ion{C}{IV} & &  & &  & \\ 
    &  \scriptsize{Ly$\alpha$}& \scriptsize{1548, 1550} & &  & &  & \\ \hline
2.3111        &                  & 13.25$\pm$0.03 & 20 &  & &  & \\ 
2.3103        & 14.57$\pm$0.01 &12.61$\pm$0.06  & 20 &  & &  & \\ 
\hline \hline
     & HI &\ion{C}{IV} & &  & &  & \\ 
    &  \scriptsize{Ly$\alpha$}& \scriptsize{1548, 1550} & &  & &  & \\ \hline
2.2313       & 14.47$\pm$0.02 & 12.83$\pm$0.06 & 5 &  & &  & \\ 
2.2306       & 14.20$\pm$0.01  &12.63$\pm$0.06 & 5 & &  &   &\\ 
\hline \hline
     & HI & \ion{C}{IV} & & \ion{Si}{IV} & &  & \\ 
    &  \scriptsize{Ly$\alpha$}& \scriptsize{1548, 1550} & & \scriptsize{1402}  & &  & \\ \hline
1.9267      & 14.23$\pm$0.01  & (not fitted) &  &13.40$\pm$0.02 & 32 &   &\\ 
\hline \hline
            & & \ion{Mg}{II} & & \ion{Mg}{I} & &\ion{Fe}{II}  & \\ 
            & & \scriptsize{2796, 2803} & & \scriptsize{2852} & & \scriptsize{2344, 2372, 2383, 2600} & \\\hline
1.2446    & &  12.70$\pm$0.02   & 24  &---  &---   &---   &--- \\ 
1.2440    & &  13.32$\pm$0.03   & 12 & 11.40$\pm$0.26  & 7  &12.56$\pm$0.06  &15 \\ 
1.2434    & & 13.52$\pm$0.04 & 12 & 11.93$\pm$0.03  & 21 & 13.35$\pm$0.02 & 20\\ 
1.2430    & & 13.51$\pm$0.01 & 20 & 11.80$\pm$0.05 & 14 & 12.99$\pm$0.03 & 11\\ 
1.2425    & & 13.10$\pm$0.01 & 15 & 11.75$\pm$0.06 & 5 & 12.71$\pm$0.05 & 12\\ 
 1.2413     & & 12.92$\pm$0.11 & 5 & ---  &---  & ---  &---  \\ 
\hline\hline
         & & \ion{Mg}{II} & & \ion{Fe}{II} & &     & \\
         & & \scriptsize{2796, 2803}& & \scriptsize{2600} & &     & \\ \hline
1.1387   & & 11.47$\pm$0.23 & 10 & ---  &---  &  & \\ 
1.1382   & &  12.52$\pm$0.03& 17 & ---  & --- &  & \\ 
1.1376   & &  12.94$\pm$0.01& 20 & 12.90$\pm$0.13 & 22 &  & \\ 
\hline                                
\end{tabular}
\end{table*}

\end{appendix}

\end{document}